\documentclass{osa-article}
\journal{osajournal}
\pdfoutput=1

\usepackage{amsmath,amssymb}
\usepackage{stackengine}
\usepackage{booktabs}
\usepackage{multirow}
\usepackage{cancel}
\usepackage{subfigure} %
\usepackage{listings}
\usepackage{tabularx}

\articletype{Review}

\begin{document}

\title{Smart Cameras}

\author{David Brady\authormark{1,2}, Minghao Hu\authormark{1}, Chengyu Wang\authormark{1}, Xuefei Yan\authormark{2}, Lu Fang\authormark{3}, Yinheng Zhu\authormark{3}, Yang Tan\authormark{3}, Ming Cheng\authormark{4} and Zhan Ma\authormark{4}}

\address{\authormark{1}Department of Electrical and Computer Engineering, P.O. Box 90291, Duke University, Durham, NC 27705\\
\authormark{2}Camputer Laboratory, 1699 South Zuchongzhi RD., Kunshan, Jiangsu 215347, China\\
\authormark{3}Tsinghua-Berkeley Shenzhen Institute, Tsinghua University,  Shenzhen, Guangdong 518055 P.R. China\\
\authormark{4}School of Electronic Science and Engineering, Nanjing University, Nanjing, Jiangsu 210009, P.R. China\\}

\email{dbrady@duke.edu} %

\homepage{http:\\www.disp.duke.edu} %

\begin{abstract}
We review camera architecture in the age of artificial intelligence. Modern cameras use physical components and software to capture, compress and display image data. Over the past 5 years, deep learning solutions have become superior to traditional algorithms for each of these functions. Deep learning enables 10-100x reduction in electrical sensor power per pixel, 10x improvement in depth of field and dynamic range and 10-100x improvement in image pixel count. Deep learning enables multiframe and multiaperture solutions that fundamentally shift the goals of physical camera design. Here we review the state of the art of deep learning in camera operations and consider the impact of AI on the physical design of cameras. 
\end{abstract}

\section{Introduction}
\label{sec:intro}
For the past 200 years, "the camera" has consisted mostly of two components: optics to create a physical image and sensors to capture the physical image. The history of the camera may be separated into three phases: the first 75 years when the sensor was a glass plate, the next 75 years when the sensor was celluloid film and the past 50 years when the sensor has been an electronic focal plane. While the medium has changed, however, the basic function of forming a physical image and recording that image has remained the same. The transition from chemical to electronic recording was particularly significant, however, in that the electronic image is not a physical object. The primary purpose of early electronic cameras was to read-out an electronic signal describing the physical image. The transition from chemistry to electronics thus redefined the camera from "a device that records images" to "a device that encodes images."

With this transition, a third component has been added to the camera: the image signal processor (ISP). Early electronic cameras lacked this component and simply returned a rasterized analog version of the physical image, but beginning in the 1980's digital signal processing chips became available and the main function of camera electronics became encoding the physical image into a digital signal. As electronic focal planes evolved from vacuum to solid state and finally to active pixel sensors, it became possible to implement analog to digital conversion directly on the sensor chip \cite{628824}, but most cameras today include an ISP co-processor. The "image processing pipeline" implemented by the ISP typically includes various steps, such as color demosaicing, white balance, denoising, and tone mapping~\cite{ramanath2005color }, but image data compression is the computationally intensive function of most ISPs. Increasingly advanced standards and application specific circuits (ASICs) have been developed over the past 30 years to enable on-camera image encoding \cite{pirsch1995vlsi, yin2017survey}.
Since encoding algorithms algorithms are most often implemented as ASICs, adjustable parameters have been relatively constrained. (Although there have been efforts to make the ISP pipeline more flexible \cite{heide2014flexisp}). The basic goal of ISP (and the camera) has remained to capture and encode the physical image captured by the sensor. 

This paper reviews recent developments in camera design that lead to a third definition, e.g. a camera is now "a device that calculates images." This new definition emerges from the complementary fields of computational imaging~\cite{mait2018computational} and computational photography, which combine physical camera design with computational processing to improve estimation of the physical scene. As personal and embedded computers first became available, researchers already started to explore joint design of physical and digital camera components \cite{cathey1984image}. Despite many proposals in this regard, however, for many years the impact of computational imaging in real cameras was  modest. 
Most cameras have remained focused on encoding analog physical images and modification to the physical design of cameras to support computational imaging has until recently been rare.

This situation has changed dramatically in the past decade as clearly computational methods such as multiframe high dynamic range \cite{hasinoff2016burst}, super resolution and high depth of field have become common place and as multicamera arrays have become common in mobile devices. The sudden success of computational imaging after many years of modest development is driven by three factors: the maturity that comes from 30 years of computational imaging system design, the continuing sophistication and miniaturization of embedded processing and finally the recent explosive growth of artificial intelligence technology. The first two of these factors have developed progressively over many years. The third factor has emerged over just the past 7 years and is the primary focus of this review. 

The main point of this paper is to discuss how artificial intelligence (AI) technology changes the physical nature of the camera. To set the context for subsequent discussion, section \ref{sec:history} places recent AI developments in the context of the long and fitful history of computational imaging. Section \ref{sec:neuralNets} then presents a basic review of artificial neural networks and their relationship to cameras. As discussed in that section, AI methods and their impact on physical design are a sufficiently large step beyond the computational imaging methods developed over the past several decades to justify the new name of "smart cameras." To date, considerable interest in the combination of AI and cameras has focused on the ability of AI to understand and act on image data, for example by recognizing people or objects. These systems assume an input normal image streams from traditional cameras and use images or video to analyze scenes. Such capabilities are not the focus of this review. Rather we are interested in the application of AI within the function of the camera itself and in the joint design of AI and physical camera components.

Generally speaking, AI within the camera is used to (1) encode image data structures, (2) control camera data acquisition parameters, and (3) generate images and video from diverse data sources. Section \ref{sec:ps} previews how integration of AI within these functions impacts the physical structure of cameras. We then review specific neural processing systems used to achieve these functions in sections \ref{sec:ds} through \ref{sec:if}. Section \ref{sec:conc} returns to physical camera design in light of the results of these sections to discuss present and emerging opportunities for novel camera designs. AI integration in cameras is a very new phenomena and is developing explosively. While we are certain that significant new insights will emerge over the next several years, we hope that this review will be useful in creating a common understanding for camera designers and software practitioners.

\section{Computational Imaging and Computational Photography}
\label{sec:history}
Dating to the invention of optics and the {\it camera obscura}, humans have been developing cameras for nearly a millienium \cite{mait206}. Nature, of course, has been developing visual systems for billions of years, culminating in the human eye and visual cortex. Over most of their history, artificial cameras relied exclusively on analog optical processing for image formation. Over the past several decades, however, digital post processing has increasingly been used to augment physical image formation. To date, digital processing has primarily relied on conventional logic and algorithms. 

"Computational imaging" consists of integrated design of physical data capture and digital image estimation systems \cite{mait2018computational}. Computational imaging emerged over the past century, originally in the development of inherently computational systems such as radar and x-ray tomography and eventually, with the development of solid state sensors, extending to cameras \cite{cathey1984image}. Computational imaging mostly emerged from the physical design community and spanned all forms of imaging systems. The parallel field of "computational photography" emerged from the computer vision community over the past 15 years to focus specifically on computational imaging as applied to visible cameras \cite{raskar2009computational}. While all digital cameras necessarily rely on algorithms to control focus, exposure and other parameters, computational photography has focused on building platforms wherein such parameters may be controlled by sophisticated algorithms integrated with image estimation \cite{adams2010frankencamera}. For present purposes the distinction between computational imaging and computational photography is not important, we draw on examples from both communities in reviewing the history, achievements and challenges of computational imaging in cameras here. Our primary goal is to highlight lessons learned and the opportunities for progress by combining computational imaging and neural processing. 

While computational imaging consists of joint design of image capture and image estimation, one may still consider these two components as distinct problems. Image capture consists of design of the "forward model," associating scene parameters with measurements. Design of the forward model is a coding problem, within resource constraints the designer selects measurements to be made by the camera system. The camera forward model is always linear in scene luminance, meaning that the forward model can always be expressed in the very simple form
\begin{equation}
    \label{eq:forwardM}
    {\bf g}={\bf H f}+{\bf n}
\end{equation}
where ${\bf g}$ is measured data, ${\bf f}$ is the optical luminance and ${\bf H}$ is a matrix describing the forward model. ${\bf n}$ is the inevitable noise associated with measurement. Image estimation is the inverse problem, e.g. estimation of ${\bf f}$ given ${\bf g}$. Until very recently, the many different inverse algorithms that have been proposed and demonstrated could be grouped into just three categories:
\begin{enumerate}
    \item Linear estimation
    \item Constrained estimation and
    \item Bayesian estimation.
\end{enumerate}
Linear estimation, using for example least squares, Wiener filtering, or Tikhonov regularization, estimates ${\bf f}$ according to 
\begin{equation}
    {\bf f}_{est}={\bf \hat H}{\bf g},
\end{equation}
where ${\bf \hat H}$ is an inversion matrix. Constrained estimation estimates ${\bf f}$ according to 
\begin{equation}
\label{eq:constrained}
    {\bf f}_{est}={\rm argmin}\left [ |{\bf H f}_{est}-{\bf g}|^2 +\sigma({\bf f}_{est}) \right ],
\end{equation}
where $\sigma ({\bf f})$ is an objective function expressing some prior constraint on acceptable solutions. Typical priors include the $l1$-norm of ${\bf f}$ or the total variation norm \cite{rudin1992nonlinear}, both of which force "sparse solutions." The goal of Bayesian estimation is to select the "most likely" object state given the measurements. Bayesian methods have most commonly employed estimation-maximization algorithms \cite{richardson1972bayesian,lucy1974iterative,dempster1977maximum}, which treat the current estimate of the scene and the measurements as distributions which can be analyzed using Bayes' theorem.   

In an ideal case, computational imaging consists of selecting ${\bf H}$ within physical constraints so as to maximize the performance of the image estimation algorithm. For example, the mean square error between ${\bf f}_{est}$ and ${\bf f}$ may be selected as a system performance metric. If linear estimation is selected as the inversion strategy and the physical constrain on ${\bf H}$ is that its weights in $[0,1]$, for example, then the Hadamard S-matrix is  known to be an optimal choice for the forward model \cite{harwit2012hadamard}. More realistic physical systems and more advanced inversion algorithms will lead to different imaging system designs. 

Linear inversion methods were dominant from the origins of computational imaging in radar and tomography until the mid 1970's. Since constrained and Bayesian estimation both rely on computationally intensive iterative methods, linear estimation remained the method of choice in commercial tomography systems until very recently \cite{beister2012iterative}. In conventional cameras, iterative methods are typically only applied in post-processing. Initial still and video feeds tend to rely exclusively on linear processing. The deficiencies of linear estimation in photographic systems are, first, that since the forward model weights (e.g. the elements $h_{ij}$ of ${\bf H}$) are constrained to be nonnegative, the forward model itself tends to be ill-conditioned, meaning that linear inversion tends to be noisy or biased or both \cite{brady2009optical}. Second, despite the fact that the forward model itself is linear, the scene estimation problem contains nonlinear elements that are not accounted by linear estimation. For example, in three dimensional scenes foreground objects obscure background objects. Multiple observations or prior knowledge may be used to accurately estimate the full 3D scene if nonlinear analysis is used. Sensor saturation may also add nonlinearities that cannot be accounted by linear methods. 

As reviewed in \cite{mait2018computational}, computational imaging based on constrained or Bayesian estimation has been applied in three situations: when isomorphic or well-conditioned measurement is impossible (as in phase retrieval problems), when dimensionality mismatch between sensors and scenes renders isomorphic measurement inconvienent or impossible (as in spectral and tomographic imaging) and when the cost of fully conditioned measurement is excessive. Unfortunately constrained optimization based on Eq. \ref{eq:constrained} is a very blunt instrument for image estimation in these situations. The total variation regularizer, for example, tends to over smooth images and remove high resolution details \cite{buades2005review}. Much excitement regarding the use of Eq. \ref{eq:constrained} in computational imaging arose from proofs that for $l1$ regularization on randomly coded undersampled data, e.g. "compressively sensed data," constrained optimization will exactly recover appropriately sparse signals \cite{donoho2006compressed, candes2006stable, candes2008introduction}, but in practice there is little evidence that photographic images sampled by physically reasonable cameras satisfy sparsity and sampling constraints associated with these proofs. 

Returning to the issue of forward model design, many novel approaches to physical optical design have been proposed for image capture. Most such proposals have turned out to be noncompetitive with conventional focal cameras. We explain why previous efforts failed in this section and why the combination of novel physical design and neural processing is succeeding in the next section. 

The role of image capture hardware is to transform the analog optical luminance into digital numbers, in Eq. \ref{eq:forwardM} ${\bf f}$ is an analog physical field and ${\bf g}$ is an array of digital numbers. More detailed discussion of how fields over continous space, ${\it f}({\bf x})$ are correspond to discrete samples is presented in \cite{brady2009optical}. The relatively few mechanisms available in a camera for coding the forward matrix ${\bf H}$ consist of 

\begin{enumerate}
    \item Lens design, or {\it pupil engineering},
    \item Focal plane modulation
    \item Dynamic sampling and
    \item Multiaperture sampling
\end{enumerate}

Some strategies beyond this list considered in the computational imaging literature, such as coded aperture imaging, involve such extremely poorly conditioned forward models that it is not reasonable to consider them in cameras. Others, such as structured illumination, are of high utility in specialized fields like microscopy or lidar but of less interest in photographic cameras. 

Lens design is, of course, the very starting point of camera design. Normally, the goal of lens design is to form a well-focused image on a focal plane. Part of the magic of a lens is that it can map all the incident optical energy from a single object point to a single measurement pixel in the image plane. Any other coding technology necessarily multiplexes object points together. If photography consisted entirely of monochromatic mappings between 2D planes, it would be impossible to beat a well-focused lens for information fidelity. However, since real scenes are 3D and involve spectral information, numerous strategies that involve the deliberate introduction of lens aberration have been attempted. As a group these strategies are called {\it pupil engineering}~\cite{ojeda2015tuning}. Pupil engineering most famously consists of the introduction of depth invariant blurs to improve depth of field or 3D imaging~\cite{Ojeda-Castaneda:88,bradburn1997realizations, chi2001electronic}, but can also be tuned to improve color, depth and other image quality metrics~\cite{greengard2006depth, prasad2003engineering, levin2007image, muyo2005decomposition, wach1998control}. Early studies of pupil engineering emphasized linear estimation methods, but deconvolution with such codes is improved using constrained priors and/or Bayesian methods. In practice, however, pupil coding techniques have yet to achieve widespread acceptance in imaging systems, although it is certainly common to balance depth of field and MTF in basic lens design. In our opinion, pupil coding has not achieved popularity because its advantages are too easily obtained using dyanamic sampling and multiaperture methods, which do not entail the loss of resolution implicit in the introduction of deliberate blur.

Focal plane coding, in contrast, has been popular since the very inception of digital photography. Focal plane coding consists of modulating the analog image optically or electronically in an image plane. The Bayer red-green-blue color filter array is the most famous example of focal plane coding \cite{bayer1976color}. Since the color planes of an image are highly correlated, interlaced sampling with post capture demosaicking has long been considered superior to multi-sensor color sampling~\cite{gunturk2005demosaicking}, although in the age of straight-forward multisensor fusion multicamera solutions are increasingly attractive~\cite{zhao2017heterogeneous, shogenji2004multispectral}. Image plane color filter arrays have been extended by various mechanisms to hyperspectral imaging using compressive sampling \cite{cao2016computational, arce2014compressive}. Focal plane modulation also forms the heart of spatial compressed sampling methods~\cite{takhar2006new}, although such methods implemented optically suffer from poor forward model conditioning \cite{brady2009optical}. Better conditioning can be achieved by implementing compressed sampling electronically in the sensor plane, where negative weights can be included \cite{oike2012256, zhang2010cmos, treeaporn2012space, neifeld2003feature}. Similar to color filter arrays, light field cameras use focal plane modulation to implement interlaced sampling for focal depth \cite{ng2005light}, but due to the relatively severe loss of transverse resolution \cite{brady2011coding} multiaperture and dynamic sampling is generally preferred for this application. As with pupil engineering, both linear and nonlinear estimation algorithms are employed with focal plane modulation. Linear algorithms are faster and computationally less intense, nonlinear algorithms produce better quality images. In practice, real-time display may use linear estimation and nonlinear estimators may be reserved for secondary image analysis. 

Dynamic coding consists of changing sampling parameters within a frame capture time or from frame to frame, for example by scanning the focal state, changing exposure time or modulating a temporal filter. Conventionally, of course, the image reported by a camera consisted of the physical image captured at a specific time. With computational imaging, however, it is increasingly common to treat a sequence of captured measurements as a common data set and then use an algorithm to jointly estimate the state of the scene at one or more points time. The "flutter shutter" system described in \cite{raskar2006coded} was an early demonstration of this approach, \cite{bando2011motion, llull2013coded} and \cite{hitomi2011video} present similar results using novel coding and estimation schemes. As mentioned above, these coding schemes can achieve many of the advantages of pupil engineering without encountering the same loss of resolution. For example, \cite{llull2015image} demonstrates point spread function engineering by moving the focus and position of a lens during exposure. Like pupil engineering, this approach can be used to code for depth or create depth invariant PSFs, but unlike pupil engineering it can be adapted to scene requirements or turned off when no longer needed. 

Multiaperture sampling is the final piece of the image coding toolbox \cite{brady2018parallel, shankar2006multiaperture, bhakta2010experimentally, collins2001algorithms, wilburn2005high, pollock2015multi}. While stereo cameras have long been of interest, computational multiaperture imaging systems first achieved popular interest through TOMBO-style arrays modeled on insect eyes to achieve thin profiles \cite{tanida2001thin,duparre2005thin, venkataraman2013picam, shankar2008thin, druart2009demonstration, portnoy2009design }. This approach focused on digital super-resolution from array images to reduce effective pixel size, but became less popular as sensor pixel sizes neared the diffraction limit. Multiaperture approaches for wide field high resolution imaging \cite{brady2009multiscale, argus, cossairt2011gigapixel, brady2012multiscale, brady2018parallel} have been increasingly popular. Similarly, as mentioned above, multiaperture systems present an alternative approach to color \cite{zhao2017heterogeneous, shogenji2004multispectral} and video sampling \cite{shankar2006multiaperture}. Array image data fusion is the primary challenge associated with multiaperture imaging, but as improved registration and neural methods have made this problem easier to resolve, multiaperture systems are increasingly attractive. 

To summarize this section so far, entering about 2016 computational photography included three classes of inverse algorithms: linear estimation, convex optimization and Bayesian estimation, three main applications: tomography, ill-posed estimation and compressed sensing and four main coding tools: pupil engineering, focal plane modulation, dynamic sampling and multiaperture imaging. The main point of this review is to analyze what changed with the explosion of neural processing since 2012, especially with the increasing use of neural processing to replace the ready, aim and shoot functions of cameras since 2016. As we review in subsequent sections, literally every aspect of camera function has been impacted by neural algorithms over the past three years. Conventional linear and nonlinear estimation algorithms remain part of the computational imaging arsenal, but now they are tools that might be used as a subset of an AI system rather than standalone components. Similarly tomography and compressed sensing are imaging applications, but virtually every imaging system, from the simplest web cam to research instruments, is improved by the use of neural image processing.  Physical coding tools remain the same, but optimal strategies for applying them are changed radically. 

This makes sense, of course, if one regards neural algorithms as simply a new class of software. Cameras are increasingly computational devices, it makes sense that their software should use the latest tools. Neural methods automate a large portion of algorithm and software development. But neural methods also fundamentally change the landscape of computational imaging and computational photography \cite{Barbastathis:19}. 

 Linear inversion has been highly developed for over 75 years through tools like Wiener filter, Tikhov regularization and Hadamard transforms. The main goal of a linearly inverted system is to make sure that the desired image lies within the range of the measurement matrix. Bayesian methods have been popular for nearly 50 years and are best regarded as a regularization and super-resolution technique for slightly ill-conditioned linear models. Constrained optimization based on TV and $l1$ priors became increasingly popular over the past 30 years. Based on sparsity priors, constrained optimization allowed for the first time reliable estimation of features in the null space of the forward model. As each approach emerged, the basics of sensor design changed. For example, linear estimators favor well-conditioned complete forward models like the Hadamard transform \cite{harwit2012hadamard}, but for nonlinear estimators exceed the SNR of Hadamard estimators using ill-conditioned feature-specific forward models \cite{brady2009optical, neifeld2003feature}.
Significantly for next generation design, 
neural methods avoid the physically implausible requirements of random sampling associated with compressed sensing. Beyond forward model conditioning constraints, the detailed structure of sampling does not appear to be crucial using these methods. Learned image estimation algorithms can account both for complex image priors and systemic nonlinearities without the sparsity requirements of compressed sensing \cite{bora2017compressed}. 
 
 Artificial neural networks are not just a new class of algorithms. Rather, neural processing is uber-algorithm that can adapt to specific situations. For example, given that the performance of traditional algorithm depends on the scene and on hyperparameters, one can train a neural network to adjust conventional linear or nonlinear algorithms to match specific scenes. One can also train networks to detect the nature of images or parts of images, such as faces, and adjust image processing to match the object \cite{kim2019facial}. 

The many ways and mechanisms by which neural methods impact image estimation and computational imaging is, of course, under very active investigation and will understood better given more time. The potential depth and subtlety of neural proprocessing is illustrated, for example, in the case of "deep image priors," \cite{ulyanov2018deep} which use untrained random weight neural networks as objective functions in Eqn. \ref{eq:constrained} to generate surprisingly accurate images. Such systems are an example of "generative networks," which can in fill missing image data or even generate photo realistic "fake data." In contrast with previous image estimation algorithms, a neural algorithm can in principle retrieve any image that can be logically inferred from measured data, meaning that camera designers have the opportunity to "shoot the moon" \cite{moon:2019}. Within this in mind, we briefly review the history of neural networks for smart cameras in the next section. 

\section{Cameras and Artificial Neural Networks}
\label{sec:neuralNets}

Neural networks and machine learning represent an alternative strategy for algorithm development based on analogy with biological neural processors. Artificial neural computing was proposed \cite{mcculloch1943logical} two years before Von Neumann's 1945 design of stored program computers \cite{von1993first}. Research studies of artificial neural computing has continued in fits and starts since that time, including the development of the perceptron in the 1960's  \cite{rosenblatt1958perceptron} as the basic structural component and the development of backpropagation \cite{rumelhart1986learning} in the 1980's to autonomously program neural processors. The practical impact of neural systems over this time span was very modest, however. The construction of processing hardware on a suitable scale has been one of the primary challenges to practical neural computing. 
 
 Absent compelling evidence of the utility of artificial neural networks, investment in the development of application specific hardware was modest. Nevertheless, diverse electronic and optical approaches were demonstrated \cite{misra2010artificial}. From one perspective, optical processing may seem particularly attractive for the camera data processing because the hardware needed for optical neural networks is similar to camera hardware and optical neural networks and cameras both implement massively parallel transformations. Many years ago one of us followed this logic as part of a team developing optical neural hardware \cite{psaltis1990holography}. This approach has recently regained popularity \cite{chang2018hybrid, lin2018all}, but where the neural network algorithms of the 1980's turned out to be 25 years ahead of their time it seems likely to us that the optical neural processors will need at least another 25 years. 
 
 In any case, electronic neural image processing has recently become central to camera operation. The primary breakthrough came through the development of hardware capable of efficient neural processing, but the hardware in question was the graphical processing unit (GPU) \cite{oh2004gpu}, which was of course not developed with neural processing in mind. In 2012 a GPU-based deep network was used to achieve a revolutionary breakthrough in computational image classification \cite{krizhevsky2012imagenet} and computational neural image analysis and processing applications immediately began to grow explosively. 
 In recent years, hardware specifically designed for neural image processing or at least consistent with neural image processing has become readily available. Early examples include the Movidius Myriad \cite{moloney2014myriad} and the NVidia Tegra \cite{ditty2014nvidia} chip families. Recently even standard image signal processing (ISP) chips include neural co-processors and, of course, many companies are developing increasingly advanced chips. At the time of this writing neural image analysis is increasingly a standard component camera component, especially in mobile devices \cite{google2018, dempsey2018teardown} and proposals have emerged for ISP chips designed specifically for neural processing \cite{Buckler_2017_ICCV, 8803607}.
 
 These electronic platforms, rather than massively parallel optical platforms or large scale electronic processors, are sufficient for camera data processing because convolutional neural networks (CNNs) \cite{lecun1990handwritten, lecun2010convolutional} emerged as the primary neural image processing architecture. A convolutional network takes advantage of the fact that the data structure of images is shift invariant. Such networks require only a very small number of neural weights and can be efficiently implemented in hardware. 
 
 Where bursts of neural computing research activity in the 1960's and 1980's required specialists to painstakingly develop special software and hardware for each experiment, extraordinarily easy to use neural computing development platforms have emerged over the past 5 years. Platforms such Keras \cite{gulli2017deep}, TensorFlow \cite{abadi2016tensorflow}, Caffe \cite{jia2014caffe} and PyTorch \cite{paszke2017pytorch} allow nonspecialists to define and train neural models with just a few lines of code. With these tools, neural computing emerges as simply a new form of software. Where conventional image processing software development focused on finding efficient and effective algorithms for tasks such as tone mapping, color correction, demosaicking, etc., neural processing emerges as a "master algorithm" \cite{domingos2015master} that can solve all of these problems. Rather than writing the algorithm themselves, programmers now need only present the problem to the master algorithm and let the algorithm create the code to solve the problem. Of course, a problem arises in the "present the problem phase." Rather than describing the problem in words, the programmer must present the master algorithm with examples of inputs and outputs for the problem, in many cases thousands of such training pairs are required to fully train the network. 
 
\begin{figure}[ht]
  \centering
  \subfigure[Demosaicing code.]{
    \label{fig:example_code} %
    \includegraphics[width=0.92\linewidth]{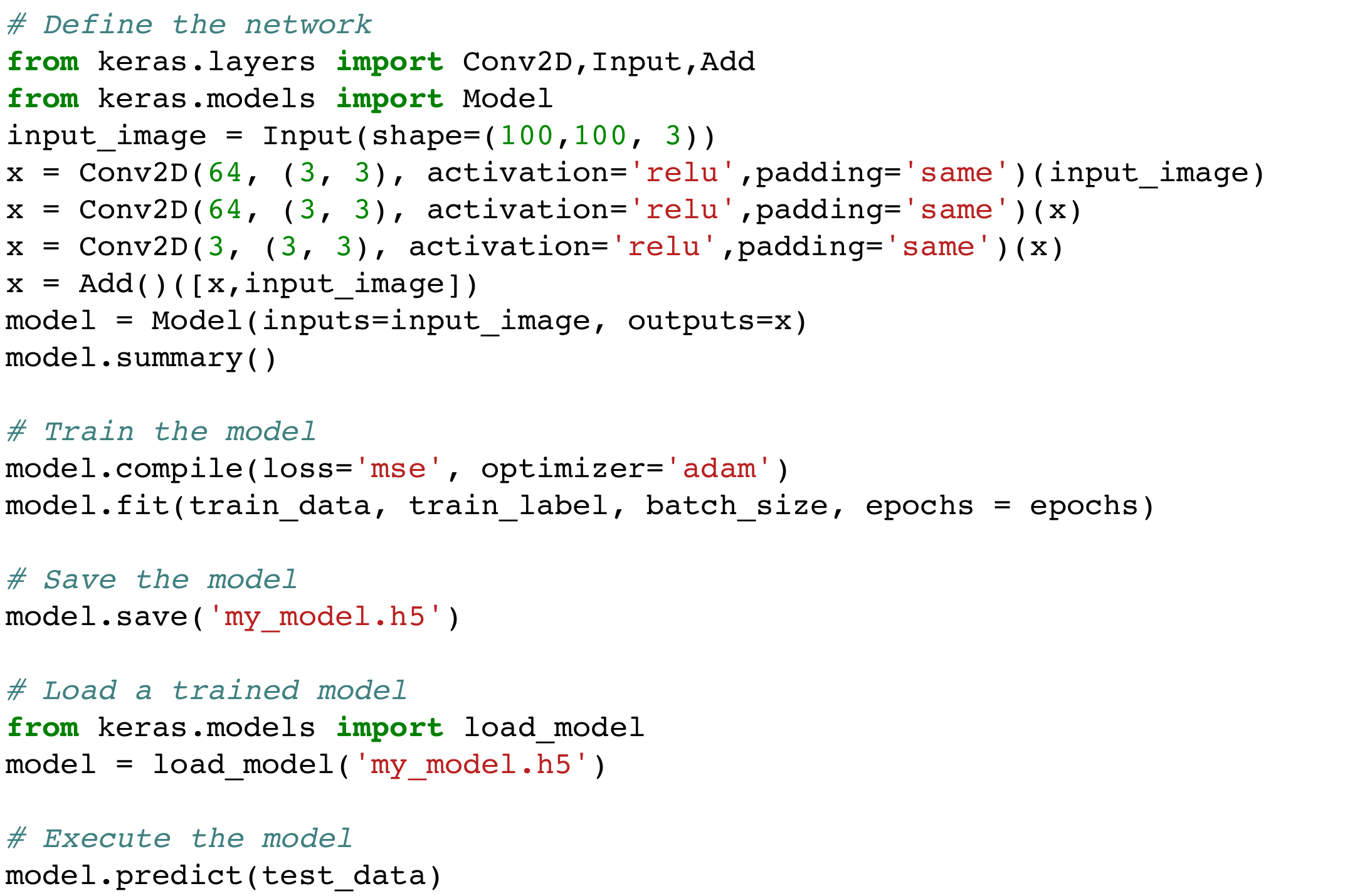}}
  \hspace{0.1in}
  \subfigure[Illustration of the sample network.]{
    \label{fig:example_illustration} %
    \includegraphics[width=0.98\linewidth]{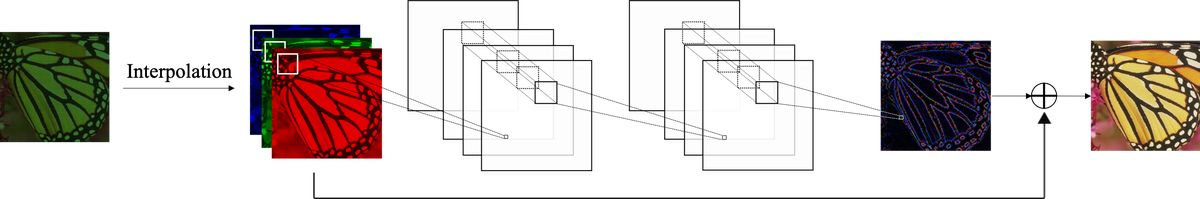}}
  \caption{Demosaicing example}
  \label{fig:example} %
\end{figure}

 As an example, Fig. \ref{fig:example_code} is the Keras code for training a neural network to demosaic RGB images, and the structure of the network is illustrated in Fig.~\ref{fig:example_illustration}. The network takes initially interpolated raw data as input, a common practice in CNN-based demosaicing methods, and outputs the RGB image. To define the network, one only needs to specify the dimension of each convolutional layer and the corresponding activation functions, leaving the computation to the hardware. The trained model can be saved as a single H5 neural model file and simply reloaded and executed. In the deployment phase, only one line of code is required and no more definition is necessary. The line 
 \begin{verbatim}
     model.predict(test_data)
 \end{verbatim}
 takes interpolated raw images in the array "test\_data" and outputs demosaiced RGB images.

The true magic in this snippet of code occurs in the "model.fit" line, which uses interpolated raw images in the array "train\_data" and reference ground truth RGB images in the array "train\_label" to train a neural network to associate Bayer and RGB images. To solve the demosaicing problem, and similarly denoising, super-resolution, etc., one can use publicly available image datasets for training. By manually mosaicing the RGB images, the mosaic-RGB pair constitutes one training sample, and one can generate a large training dataset from existing image dataset such as Kodak \cite{kodak} or McMaster \cite{zhang2011color}.

 In the development of advanced tools like Keras and Pytorch neural software has become institutionalized and easy to develop. But what does this have to do with cameras? After all, cameras existed for 150 years without any software at all and satisfactory image capture and processing software was developed in the preneural phase of image processing. It turns out, however, that smart cameras, e.g. cameras with integrated neural processing obtain revolutionary improvements over conventional systems, as we review in subsequent sections. "Ready," "aim" and "shoot" are basic functions of conventional cameras. In a conventional camera each of these steps requires human intervention, a photographer that sets focus and exposure parameters, points the camera and captures the image. In recent years these functions have become increasingly automatic through autofocus and autoexposure algorithms, but the basic goal of such algorithms has remained consistent with the "ready, aim, shoot" mantra, meaning that the goal is to set parameters so as to maximize the quality of the captured physical image. Smart cameras improve these algorithms by making better choices, but on a more basic level smart cameras can completely overcome the "ready, aim, shoot" mind set. Instead they can simply and efficiently capture all available optical information and leave image composition to post processing. 
 
 The "read, aim, shoot" mind set assumes that the goal of a camera, as discussed in the introduction, is to capture and encode a physical image, e.g. the image captured at the moment of shooting. Conventional cameras encompass numerous settings for pan, tilt, zoom, exposure, frame rate, focus and color sampling and processing. These settings ideally are adjusted to maximize measures of image or information quality, but in conventional systems adjusting settings to capture some aspect of a scene necessarily indicates that other aspects of the scene are not well captured. For example, focusing on the foreground implies that the background is out of focus, setting exposure for the foreground may mean that the background is over exposed, etc. The set of all possible images that could be captured of a scene is sometimes called "the light field." A conventional focal camera captures just a slice of the light field. Smart cameras, in contrast, may combine intelligent data capture and processing to tomographically capture and estimate the multidimensional light field. As discussed in the introduction, with the goal of estimating the light field the camera transitions from an image capture device to an image estimation device. 
 
 A smart camera achieves this goal in three ways: first by using artificial intelligence to dynamically control capture parameters, such as focus, exposure, frame rate, pointing and zoom, second by using AI based encoding to reduce the cost of measurement and thereby allow more complete light field sampling and third by efficiently and accurately estimating the light field from measured data. Of course, algorithms for each of these goals have always been essential to digital cameras. One may reasonably ask, in what way is neural focus control, neural compression or neural image estimation a significant advance on conventional autofocus, compression or image processing?
 
 These three problems map directly onto the three major problems in machine learning: reinforcement learning, unsupervised learning and supervised learning. Reinforcement learning applies to problems for which the solution is unknown but for which one can compare one solution against another. Game play is the canonical example, we don't know the best move in any state of a game, but we can compare the outcome of different moves. For a camera, we don't know the best focus, exposure or field of view at any given time, but we can compare the quality of the media achieved with different control strategies. Unsupervised learning refers to situations where one seeks to find patterns in data without specifying the identity of the patterns. For example, we may wish to know which groups of foods are associated with a healthy diet. In the case of cameras, unsupervised learning is the basis of compression algorithms, which seek to find image features that most efficiently describe image data. Supervised learning refers to situations where the desired input and output data pairs are known. Classic machine learning examples consist of annotated pairs of images and semantic descriptions, such as associating identity with facial images. In cameras, supervised training pairs often consist of images transformations where the forward transformation is easily described but the inverse transformation is hard. Demosaicking is such an example, the forward tranformation from an RGB image to Bayer data is trivial, but the best algorithmic transformation from Bayer data to RGB is unknown. Determining focus from blurry images is another example, it is easy to describe the blur transformation but hard to explain how the blurry image specifies the best focal state. 
 
 Conventional algorithms, still in use in the vast majority of cameras, are based on arbitrary {\it ad hoc} solutions to these machine learning challenges. Camera control, e.g. autofocus typically attempts to maximize a contrast or sharpness metric over an image or region of interest. Auto exposure similarly seeks to maximize dyanamic range distribution according to some {\it ad hoc} metric. Compression algorithms use an arbitrary DCT or wavelet basis to describe images. Image formation algorithms for denoising or demosaicking typically optimize useful, but arbitrary, metrics such as total variation or sparsity. While such metrics can be shown to improve qualitative image quality, they tend to destroy unusual and important image features.
 
 Neural algorithms, in contrast, can be intelligent in just the way that an intelligent photographer might be intelligent. A neural algorithm can decide how much time to spend on the foreground and how much time to focus on the background to optimize estimation of all in focus or 3D image. A neural algorithm can recognize the difference between signal and noise based abstract patterns rather than arbitrary metrics.
 In these distinctions we arrive at the critical difference between "computational imaging" and "smart cameras." The field of computational imaging as it has developed so far focused on joint design of the image forward model and mathematical methods to invert the forward model to improve system performance. As such, computational imaging really focuses just on aspects of the imaging system {\it per se}. A smart camera, in contrast, is the set of software, hardware and data used to produce media from visible scenes The software may use AI to reconfigure the hardware, may look up related pictures on the internet to improve processing, in short may do anything that a "brain" can do to produce an image. A smart camera can combine data from cameras at different locations, from cameras capturing different spectral ranges, from the combination of cameras and radar \cite{lekic2019automotive}, etc. 
 
 In practice, an artificial neural network is used as a black box implementing an abstract transformation. In the example shown above, the transformation is from raw Bayer data to the demosaicked RGB image. In an conventional camera, this transformation is implemented using an interpolation or optimization algorithm. Using a neural network, in contrast, the transformation may be programmed by training the network with examples of raw Bayer images and their corresponding RGB counterparts. The programmer hopes that this trained association generalizes to implement a demosaic function that is superior to conventional algorithms, for example an intelligent demosaic function could recognize edges and avoid the "zipper" effect of conventional systems or even use a different approach for hair-like features or faces. However, the inner workings of the function are not known in any way other than knowledge of the network structure and weights. The basic features of the network, e.g. the depth, number of neurons, loss function and activation function are determined from previous experience and experimentation. 
 
 A smart camera uses diverse sets of neural algorithms and conventional algorithms and software. Basically the neural components are functions forming part of the overall camera operating system. These functions could range from high level human interface functions allowing the user to ask the camera to brighten the image or zoom in, to low level functions using in denoising, demosaicing, tone mapping, etc.

\section{Physical structure of smart cameras}
\label{sec:ps}

Despite over three decades of computational imaging research, the basic physical structure of cameras been mostly unchanged for the past 150 years. As discussed in section \ref{sec:intro}, the ISP has joined the lens and the sensor as a fundamental camera component, but the basic function of the lens and the sensor are the same as they ever were. Recently, however, some changes have begun to appear. Most mobile phone cameras now have multiple apertures and there is growing recognition multiaperture design can enable substantial improvements in field of view, resolution \cite{brady2018parallel}, dynamic range, color \cite{shogenji2004multispectral}, frame rate \cite{shankar2010compressive} and depth of field. Many systems now also incorporate novel temporal sampling strategies, such as multiframe sampling for high dynamic range  \cite{sen2018overview} or focal stacking.  

As illustrated by the accelerating preference for camera arrays and burst sampling in mobile devices \cite{DBLP}, integrated AI technology creates significant opportunities for new physical designs. AI algorithms are particularly significant in that they are not just image formation algorithms, rather they are integrated systems for camera control, data management and scene estimation. Sections \ref{sec:ds} through \ref{sec:if} of this review present work to date in developing neural algorithms for these applications. Prior to this discussion, it is helpful to explain how the physical structure of the camera may be adapted to feed these algorithms. In considering smart cameras it is important to note that while AI is helpful in improving image quality, compression and control for traditional cameras, the traditional camera does a reasonable job of capturing images and video of the traditional type. The revolutionary aspect of smart cameras lies in the potential for dramatic improvements in the quality and quantity of camera media. We use "media," rather than "picture" or "video" to emphasize that the data captured by the smart camera may be difficult to classify simply as a traditional still image or video. Here "dramatic improvement" means increasing effective pixel count from the megapixel to the gigapixel range, increasing dynamic range from 8 bits per pixel to 32 or 64 bits, as well as radically increasing depth of field, range resolution and color fidelity. "Dramatic improvement" also suggests the possibility of new data structures, such as "light field images." A light field image may be refocused, but may also allow view point translation and zoom. To obtain such improvements, physical design utilizes the coding mechanisms discussed in section \ref{sec:history}, e.g. lens and focal plane design, dynamic and multiaperture sampling. Physical design also encompasses the design of the post-digitization electronic computation platform. This section considers these coding and computing mechanisms in more detail to inform discussion of neural architectures in subsequent sections. 

As discussed in the introduction, the modern camera consists of optics, sensors and ISP. Design of these components in any particular camera is interdependent and is also dependent on coding choices. In a traditional camera, the most basic coding choice is the assumption that optics forms "the image" and the sensor samples the image. With smart cameras, this assumption is no longer valid. Based on results reviewed in section \ref{sec:if}, one may assume that a smart camera can fuse multiple aperture data into an integrated image, video or world-model. One may also assume that the estimated media may draw on data collected over time, rather than a single frame. These assumptions represent a major shift in design philosophy; image registration, stitching and fusion has long been one of the most significant challenges in image processing, but neural processing shifts traditional thinking about this challenge. These assumptions also have profound implications for physical camera design. Obviously one can use multiple apertures to stitch different fields of view into a panoramic image. As discussed in \cite{brady2018parallel}, narrow field design simplifies lenses, meaning that for a given field of view and resolution the smallest size and lowest cost solution may be a multiapeture array. More generally the assumption that we can easily fuse multiple apertures calls into question the need for zoom lenses, the use of interlaced color filter arrays and the structure of image media. Similarly, the assumption that data capture is not local in time calls into question conventional focus, exposure and color sampling strategies. The rest of this section discusses the implications of these assumptions for lens, sensor and ISP design. 

We begin with lens design. Generally speaking the lens should be designed to minimize aberration and maximize modulation transfer across the spectral range and field of view. The lens modulation transfer should be approximately matched to the sensor pitch. Distortion is not of particular concern because it can be digitally compensated by the ISP. A lens system also includes temporal coding in the form of focus control, image stabilization, pointing and zoom. 
Recognizing that conventional cameras are generally fully capable of capturing diffraction limited 2D images and of adjusting focus, smart camera lens systems that improve on conventional design must (1) resolve substantially more transverse information than conventional cameras and (2) adapt mechanical parameters such as focus, pointing and stabilization much faster and more intelligently than conventional cameras.

In considering the potential for smart cameras to achieve this goal, it is important to understand  revolutionary advances in lens technologies over the past two decades. While physical diffraction limits suggest that lenses should resolve wavelength scale features, removable photographic lenses have never been capable of producing point spread functions at wavelength scale. Indeed, the very best removable lenses today struggle to produce features at frequencies below 100 lp/mm \cite{sony135mm}. Over the past decade, however, mobile phone cameras have emerged that use $2\mu $ pixel pitch with lenses that support modulation transfe(Copy)r beyond 300 lp/mm \cite{ji2013design, li2011design}. In building multiscale gigapixel cameras, our team demonstrated that small scale microcameras could be used in arrays to utilize f/2 to f/3 lenses at with >300 lp/mm resolution \cite{tremblay2012design, brady2012multiscale}. Conventional removable lens systems are much too large to support 10-100x increases in processed pixel count that smart cameras demand, but multiscale or discrete arrays of 1-2 micron pixel pitch microcameras with focal lengths varying from 3 to 50 mm can resolve 100 megapixel to several gigapixel fields with modest size, weight, power and cost \cite{brady2018parallel}. Of equal significance, such arrays can use the fast and reliable focus mechanisms developed for mobile camera systems \cite{gutierrez2007auto, 2019arXiv190906451P}. 
With the shift in lens design toward smaller pixels and faster f\#, lens systems for smart cameras may be expected to appear like insect eyes, with clusters of lenslets of various sizes sampling various focal states, colors and frame rates. AI may be used to locally adjust sampling across the field, including pointing and zooming lens resources, to assure maximal media quality and minimal resource expenditure. 

Turning now to the sensor module, the  drawback of small pixel camera modules relative to traditional full frame interchangeable lens systems is that the total light collection capacity and sensor dynamic range is reduced by the ratio of the pixel area, e.g. a 1 micron pixel collects 100x less light than a 10 micron pixel.  However, with smaller lenses and smaller chip area, the smaller camera modules enable multiple modules to regard the same area to improve light collection. More significantly, each module may observe with different sampling characteristics to improve quantum efficiency and information specificity. Where current cameras use achromatic lenses and color filter arrays to interleave scene information, smart camera may more efficiently use blue microcameras with pixel structure and lens design adapted to the blue spectrum and red microcameras with pixel structure and lens design adapted to the red. 

Sensors may also apply variable frame rates within an array camera to adjust to scene statistics and to enable wide dynamic range. Effects currently requiring advanced pixel design, such as logarithmic sampling, may as effectively be achieved using multiple microcameras. The major challenge is collecting as much information as possible with as little size, weight power and cost as possible. Multilayer integration, with the sensor stacked with ISP and memory layers, allows radical reductions in size and power \cite{haruta20174, mukhopodhyay2018camel}. Recent studies even integrate neural processors into such systems \cite{amir20183, mudassar2019camera}. In the longer term one can expect sensors that directly integrate compression and neural sampling in the focal plane architecture, closing a loop back to artificial retinas developed in an earlier cycle of neural computing \cite{douglas1995neuromorphic}. In short, beyond noting that AI multisensor processing opens paths to numerous new architectures, it is difficult to predict what sort of sensor designs will arise in smart cameras. Certainly one can expect more sensor diversity integrated into array cameras. 

Turning finally to ISP architecture for smart cameras, as noted in section \ref{sec:intro}, ISP processing chips have already evolved considerably from still image JPEG compressors to the modern streaming video chips \cite{pirsch1995vlsi, yin2017survey, pirsch1998vlsi}. In addition advanced tensor processing \cite{moloney2014myriad} and system on module systems \cite{ditty2014nvidia}, recent conventional ISP's have included tensor processing units for computer vision \cite{cv22}. Moving forward one expects smart camera electronic hardware to increasingly reduce the separation between the focal plane and the ISP and to increasingly replace the simple ISP pipeline with neural processing. This is particularly important in smart cameras because where conventional ISP codes the image data stream into a standard media format, such as h.264 or h.265, a smart camera captures unstructured information that (1) will not be directly displayed and (2) cannot be efficiently represented as a standard data stream. The next section of this paper explains how what these novel data structures might be like.

\section{Data structures}
\label{sec:ds}
The data rate into a camera is the product of number of pixels per frame and the frame rate, corresponding for example to 2 gigabits/second  for a 8 bit 4K video at 30 frames per second. In film and analog cameras, the utility of the camera is to translate this data from optical form to serial frames or electrical signals. With the development of solid state focal planes and digital ISP's however, the primary function of the camera has become transcoding the optical data stream into a compressed digital format. Video compression ratios of 100-1000 combined with ever decreasing memory costs have reduced cost per stored pixel by 4-5 orders of magnitude relative to film and enabled near ubiquitous video recording. 

In some situations, such as high-end photography, broadcast and movie media recording, uncompressed media are still recorded. In these situations, the raw format media is subsequently processed off camera and then compressed before distribution to online users. In the vast majority modern systems, however media captured by a camera is immediately converted to compressed digital formats, such as JPEG~\cite{JPEG}, JPEG 2000~\cite{JPEG2K} for images, and H.264/AVC~\cite{H264AVC} or HEVC~\cite{HEVC} for videos. Once encoded, the media becomes independent of the capturing system and can conveniently and efficiently be distributed for analysis and viewing. 

\begin{figure*}[htb]
\begin{center}
\includegraphics[width=0.9\textwidth]{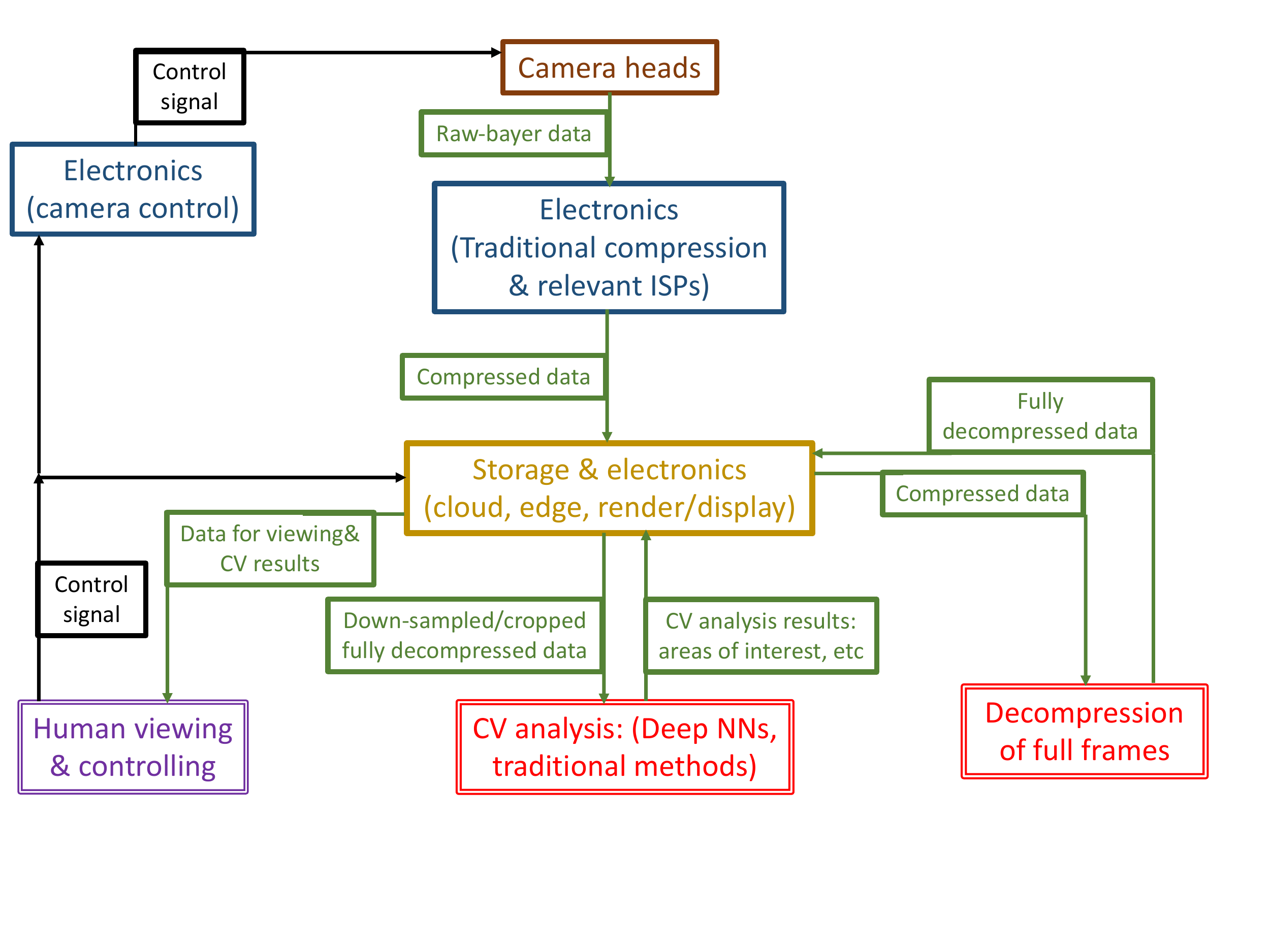}
\caption{\label{fig:Data_struct_trad} Graphical illustration of traditional data management and structure.}
\end{center}
\end{figure*}

The traditional pipeline for camera data  management is illustrated in Figure~\ref{fig:Data_struct_trad}.
Color filtered (e.g. bayer data) captured by the camera heads is processed in raw format for demosaicking, white balance, denoising, tone mapping and color conversion prior to compression to a standard format. Subsequently, the compressed data stream is stored or transmitted to cloud, edge and/or render/display devices. The electronics in the camera heads must be able to carry out all ISP. This approach is efficient if one assumes that 
\begin{enumerate}
    \item raw data is needed for ISP
    \item all image data will subsequently be required for display and analysis
    \item the cost of edge processing is not higher than the cost of cloud processing
\end{enumerate}

Beyond the storage and transmission layer, the traditional model sends standard compressed streams to displays, where it is fully decompressed for human viewing or it can feed the compressed datastream into computer vision systems. The input dimensions of most deep neural networks for computer-vision (CV) analysis are small, such as $256 \times 256$ or $448 \times 448$. Mega-pixel-level images need to be down-sampled and/or cropped before being analyzed by these networks. The down-sampled and/or cropped fully-decompressed data can serve as input of the CV-analysis deep NNs.
Human operators may send out control command to electronics that control the camera heads to tune their configurations, and send control command to electronics to carry out and/or obtain results of desired CV analysis. 

Image compression aims to reduce signal redundancy and represent the original pixel samples (in RGB or other color space) using a compact and high-fidelity format. Conventionally, transform coding (e.g., JPEG, JPEG 2000) or hybrid transform/prediction coding (e.g., intra coding of H.264/AVC and HEVC) is utilized. Typical transforms include discrete cosine transform (DCT)~\cite{DCT}, wavelet transform~\cite{JPEG2K}, and so on. Transforms are usually with fixed basis, that are trained in advance presuming the knowledge of the source signal distribution. Alternatively, intraframe data prediction leverages local~\cite{lainema2012intra} and global correlations~\cite{IBC} to exploit  redundancy. Since intra prediction can be expressed as the linear superimposition of casual samples, it can be treated as an alternative representation of a transform. Compression is then achieved via applying the quantization on transform coefficients followed by an adaptive entropy coding. Thus, image compression simply consists of concatenating ``transform'', ``quantization'' and ``entropy coding'' consecutively.

Replacing traditional image compression methods with machine learning-based compression compression algorithms can improve system performance in various ways. First, and most simply, compression algorithms are characterized by rate-distortion functions that describe signal quality as a function of compression ratio. Various studies have shown that deep neural net (DNN) compression may achieve better rate-distortion performance than conventional algorithms ~\cite{balle2018variational,rippel2017real,mentzer2018conditional}.

Most recently proposed machine learning based image compression algorithms~\cite{balle2018variational,rippel2017real,mentzer2018conditional} leverage the autoencoder structure, which transforms raw pixels into compressible latent features via stacked convolutional neural networks (CNNs).  These latent features are quantized and entropy coded subsequently by exploiting the statistical redundancy. Recent works have revealed that compression efficiency can be improved when exploring the conditional probabilities via the contexts of autoregressive spatial neighbors and hyperpriors~\cite{mentzer2018conditional,li2017learning,balle2018variational,liu2018deep}. Typically, rate-distortion optimization~\cite{sullivan1998rate} is fulfilled by minimizing  Lagrangian cost $J = R + \lambda D$, when performing the end-to-end training. Here, $R$ is referred to as {\it entropy rate}, and $D$ is the {\it distortion} measured by either mean squared error (MSE) or multiscale structural similarity (MS-SSIM)~\cite{wang2003multiscale}. A noticeable explorations have been made recently to better exploit the correlations by including nonlocal processing~\cite{liu2019non,liu2019practical}, explicit or implicit attention masks~\cite{mentzer2018conditional,liu2019non}. 

Neural  compression can also be extended to perform  video compression, where additional modules are introduced to handle the temporal correlations. For example, temporal redundancy can be exploited by prediction and residual, where prediction can use block-based motion search or compensation~\cite{chen2017deepcoder}, or optical flow-based compensation~\cite{lu2019dvc,liu2019neural}, and residual coding can reuse image compression structure~\cite{liu2019non}.

\begin{figure}[t]
    \centering
    \subfigure[]{\includegraphics[scale=0.35]{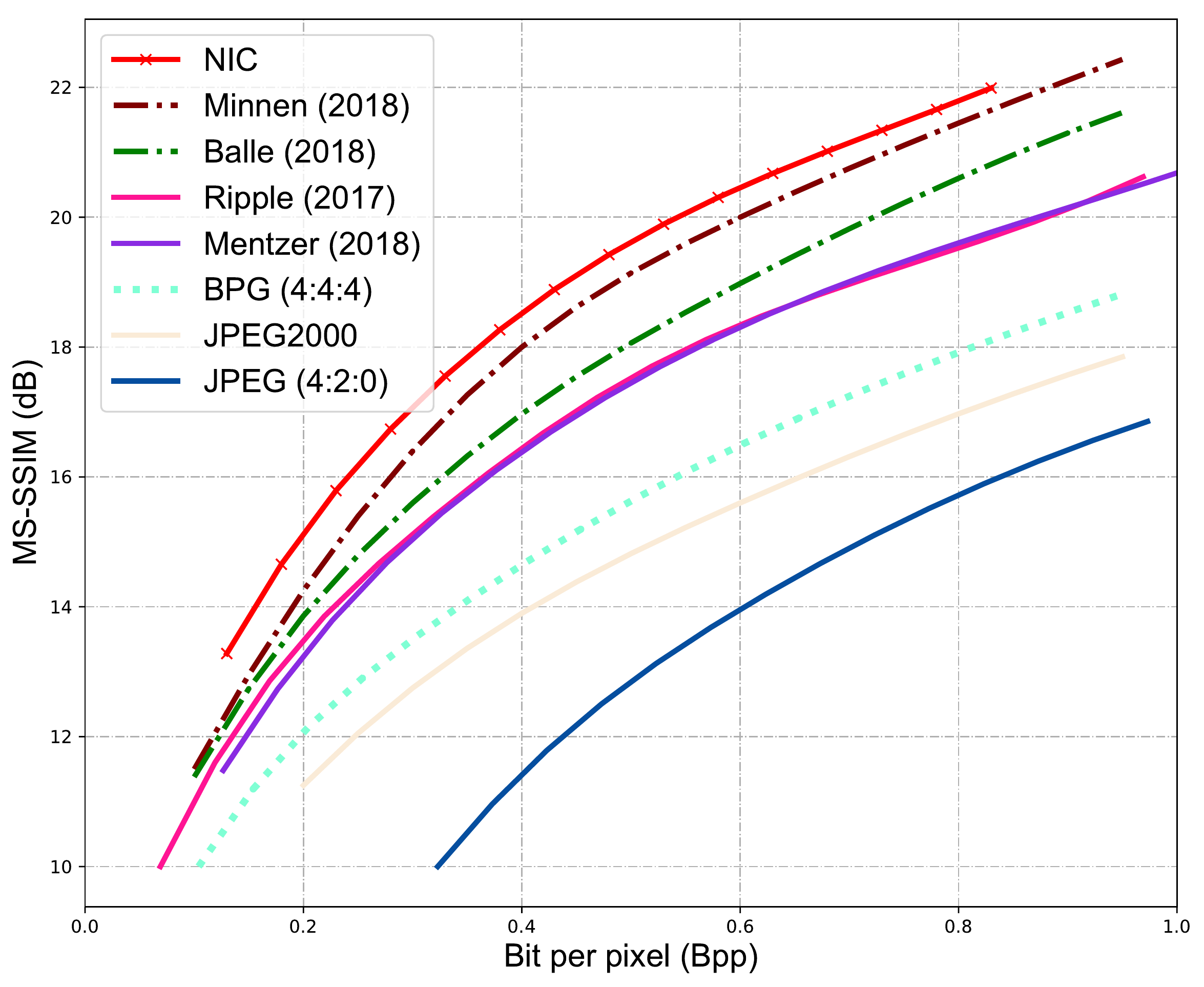}}
    \subfigure[]{\includegraphics[scale=0.35]{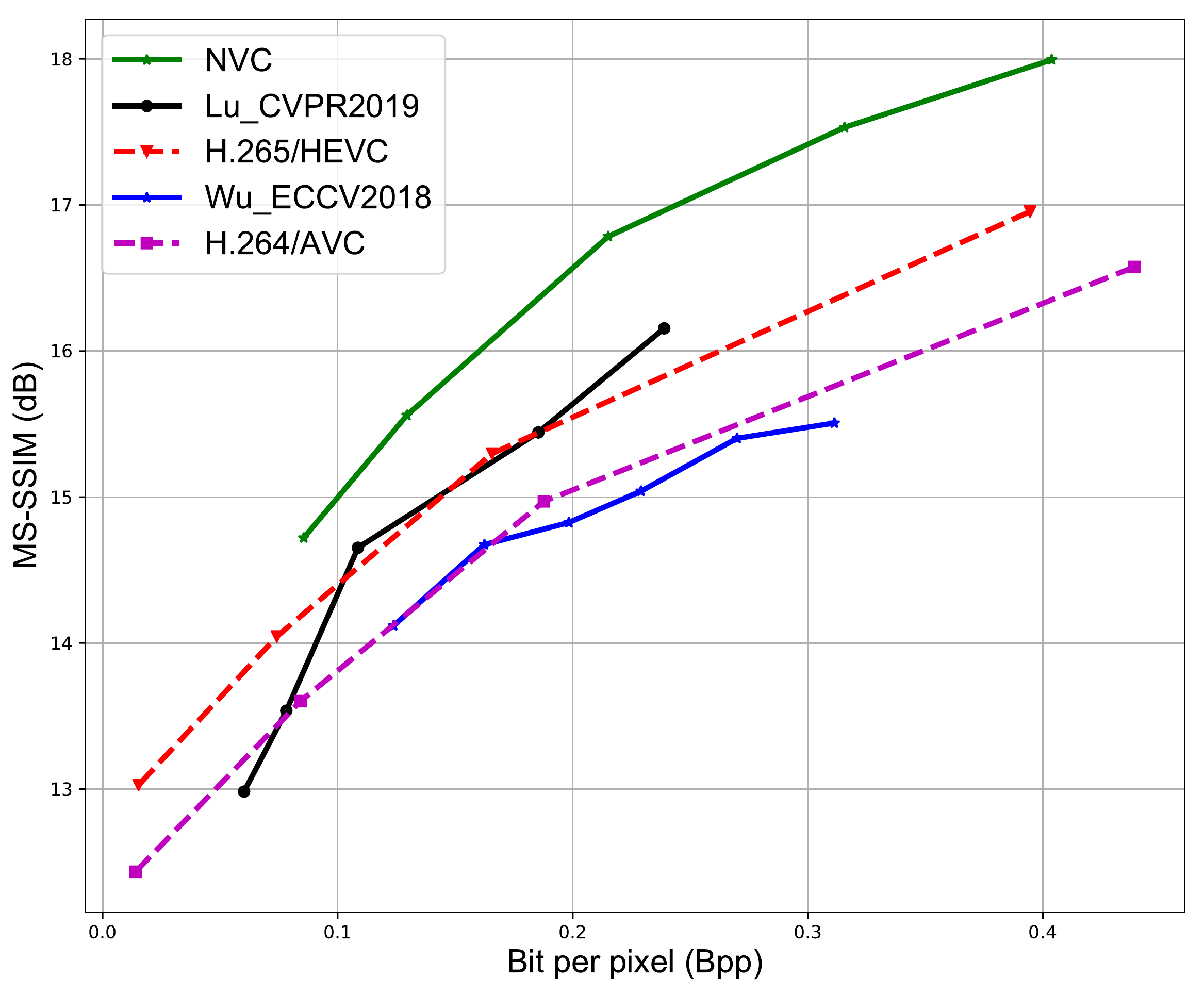}}
    \caption{Illustration of neural image and video compression. (a) neural image compression (NIC); (b) neural video compression (NVC).}
    \label{fig:neural_image_video}
\end{figure}

Figure~\ref{fig:neural_image_video} illustrates the rate-distortion performance of neural image and video compression, in comparison to conventional image compression methods, such as JPEG, HEVC Intra Picture Coding (a.k.a., BPG\footnote{https://bellard.org/bpg/}), conventional video compression algorithms, such as H.264/AVC, HEVC, and recently emerged other learned image/video coding schemes~\cite{balle2018variational,mentzer2018conditional,liu2019neural,lu2019dvc}. It shows that neural compression offers promising efficiency for applications.

Beyond simple improvements on rate-distortion performance, smart data management can be used to optimize other critical system parameters. Currently computational data power per pixel is a key limiting factor in camera performance. Neural processing can reduce this power in numerous ways. For example,
neural processing can perform ISP and vision tasks (such as classification, retrieval, etc) on compressed compressed data streams without decoding the bitstream to pixel domain, radically reducing effective computational load per pixel~\cite{shen2018codedvision}. 

\begin{figure*}[htb]
\begin{center}
\includegraphics[width=0.9\textwidth]{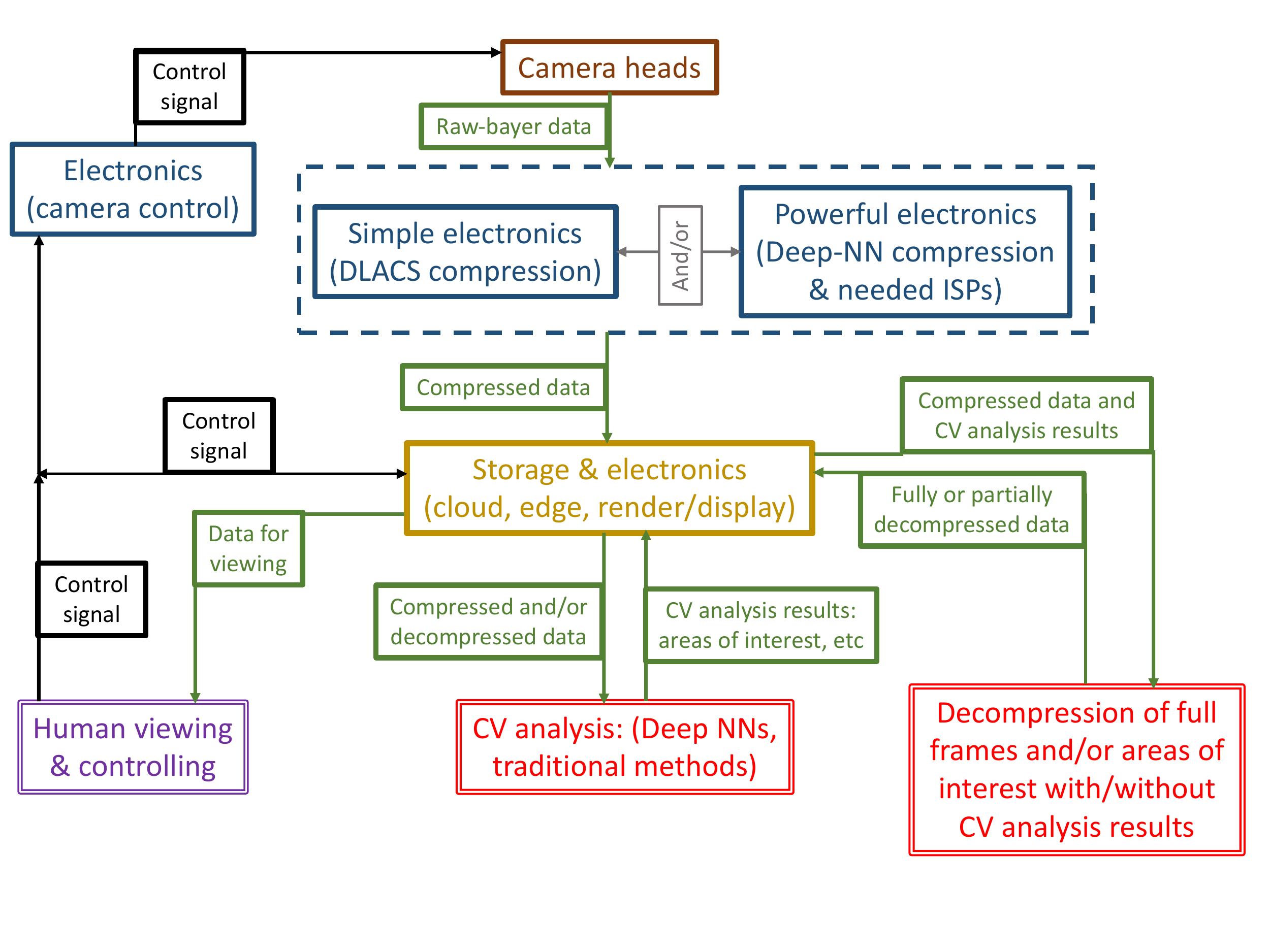}
\caption{\label{fig:Data_struct} Graphical illustration of proposed data management and structure of smart camera.}
\end{center}
\end{figure*}

Figure~\ref{fig:Data_struct} illustrates how the image processing pipeline of \ref{fig:Data_struct_trad} may be revised for smart cameras. In contrast with the traditional system, which really is a pipeline of image data from the camera to the display, the smart camera system is a network of microcamera sampling resources, cloud layer resources and display and analysis resources. The smart camera structure specifically reverses the three assumptions that defines traditional ISP
\begin{enumerate}
    \item Because neural processing can efficiently implement image processing on coded data streams, the smart system compresses first to reduce the camera head data load, and delays image processing to the cloud layer,
    \item Because any given display or analytical module is only likely to need a subset of all captured data, coding for the display is delayed to the last step of transmission from the cloud to the display and
    \item Because (1) not all data is likely to be processed and (2) cloud power is less expensive than edge power, ISP may be delayed to the cloud layer. 
\end{enumerate}
In recognition of the fact that the balance between camera head and cloud processing will vary between applications, Fig.~\ref{fig:Data_struct} includes alternate paths for very low head power sampling using compressive measurement and more conventional on camera ISP. In practice, what the camera head, cloud and display layers may actually all be in the same device, but even in this case the cloud layer saves power by only selectively processing captured data for display or analysis. Certainly, in array cameras one is likely to seek to minimize power expenditure at the microcamera level. 

Based on simple sampling load and no ISP,  electronics for low-power compression may be of much smaller and lower power, compared with those for traditional and deep-NN based compression methods,.
The compressed data is transmitted to storage in cloud, edge devices and/or render/display devices. While full decompression can be carried out in these or connected devices for human viewing, the compressed data may also be directly used for CV analysis by the dedicated devices. Based on the CV-analysis results, the compressed data containing the areas of interest may be decompressed for viewing while other parts of the data need not be decompressed. Based on different cases and/or purposes of using smart camera, the computation consumption can be flexibly distributed on electronics connected to the camera heads, in the cloud, in edge, render and/or display devices. 

\begin{figure*}[htb]
\subfigure[No compression.]{
\begin{minipage}[t]{0.3\linewidth}
\centering
\includegraphics[width=0.9\textwidth]{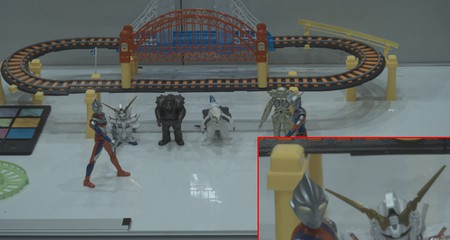}
\end{minipage}%
}%
\subfigure[$\textrm{Ratio} = 1/48$, $\textrm{PSNR} = 37.30$, $\textrm{SSIM} = 0.924$.]{
\begin{minipage}[t]{0.3\linewidth}
\centering
\includegraphics[width=0.9\textwidth]{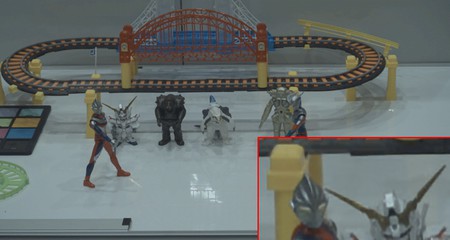}
\end{minipage}%
}%
\subfigure[$\textrm{Ratio} = 1/720$, $\textrm{PSNR} = 33.02$, $\textrm{SSIM} = 0.886$.]{
\begin{minipage}[t]{0.3\linewidth}
\centering
\includegraphics[width=0.9\textwidth]{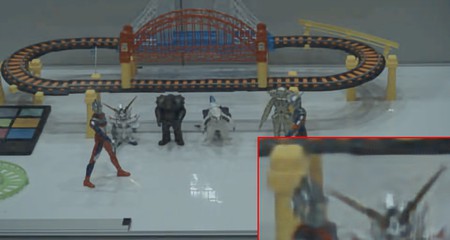}
\end{minipage}%
}%

\subfigure[No compression.]{
\begin{minipage}[t]{0.3\linewidth}
\centering
\includegraphics[width=0.9\textwidth]{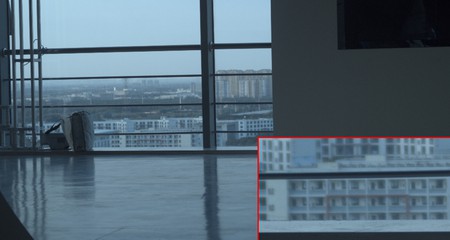}
\end{minipage}%
}%
\subfigure[$\textrm{Ratio} = 1/48$, $\textrm{PSNR} = 38.66$, $\textrm{SSIM} = 0.943$.]{
\begin{minipage}[t]{0.3\linewidth}
\centering
\includegraphics[width=0.9\textwidth]{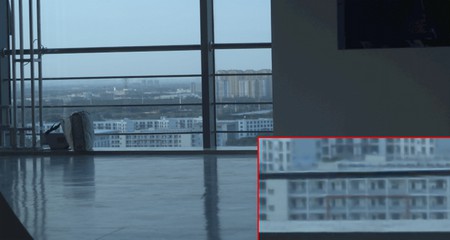}
\end{minipage}%
}%
\subfigure[$\textrm{Ratio} = 1/816$, $\textrm{PSNR} = 35.04$, $\textrm{SSIM} = 0.919$.]{
\begin{minipage}[t]{0.3\linewidth}
\centering
\includegraphics[width=0.9\textwidth]{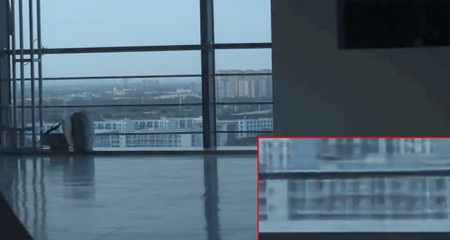}
\end{minipage}%
}%

\subfigure[No compression.]{
\begin{minipage}[t]{0.3\linewidth}
\centering
\includegraphics[width=0.9\textwidth]{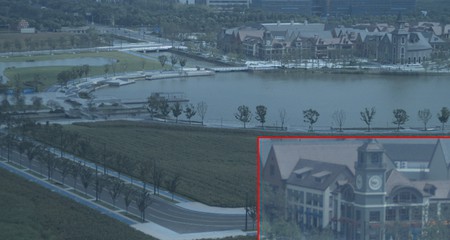}
\end{minipage}%
}%
\subfigure[$\textrm{Ratio} = 1/48$, $\textrm{PSNR} = 35.98$, $\textrm{SSIM} = 0.893$.]{
\begin{minipage}[t]{0.3\linewidth}
\centering
\includegraphics[width=0.9\textwidth]{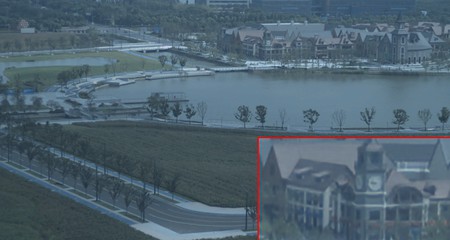}
\end{minipage}%
}%
\subfigure[$\textrm{Ratio} = 1/672$, $\textrm{PSNR} = 32.30$, $\textrm{SSIM} = 0.835$.]{
\begin{minipage}[t]{0.3\linewidth}
\centering
\includegraphics[width=0.9\textwidth]{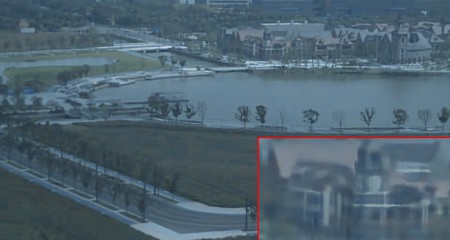}
\end{minipage}%
}%
\caption{\label{fig:DLACS_examples}(color online).  Example 4K RGB images from non-compressed raw-bayer data with ISPs (left column), DLACS-only compressed raw-bayer data followed by decompression and ISPs (middle column), DLACS-JPEG-hybrid compressed raw-bayer data followed by decompression and ISPs (right column). Figures from ~\cite{DLACS_clab}.}
\end{figure*}

For suppressing the computation on the encoder side and the size and cost of electronics connected to the camera heads, we recently proposed  deep-learning-aided compressive sampling (DLACS) to  simplify the encoding process and while maintaining acceptable rate distortion performance~\cite{DLACS_clab}.
DLACS  directly compresses captured raw-bayer data with low-bit-depth compressive-sampling masks.  This single layer operation may be the only intraframe method used on the camera head, The compressed data can be used without decompression for camera control and scene analysis. The compressed data can be decompressed by with a deep residual CNN based on region of interest guidance either on the camera or in the cloud. This deep-CNN-based decompressor also allows change of input dimensions with the same trained parameters, and thus allows for partially decompressing part of the compressed data with defined XY-plane boundaries.
DLACS is able to reduce camera-head electronic power by more than 20 times compared with intra-frame compression using JPEG, while reaching similar reconstruction quality as JPEG.
Compressed output from DLACS can also be further compressed and/or processed by other methods, such as entropy coding, JPEG or inter-frame compression methods.
Figure~\ref{fig:DLACS_examples} presents comparisons of reconstruction quality and compression ratios of three typical examples, using DLACS only and combining DLACS with JPEG in a hybrid manner. 
More details of comparisons between compression ratio, reconstruction quality and computation complexity between DLACS and traditional methods can be found in ~\cite{DLACS_clab}.

\begin{figure*}[htb]
\subfigure[Raw-bayer]{
\begin{minipage}[t]{0.45\linewidth}
\centering
\includegraphics[width=0.9\textwidth]{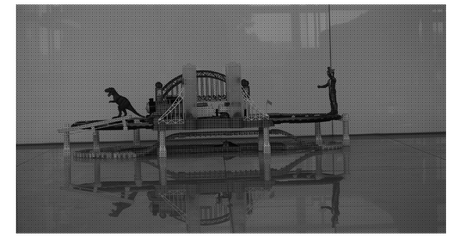}
\end{minipage}%
}%
\subfigure[RGB]{
\begin{minipage}[t]{0.45\linewidth}
\centering
\includegraphics[width=0.9\textwidth]{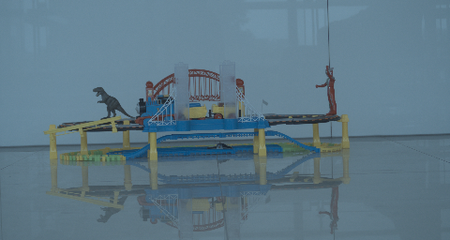}
\end{minipage}%
}%

\subfigure[DLACS-compression output]{
\begin{minipage}[t]{0.9\linewidth}
\centering
\includegraphics[width=0.99\textwidth]{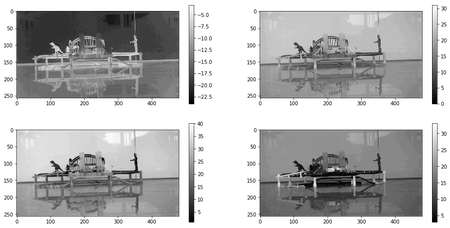}
\end{minipage}%
}%
\caption{\label{fig:RB_RGB_comp} Panel (a): an example raw-bayer image of dimension $[2048, 3840, 1]$. Panel (b): the corresponding RGB image of dimension $[2048, 3840, 3]$. Panel (c): four sub-panels presenting four channels of DLAC-compression output, compressed data in each channel is an 8-bit-integer array of dimension $[256, 480, 1]$.}
\end{figure*}

It was proposed that, it is reasonable and power-saving to capture only low-resolution images for computer-vision analysis, and capture high-resolution data only for viewing purpose~\cite{Buckler_2017_ICCV}.
In addition to suppressing computation and power in the compression process, DLACS also provides an opportunity to achieve power saving for a system both capturing and analyzing image data, in a simpler manner compared with the proposed method in~\cite{Buckler_2017_ICCV}.
Compressed data from DLACS, while are of small dimensions, visually resemble the high-resolution raw-bayer and/or RGB as presented in Figure~\ref{fig:RB_RGB_comp}.
Because in the DLACS compression process, local pixels in no-gap-no-overlapping areas of the original raw-bayer image are combined together by the CS masks, the positions of contents in the compressed data and decompressed images are linearly related, and finding boundaries of areas of interest in compressed data will directly allow determination of those in decompressed images.
With defined boundaries from CV analysis of the compressed data, areas of interest may be isolated and be decompressed for viewing.

\section{Camera Control}
\label{sec:cont}
The basic assumption of a conventional camera is that, whether in still or video format, the "image" presented by the camera corresponds to the physical image captured by a focal plane. This means that camera controls, such as focus, exposure, zoom and view angle, are adjusted at all times to maximize the quality of the instantaneous frame. For a smart camera, in contrast, there may be no concept of the instantaneous image. Measurement data captured by the camera can be recast to meet diverse viewer or analytical needs. When an image is formed, it may be estimated from an ensemble of data captured from various apertures at various times. This approach has already been adopted in "burst imaging" for multiframe HDR and focus algorithms \cite{hasinoff2016burst} and in  parallel camera systems \cite{brady2018parallel}. To date, however, such systems have still been based on traditional control strategies. 

Smart camera control extends to broader scenarios. Smart control can be divided into four cases. First, as with traditional camera, the goal of control may be to find  optimal settings for capture of the instantaneous image. Even in this case, AI extends the nature of traditional systems by allowing the camera itself to determine region of interest or other content-based control strategies. The second case assumes a static 3D high dynamic range scene. In this case, the goal of control is to determine a sequence of system states to be captured and synthesized to estimate the 3D, light field, high depth of field and/or HDR image. In contrast with a traditional focal stacking, super-resolution or HDR burst imaging approaches that capture a series of images with uniform settings over the focal or exposure range, a smart camera uses AI-based scene analysis to determine the number of images required and the settings for each. The third case considers dynamic scenes when the goal of control is to produce a trajectory of system states that best capture the scene. In the fourth case, array cameras are considered, where the control becomes to a collaborative task, control strategy can include design of the array itself and the goal is to produce a set of state trajectories with coordination across the array. 

Research and deployment of AI for camera control is still in its infancy, this section reviews published work in this area as well as suggesting areas of fruitful further development. Since dynamic control strategies are likely to simultaneously integrate multiple control parameters, we divide the section into examples of traditional and neural focus and exposure control before discussing architectures for dynamic control at the end of the section.

\subsection{Focus Control}
Auto Focus (AF) may be regarded as one of the earliest manifestations of artificial intelligence and robotics. From those early systems to the present, however, AF has always been a quintessential "expert-system" in which focus metrics and optimization methods have been predetermined by human expert opinion. Traditional AF methods can be divided into two categories: active AF and passive AF~\cite{yao2006evaluation}. Active AF measures object range using optical, ultrasonic or radar illumination. Since such systems require specialized and power-hungry components, image-based passive AF is more popular. Passive AF may further be subdivided into phase detection and contrast maximization methods. Phase detection employs extra components to split local components of the image into two or more paths. The displacement between the two images is used to determine the relative position between the image plane and the focal plane~\cite{baltag2015history}. Phase detection can determine locally best focus in a single frame and thus avoids the need for iterative search~\cite{sliwinski2013simple}. Such methods, however, sacrifice resolution, system volume and sensor uniformity in order to obtain phase information. In addition, phase detection assumes that a globally consistent "best focus" exists. Contrast maximization requires no hardware beyond a uniform focal plane, but has conventionally required search over multiple captured frames to achieve satisfactory results. 
The two main components of traditional contrast maximization methods are the evaluation metric and the search strategy. Evaluation metrics are hand-crafted features that measure the focus degree, or sharpness, of the image, which determines the quality of the focused images. The most commonly used metrics are gradient-based, including absolute gradient, squared gradient, Laplacian filter, Tenengrad function, Brenner function, \emph{etc.}~\cite{santos1997evaluation, yao2006evaluation}. Other evaluation metrics can be divided into 4 categories: correlation-based, statistic-based, transform-based and edge-based metrics~\cite{yao2006evaluation}. Correlation-based metrics evaluate the correlation among adjacent pixels, such as auto-correlation, standard deviation~\cite{santos1997evaluation} and joint density function~\cite{yousefi2011new}. Statistic-based metrics exploit the statistics of an image, such as entropy~\cite{santos1997evaluation}, gray level variance~\cite{yao2006evaluation} and histogram~\cite{yao2006evaluation, guo2018fast}. Transform-based metrics analyze the frequency components of the image, and the edge-based metrics investigate the edge information~\cite{yao2006evaluation}. 

Given an evaluation metric, the task of a AF module is to locate the focus position that maximizes this metric. The search strategy determines the speed of this process. Ideally, the search strategy should use as few time steps as possible. Popular strategies include  Fibonacci search~\cite{krotkov1988focusing},  rule-based search~\cite{kehtarnavaz2003development} and  hill-climbing~\cite{he2003modified}. The number of time steps required for Fibonacci search depends only on the focus range, but a relatively long total travel distance is required.  Rule-based search requires shorter travel distance but more time steps. Translation distance in each step time depends on the estimated distance from the optimal focus.  Hill-climbing consists of two stages: out-of-focus region search and focus region search~\cite{yao2006evaluation}. In general, a large number of time steps are required for these search methods. This issue can be addressed by modeling the pattern of the metric with respect to the change of the focus position. Curve fitting methods assume an optical defocus fitting model, e.g., ROL~\cite{yazdanfar2008simple} and BPIC~\cite{wu2012bilateral}, and capture only a few images to fit the model. Typically only 4-7 time steps are required in curve fitting methods~\cite{wang2018fast}.

The major disadvantage of contrast maximization methods, i.e., time delay, is a result of the indirect nature of the evaluation metric. The evaluation metric itself provides no direct information about the focus state. Rather focal state evaluation emerges from a large number of comparisons and, accordingly, time steps. The travel distance in each time step is pre-determined by the search method. Even in the rule-based method and the hill-climbing method, where adaptive travel schemes are adopted, the distances are selected based on built-in parameter settings. 

In contrast, when a human is presented a single defocused image, they can quickly decide whether a relatively long or short distance should be traveled to approach the optimal focus. AI AF algorithms should similarly be able to approach the optimal focus position in only a few, or even one, step by analyzing the defocused image. In this regard, the curve fitting methods can be viewed as early attempts to predict the optimal focus directly from the defocused images, but they still rely on hand-crafted features. Related strategies use machine learning techniques, such as one-nearest neighbor~\cite{han2011novel}, self-organizing map~\cite{chen2010passive} and multilayer neural network~\cite{park2008fast}. These methods achieve fast AF by formatting it as a classification problem with the sacrifice in focal accuracy. So far, direct focus prediction from single image for digital cameras has not been fully studied, but the recent success of CNNs in similar problems~\cite{jiang2018transform,ren2018learning,wei2018neural} indicate the potential of AI in mapping the image to the focus position. 

\begin{figure}[t]
	\centering
	\subfigure[System Structure]{\includegraphics[width=0.72\linewidth]{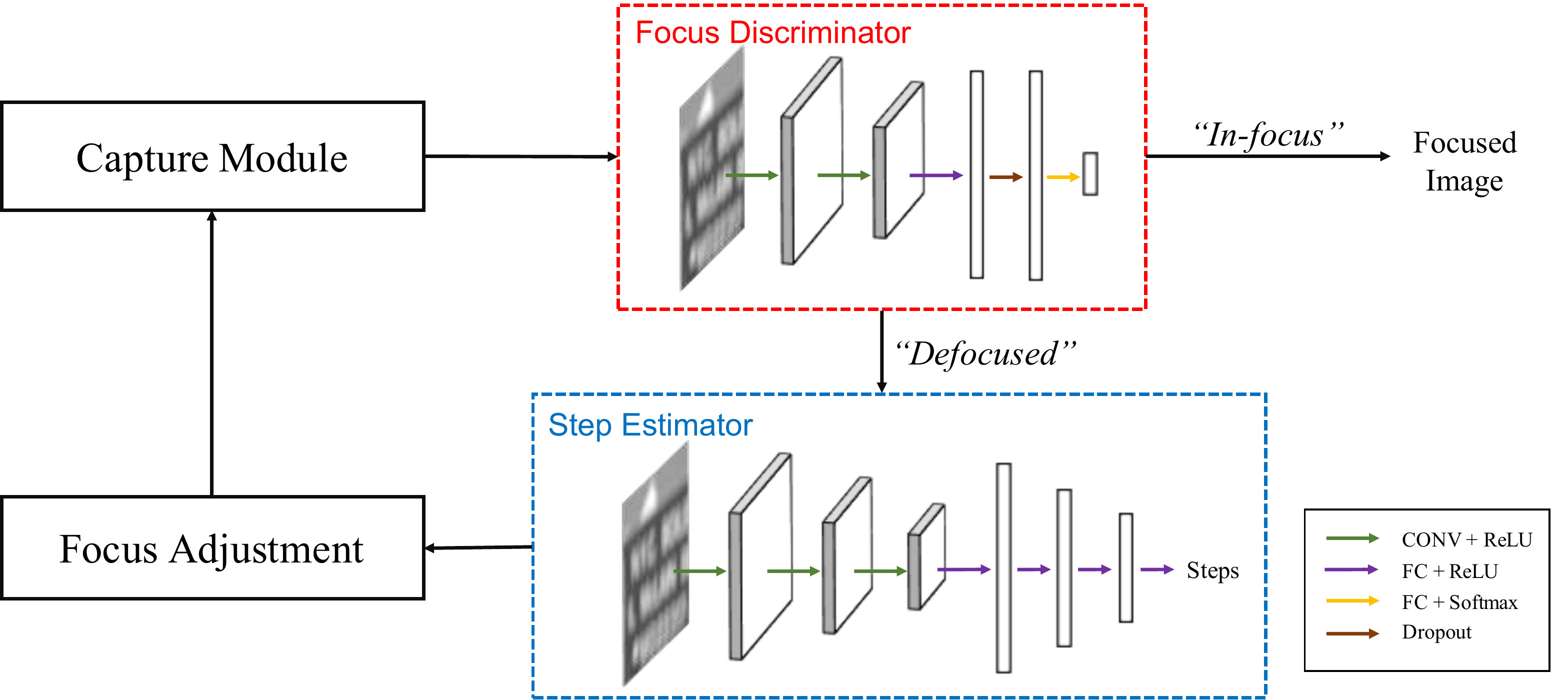}}
	\hspace{0.1in}
	\subfigure[System Implementation]{\includegraphics[width=0.23\linewidth]{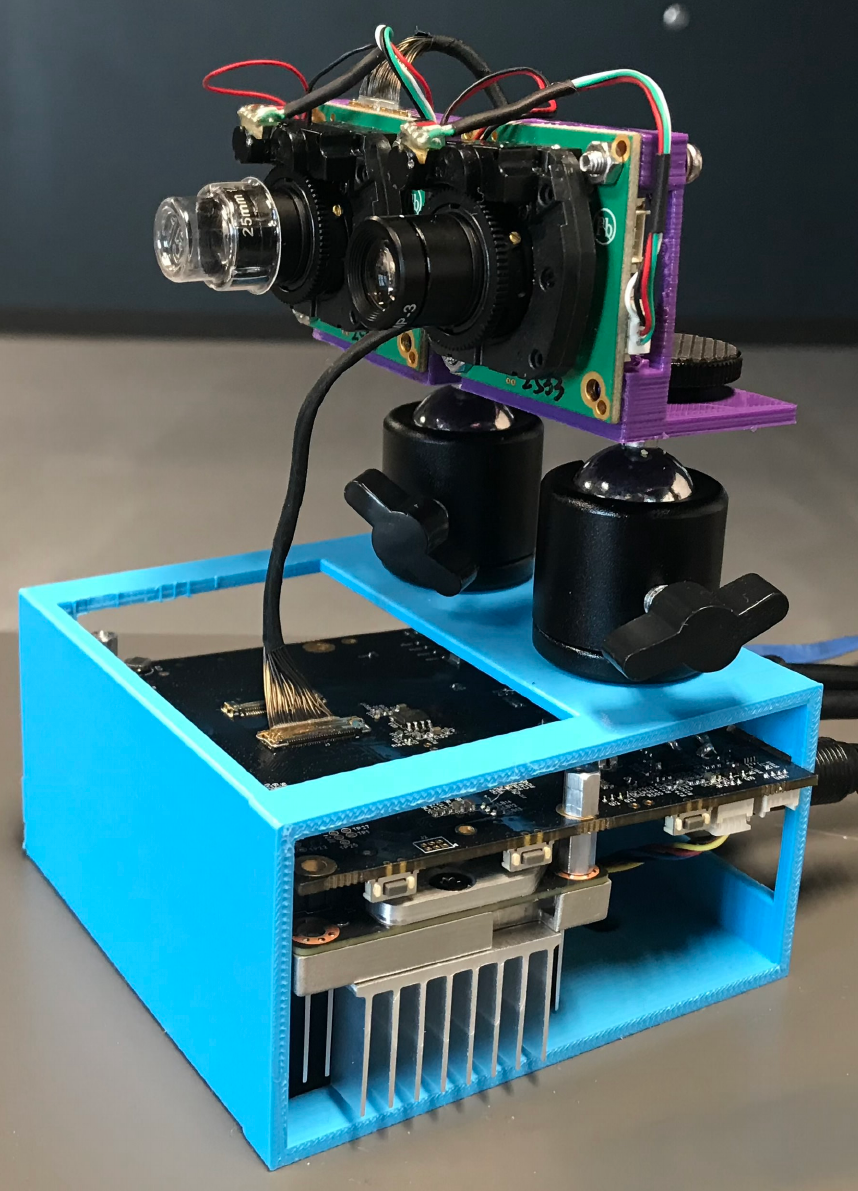}}
	\caption{Overview of the CNN-based AF system. (a) Input to the discriminator or the estimator is a block of size $512\times512$ from the image. \textbf{Focus discriminator}: The filter size/number of filters/stride for the two convolutional layers are $8\times8/1/8$ and $8\times8/1/8$. The dimension of the fully-connected layers are 10 and 2. The dropout rate for the dropout layer is 0.5. \textbf{Step estimator}: The filter size/number of filters/stride for the three convolutional layers are $8\times8/4/8$, $4\times4/8/4$ and $4\times4/8/4$. The dimension of the fully-connected layers are 1024, 512, 10 and 1. (b) The network configuration is tested on a camera module with Evetar lens (25mm, F/2.4) and CMOS image sensor (Sony IMX274 4K).}
	\label{fig:AF_example}
\end{figure}

An example CNN-based AF solution is described here. The system, illustrated in Fig.~\ref{fig:AF_example}, consists of two main parts, namely a step estimator and a focus discriminator. Once an image is captured, the focus discriminator first determines whether the image is in focus and the system outputs the image if it is well focused. If the image is defocused, the step estimator estimates the displacement of focus from this image and the focus adjustment module, such as a stepping motor, changes the focus position accordingly to capture the next image. Table~\ref{tab:af_compare} compares the CNN-based method with two traditional searching methods. There are two advantages of the CNN-based method: 1) the number of updates is significantly reduced compared with traditional methods, and 2) this method does not require any specific starting position while traditional methods typically starts at one end of the focus range.

\label{tab:af_compare}
\begin{table}[ht] 
{\footnotesize
\caption{Efficiency comparison between the Fibonacci search~\cite{krotkov1988focusing}, the Rule-based search~\cite{kehtarnavaz2003development} and the CNN-based method. The Tenengrad \cite{yao2006evaluation} was used as the evaluation metric for the two searching methods. Eight plain images and eight objects are tested, and the numbers represent the number of updates required to find the correct focus position. For CNN-based method, each sample was tested 4 times starting at random focus position, and the average times are recorded here.}

\begin{minipage}{0.49\textwidth}
\centering
\begin{tabular}{cccc}
\toprule
\textbf{Image}&\textbf{Rule-based}&\textbf{Fibonacci}&\textbf{CNN} \\
\midrule
Street 1  &  63  &  13 & 1.25  \\
Street 2  &  44  &  13 & 2.25  \\
Building 1  &  37  &  13 & 2  \\
Building 2  &  71  &  13 & 1.25  \\
Tiger  &  50  &  13 & 1.25  \\
Eagle  &  46  &  13 & 1.5  \\
Scenery  &  62  &  13 & 1.25  \\
Indoor  &  87  &  13 & 1.75  \\
\bottomrule
\end{tabular}
\end{minipage} \hfill
\begin{minipage}{0.49\textwidth}
\centering
\begin{tabular}{cccc}
\toprule
\textbf{Object}&\textbf{Rule-based}&\textbf{Fibonacci}&\textbf{CNN} \\
\midrule
Toy  &  46  &  13 & 1.75 \\ 
Box 1  &  47  &  13 & 1.25 \\ 
Box 2  &  51  &  13 & 1.25 \\ 
Cup   &  64  &  13 & 1.5 \\ 
Bag  &  39  &  13 & 2.25 \\ 
Plant  &  62  &  13 & 1.75 \\ 
Book  &  40  &  13 & 1.5 \\ 
3D printing  &  43  &  13 & 1.25 \\ 
\bottomrule

\end{tabular}

\end{minipage}
}
\end{table}

Thus far we have discussed AF under the assumption that there exists a static best focus. In practice the AF problem is further complicated by the fact that the real world is 3-dimensional and dynamic. The typical strategy for resolving this issue is to focus a region of interest, commonly the central region of the image~\cite{han2011novel}, faces in the scene~\cite{rahman2008real} or manually selected regions. There are also object-detection-based AF methods which analyze sub-focusing-windows to select the object and the region to perform AF~\cite{han2011novel,lee2008enhanced}. While the methods vary, the essence is that the optimum depends on contrast as well as human interpolation of the image. The concept that optimal focus is subjective naturally leads one to wonder if AI algorithms might be superior to traditional methods in the definition of "optimal". With AI's ability to analyze image content having been examined in other problems, such as saliency detection~\cite{cong2018review}, semantic segmentation~\cite{long2015fully}, social image understanding~\cite{li2016weakly}, image captioning and visual question answering~\cite{anderson2018bottom}, it is reasonable to believe the answer is yes. The more general approach is to treat autofocus as part of a dynamic control problem, which we discuss at the end of this section. 

\subsection{Exposure Control}

The need for exposure control is an artifact of the frame-based linear dynamic range sampling structure of modern focal plane arrays. While one expects logarithmic-response pixel-based read-out structures may eventually become practical, in the mean time one must use algorithms to set exposure. The main objective is to capture images with appropriate dynamic range. In the film camera age, exposure required manual settings, and expert experiences played an crucial role in capturing well-exposed images. With the digitization of cameras in 1970's~\cite{adcock1977electronic}, feedback in camera was enabled which allowed the camera itself to analyze the brightness of the image and adjust the exposure accordingly. Exposure control has been a heated research topic in industry. Searching on "camera auto exposure" in Google patent search returns 60,500 results. Some early AE methods are explained in \cite{johnson1984photographic, muramatsu1997photometry, sampat1999system, cho1999fast, kuno1998new, kao2005integrated}. While these basic methods suffice for simple scenarios with stationary illumination, performance with more complicated environments is unsatisfactory due to the lack of environmental knowledge or uneven illumination conditions~\cite{shim2014auto}. As with focus, AI technology enables exposure control and multi-exposure fusion at a greatly increased level of sophistication. Here we briefly review traditional methods and contrasts them with emerging AI strategies. 

As with the evaluation metrics in AF, some assessment tools have been applied in determining the necessary exposure level, such as histogram analysis and information analysis. \textbf{Histogram analysis} builds on the experience that a well-exposed image should appear in the middle region of the histogram. \cite{nourani2007automatic} combined the histogram analysis with the mean sample value to balance the tonal distribution in the image. \cite{neves2009autonomous} applied histogram analysis to adjust exposure while the black and white regions are known in advance. \cite{montalvo2013acquisition} analyzed the color histogram of the image and adjusted the focus to match the histogram of an reference image. \textbf{Information analysis} derives from the fact that a well-exposed image should contain sufficient details about the scene. The tools to analyze the information in the image include gradient analysis and entropy analysis.\cite{shim2014auto} applied the gamma corrections to synthesize different exposure level to find the proper exposure setting that maximized the amount of gradient information in an image, and a more robust gradient-based metric was proposed in \cite{zhang2017active}. Rahman {\it et al.} \cite{rahman2011using} showed that the entropy of an image varied with the exposure setting and therefore could be used as an assessment method. 

As with focus control, global analysis fails to deliver satisfactory exposure results in many cases, such as when the illumination is not consistent in the scene. A simple solution is to apply AE to the central region of the image~\cite{kuno1998new}. In more complicated cases, determining the optimal exposure involves analyzing the content of the image. A common practice is to divide an image into regions and compute the proper exposure level by weighing the regions with respect to their importance. \cite{murakami1996exposure} analyzed the importance of the background and the light degree of both foreground and background before applying fuzzy logic to control the exposure. \cite{lee2001advanced} divided the image into five regions and computed the weighted illumination to emphasize the main object. \cite{liang2007auto} detected the main object through a search process. \cite{kao2006adaptive} adjusted exposure by first tracking moving objects or human faces. \cite{nourani2007automatic} used pre-defined masks to compute the exposure level. 
While empirical assumptions such as histogram distribution, are plausible, evaluation of the exposure results is a subjective task. A smart AE algorithm could therefore adopt a sophisticated mechanism to estimate what human will regard as "optimal exposure" for each given scenario. There indeed exist algorithms that predict the human perception of image quality \cite{bosse2017deep}, image quality assessment (IQA).

Ultimately single frame exposure control cannot yield fully satisfactory results. 
The dynamic range in natural scenes can exceeds 100,000:1~\cite{bandoh2010recent}. Images captured by a focal plane are typically linearly digitized to 8-12 bits (just 256-4096 values). Such low dynamic range (LDR) is insufficient to record the light levels in typical scenes~\cite{bandoh2010recent}. A direct solution is to capture a sequence of images with different exposure and fuse them to one output image \cite{Mann95onbeing, debevec2008recovering, nuske2006extending, barakat2008minimal}. This procedure is referred to as high dynamic range (HDR) imaging and is commonly deployed in commercial cameras. While merging several LDR images has been the main research topic in HDR imaging, two under-exploited control problems need to be solved before a HDR image is produced, i.e., how many images are required, and how to determine the proper exposure level for each image. In early HDR imaging systems, the exposure time for different images varied by a constant factor, but these values, as well as the number of images, were not determined by principled methods \cite{grossberg2003high, guthier2012optimal}. Later, user input was integrated in the HDR imaging process~\cite{o2006simple,grossberg2003high}. Although the performance was improved, the imaging process was no longer fully automatic. To effectively and efficiently perform HDR imaging, \cite{hirakawa2010optimal} designed a control mechanism considering the statistics of the photon arrival process, \cite{guthier2012optimal} proposed to use the desired histogram to control the shutter speed, and \cite{zhang2017active} applied a gradient ascent method to update the exposure time. \cite{huang2013intelligent, pourreza2015exposure, pourreza2015automatic} proposed to first apply the basic AE function in camera followed by different exposure settings based on the analysis of the captured image. In controlling the number of images required, \cite{barakat2008minimal} described the method to determine the minimum number of exposure settings required to capture the whole scene. 

In general, two strategies have been used in HDR imaging. In the first, the exposure settings for all the images are pre-determined, typically consisting interleaved rows or interleaved frames of high/low exposure. In the second, the exposure of the current image depends on all previous images \cite{hirakawa2010optimal}. It is not hard for one to believe that the latter adapts better to the environment by constantly updating the parameters. A smart HDR imaging system should adopt such mechanism and intelligently determine the continuation and the exposure time for each image.

 The capture scheme and the fusing algorithm are the two main components of an HDR imaging system, and they are typically designed and executed individually. The separation in the design of the two components may achieve sub-optimum for the two tasks, but the concatenation of the two optimal components does not guarantee an optimal system. Thus, the HDR imaging process should be optimized as a whole problem. The capture-then-fusing execution order neglects the fact that the fusing process may supervise the capturing process by providing feedback. For example, the current fusing result can be evaluated to decide if more images are required or what exposure level is desired for the next image based on the current result in the fusing process.  Deep leaning based IQA for HDR images, such as \cite{jia2017blind}, may be used evaluate the fused image. As with focus, however, one may expect better results if exposure control is implemented as part of a general dynamic control system.

\subsection{Dynamic configuration}
Generally speaking, the configuration space of a camera is a group of settings that control the quality and the quantity of the captured data, which include focus, exposure, frame rate, resolution, camera position and Pan-Tilt-Zoom (PTZ) parameters~\cite{piciarelli2015dynamic}. Conventional camera aims to optimize the measurement of the instantaneous physical image. This approach, however, limits the scope of digital cameras. In fact, the development of fast processing units (e.g., neural computing chips) and fast control module (e.g., voice-coil motor) has enabled the camera to do more than capturing. Instead of functioning as a simple tool to record the instantaneous data, when dynamic scenes are considered, the camera should constantly adjust its configuration to adapt to the non-static nature of the scenes. The necessity of dynamic configuration originates from the fact that the real world is high-dimensional while the camera measurement is not. Therefore, in order to best capture, estimate and present the world with limited low-dimensional measurements, a strategy to obtain these measurements is essential. In this regard, the camera becomes a self-controlled dynamic robotic system. While dynamic camera control has been under-exploited, extensive research has been conducted on other robotics related control problems, e.g., autonomous vehicle. The essence of autonomous vehicle is to output control signals to navigate without human interaction. The straightforward solution is to apply supervised learning and build end-to-end networks that outputs the control signals~\cite{Chen_2015_ICCV,bojarski2016end,bojarski2017explaining,xu2017end}, however, it requires a large number of expert data and it is not robust to complex tasks. Therefore \textbf{reinforcement learning} is more suitable for such problems. 

The two main components of reinforcement learning are the agent and the environment. The agent takes actions based on the environment and in return changes it. The environment can be modeled being in state $s\in S$ which contains the information about the current environment. For example, in the autonomous vehicles, the state is the data collected from sensors, including the speed, the distance to other vehicles, orientation, etc. Given the state $s$, the agent determines the action $a\in A$, such as brake, acceleration or turn in a vehicle, which changes the state of the system. The mapping from the state $s$ to the action $a$ is called the policy $\pi$, i.e., $\pi:S\to A$. For each time step $t$, after performing the action $a_t$, the agent receives an immediate reward $R_t$ associated with the change of the state. The goal of the reinforcement learning is to obtain the optimal policy $\pi^*$ that maximizes the expected reward, most commonly the expected accumulative reward
\[ E\left[\sum_{t=0}^{T}R_t\right]\]
or the expected accumulative weighted reward \cite{kober2013reinforement}
\[E\left[\sum_{t=0}^\infty\gamma_tR_t\right].\]

Reinforcement learning has been applied to autonomous vehicle for decades, and its performance has been greatly enhanced with the rise of deep learning. \cite{shalev2016safe} used the policy gradient iterations. This approach decomposed the policy into a learned mapping and an optimization problem, which balanced comfort and safety. \cite{sallab2017deep} integrated Recurrent Neural Networks and attention models into the reinforcement learning framework. Q-learning was applied in \cite{isele2018navigating} to navigate through intersections. 
A double deep Q-network was applied in \cite{makantasis2019deep}. To better cope with the long-term planning, \cite{paxton2017combining} combined hierarchical neural net policies and Monte Carlo Tree Search. \cite{zhu2018human} applied a deep deterministic policy gradient algorithm in autonomous car-following planning.

\begin{figure}[t]
	\centering
	\includegraphics[width=0.75\linewidth]{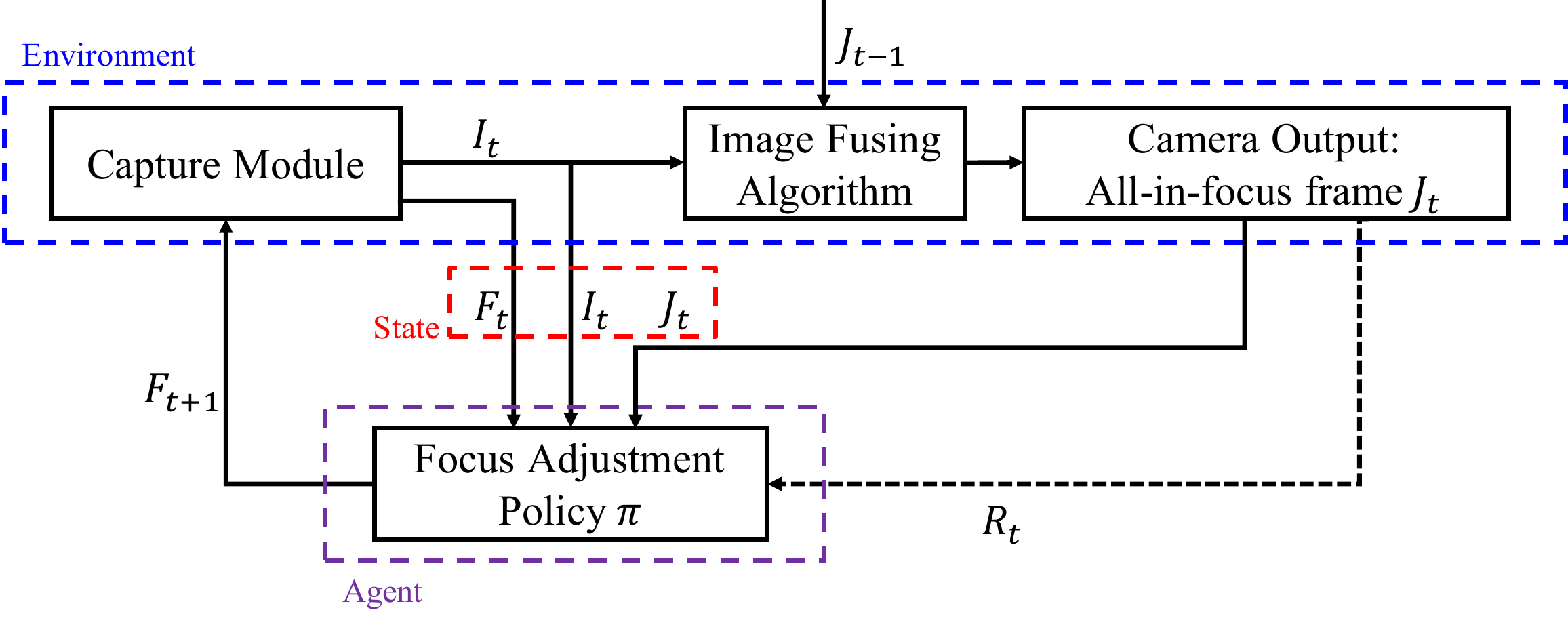}
	\caption{A dynamic focus control system. $I$ represents the measurement from the camera, $J$ represents the fused all-in-focus frame, $F$ represents the focus position, and $R$ represents the reward. The subscript $t$ denotes the time stamp. The fusing algorithm takes in the current measurement and the previous output frame and generates the new all-in-focus frame.	In this system, the processing unit acts as the agent which infers the focus adjustment from the state $\{F_t, I_t, J_t\}$.}
	\label{fig:reinforcement_focus}
\end{figure}

The similarity between the camera and the autonomous vehicle implies that reinforcement learning paradigm provides the solution to the dynamic camera control. In this paradigm, the agent is the processing unit in the camera which outputs the commands to adjust the camera configuration, and the state is the current measurement and the camera configuration. Fig.~\ref{fig:reinforcement_focus} shows an example of dynamic focus control. The ultimate goal of this system is to output the all-in-focus video. In each time step, the camera should determine the focus position for the next measurement given the current state. Both the policy and the fusing algorithm can benefit from neural networks, especially recurrent neural networks.

The dynamic configuration of array cameras further considers the resources re-allocation. Although both the hardware and software have developed substantially, cameras, especially multi-camera systems, are constrained by the limited resources, such as time, energy, bandwidth, storage, number of cameras and etc. These constraints demand an intelligent way of re-allocating resources during the capturing process, and the cameras should collaborate to achieve the functionality. A typical example is collaborative sensing in PTZ camera network. In such a network, a strategy is required to decide the orientation for each camera in order to satisfy the ultimate observation goal which can be maximum coverage, best image quality, reliable tracking, and etc. Details of different observation goals and allocation strategies are surveyed in \cite{piciarelli2015dynamic}.

\begin{figure}[t]
	\centering
	\includegraphics[width=0.9\linewidth]{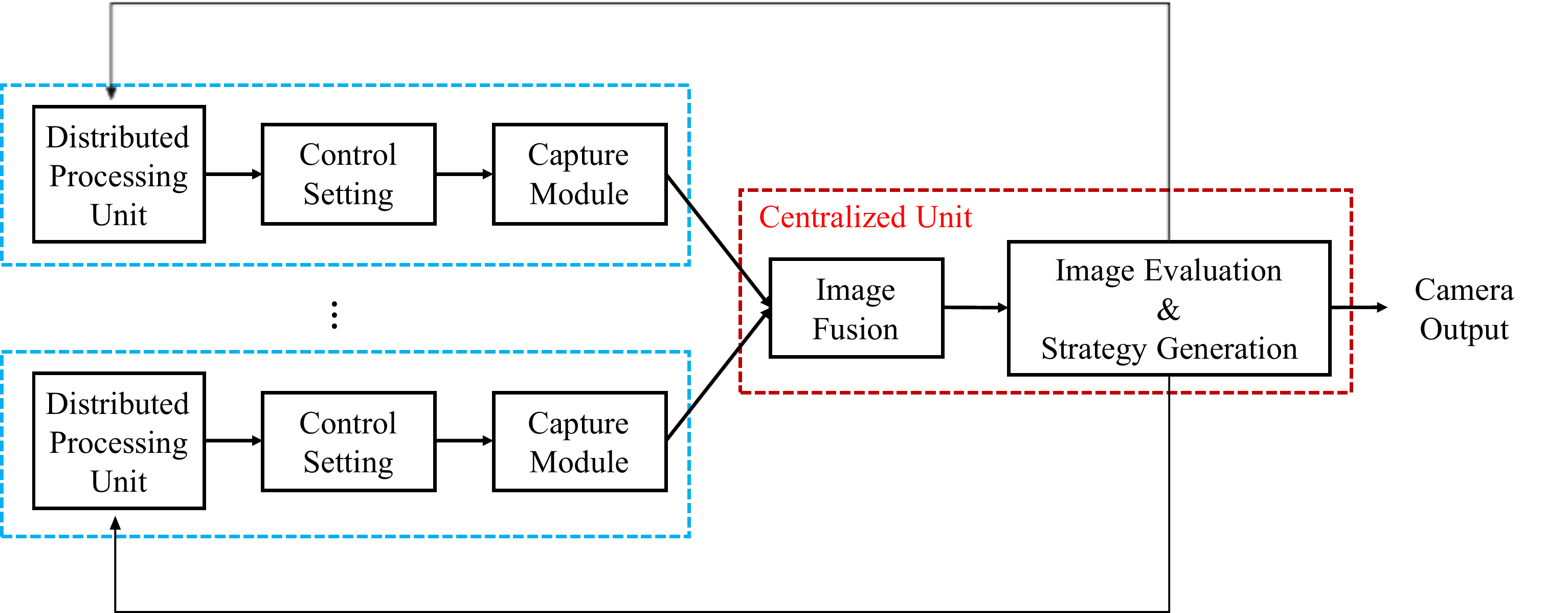}
	\caption{A distributed control processing architecture for array cameras. Each blue bounding box represents a camera, and a centralized unit (red bounding box) is required to collect and process the captured data from each camera.}
	\label{fig:general_control}
\end{figure}

Considering the highly distributed nature of the array cameras, a distributed control processing architecture is necessary. As illustrated in Fig.~\ref{fig:general_control}, each camera contains a processing unit which is supervised by a centralized unit. In the centralized unit, a fusing method generates the immediate output from the incoming frames, and this output is analyzed to determine the system configuration for the next time step. Each distributed processing unit receives the information from the centralized unit and generates the detailed updated configuration of this camera. In this structure, the centralized unit and the distributed processing units together compose a hierarchical agent.

\section{Image Formation}
\newcommand{\zyh}[1]{{\color{red}zyh: #1}}
\newcommand{\ytan}[1]{{\color{blue}ytan: #1}}
\newcommand{\lu}[1]{{\color{green}lu: #1}}
\newcommand{\mh}[1]{{\color{magenta}mh: #1}}
\label{sec:if}

While machine vision and image analysis are key applications of smart cameras,  this review focuses on  the impact of AI within traditional camera functions, specifically on the capture and rendering of physical images. Even within this context, we find the impact of AI to be revolutionary. Conventionally, even with aggressive post capture processing, the image formed by a camera closely resembles the physical light distribution on the focal plane. Smart cameras, in contrast, increasingly form images that are not directly proportional to the focal image. Such images are formed using disjoint data captured at multiple times and from multiple perspectives. Neural algorithms are the enabling component of this revolution.

Section \ref{sec:ds} showed that neural processing can improve conventional edge-based compression or, alternatively, can enable novel approaches to distributed and cloud-based data management. In this section, we similarly find that neural processing replaces and enhances traditional ISP pipelines, and also supports strategies without precedent in conventional cameras. Since smart cameras are likely to consist of microcamera arrays, we separate neural processing strategies into ``intra-camera algorithms'' aimed at improving performance and computational efficiency for traditional ISP tasks and ``inter-camera algorithms'' aimed at synthesizing media from multisensor data. Recognizing that active world-wide efforts are underway in both these categories, we attempt in this section to present a comprehensive snapshot of the current state-of-art for neural solutions to these challenges. 

As discussed in section \ref{sec:history}, image formation algorithms have evolved from linear methods through iterative optimization and Bayesian methods. Neural methods allow the incorporation of more sophisticated priors and more sophisticated transformations. The use of neural processing for traditional intra-camera ISP processes, such as demosaicing, denoising, tone mapping and upsampling produces marginal improvement in image distortion metrics, but may produce substantial improvements in perceived image quality. Recent research explores the fundamental tension between these two objectives~\cite{Blau_2018}. More broadly, neural processing ultimately enables the integration of the entire ISP pipeline into an integrated estimation process~\cite{schwartz2018deepisp}. 

Digital super-resolution is a central function of intra-camera image estimation algorithms. At its core, super-resolution is an exploration of the long standing issue of how best to display image data. While images are often displayed either at the raw capture resolution or in subsampled versions, sampling theory suggests that Lanczos or spline upsampling is appropriate for analysis of high resolution features~\cite{brady2009optical}. Numerous single and multiframe, linear and nonlinear methods have been used for image super-resolution. Recently, deep learning based super-resolution has become extremely popular. As discussed in \cite{Yang_2019} these methods improve both image distortion and perceptual quality relative to previous upsampling strategies, although there are limits to simultaneous improvement of both metrics. It is incorrect to think, however, that a 10 megapixel image upsampled from a 1 megapixel camera can be of equal quality to an optically resolved 10 megapixel image. The same methods that effectively upsample the 1 megapixel image can likely be applied to produce a high quality 100 megapixel image from the 10 megapixel version. Thus, digital super resolution is a compelling tool for display of image data, but not a replacement for high quality physical optics. 

As discussed below, single image super-resolution networks lead naturally to reference-based super resolution (RefSR), which uses a high resolution image taken from one perspective to super-resolve a low resolution image from a different perspective with overlapping field of view ~\cite{zheng2017learning}.  RefSR is the canonical problem for inter-camera image formation. In contrast with the marginal improvements that deep ISP offers for intra-camera processing, inter-camera image formation using neural processing enables revolutionary improvements in camera performance. Image stitching from multiple microcameras has been the core goal of array camera systems. Prior to deep learning methods, image stitching relied on feature mapping and image warping. These methods cannot fully account from multiview parallax. Conventional stitching relies on "transition zones" between distinct microcamera views and keeps the central part each physical camera image intact. As discussed below, RefSR-based methods transfer all microcamera hyperplanes to a global hyperplane to enable parallax-free stitching. More broadly, deep array camera processing enables logical combination of diverse data sources, such as images of different spectral ranges, radar and lidar or even object category to form images. Such processes have no analog in traditional physics-based image processing. RefSR also plays a key role in new approaches to the generation of unsampled view points from array camera data, including both spatial and temporal view point translation. 

Given the sheer volume of studies of deep learning based image formation, our present review is necessarily incomplete, but in the following subsections we attempt to present a reasonable cross section of the current state of the art. 

\subsection{AI for Intra-Camera Image Formation}
\begin{figure}[t]
	\centering
	\includegraphics[width=1.0\linewidth]{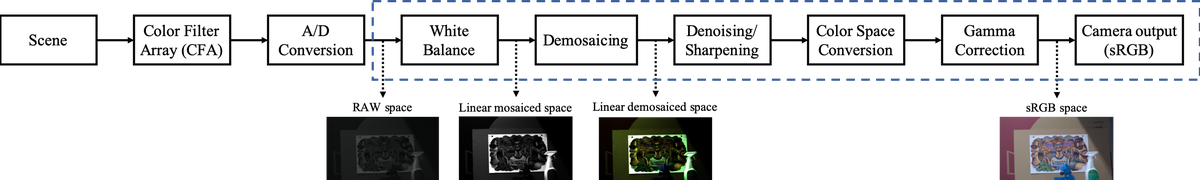}
	\caption{A camera pipeline \cite{khashabi2014joint}. The blue bounding box illustrates the functions of typical ISP.}
	\label{fig:ISP}
\end{figure}

The procedure to transform raw electrlnic sensor data into images interpretable by human eyes is referred to as image signal processing, ISP. While the detailed implementation varies in different cameras, the basic functions performed by ISP include demosaicing, denoising, white balance and contrast enhancement and tone mapping. Fig. \ref{fig:ISP} shows a typical pipeline of image signal processing in digital cameras. 

To build a reliable ISP, each function has to be carefully designed.  The performance of these operations can be significantly improved using neural processing. The following paragraphs discuss how AI facilitates these functions. However, designing different networks for each operation is still burdensome. The correlation among different tasks, such as denoising and demosaicing, encourages the design of multi-purpose networks. Heide, {\it et al.} \cite{heide2014flexisp} formulated the the functions of the ISP as an optimization problem and substituted the whole pipeline with an unified system. Schartz {\it et al.} divided the tasks of ISP into local and global operations respectively, which guided the design of an end-to-end two-step network~\cite{schwartz2018deepisp} . Similarly, Chen {\it et al.} designed a fully-convolutional network that substituted the whole ISP~\cite{chen2018learning} . Although these two networks focus on low-light images, their performance demonstrates the potential of learning based ISP. 

\begin{figure}[htbp]

\centering
  {\footnotesize
      \begin{minipage}[c]{1.0\linewidth}
    		\begin{minipage}[c]{0.258\linewidth}
    			\vspace*{6px}
    			\subfigure{
    				\stackunder[5pt]{\includegraphics[width=0.9\linewidth]{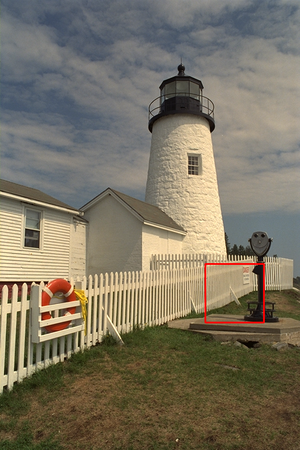}}{Original Image}
    			}		
    		\end{minipage} 
    		\begin{minipage}[b]{.738\linewidth}
    			\begin{tabular}[c]{c c c c}
    				\subfigure{
    					\stackunder[4pt]{\includegraphics[width=0.2\linewidth]{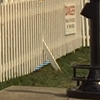}}{AHD~\cite{hirakawa2005adaptive}}}		&
    				\subfigure{
    					\stackunder[4pt]{\includegraphics[width=0.2\linewidth]{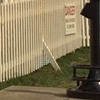}}{GBTF~\cite{pekkucuksen2010gradient}}}	 &
    				\subfigure{
    					\stackunder[4pt]{\includegraphics[width=0.2\linewidth]{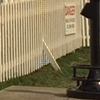}}{DLMMSE~\cite{zhang2005color}}}	 &
    				\subfigure{
    					\stackunder[4pt]{\includegraphics[width=0.2\linewidth]{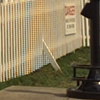}}{LDI-NAT~\cite{zhang2011color}}}	\\
    				\subfigure{
    					\stackunder[4pt]{\includegraphics[width=0.2\linewidth]{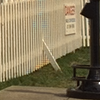}}{MLRI~\cite{kiku2014minimized}}}		& 
    				\subfigure{
    					\stackunder[4pt]{\includegraphics[width=0.2\linewidth]{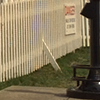}}{ARI~\cite{monno2015adaptive}}}	 & 
    				\subfigure{
    					\stackunder[4pt]{\includegraphics[width=0.2\linewidth]{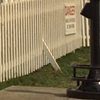}}{DRL~\cite{tan2017color}}}	 &
    				\subfigure{
    					\stackunder[4pt]{\includegraphics[width=0.2\linewidth]{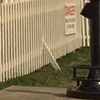}}{Ground Truth}}	
    			\end{tabular}
    		\end{minipage}
    	\end{minipage}
    \caption{The results of representative demosaicing methods, including adaptive homogeneity-directed demosaicing (AHD)~\cite{hirakawa2005adaptive}, gradient based threshold free color filter array interpolation (GBTF)~\cite{pekkucuksen2010gradient}, directional linear minimum mean square-error estimation (DLMMSE)~\cite{zhang2005color}, local directional interpolation and nonlocal adaptive thresholding (LDI-NAT)~\cite{zhang2011color}, minimized-laplacian residual interpolation (MLRI)~\cite{kiku2014minimized}, adaptive residual interpolation (ARI)~\cite{monno2015adaptive} and deep residual learning (DRL)~\cite{tan2017color}. The zoomed region of `railings' demonstrates the success of deep learning in  \cite{tan2017color} significantly reducing artifacts. \label{fig:example:demosaicing} }
	}
\end{figure}

\textbf{Demosaicing.} Simple silicon detectors are relatively insensitive to color. A camera capable of capturing full-color information, such as a 3-CCD camera, deploys a beamsplitter and multiple sensors. To reduce system volume and cost, it is common to use a single sensor covered with a color filter array (CFA). In this manner, each pixel only records the information about one color channel: red, green or blue. The inference of the missing channels for each pixel based on the neighboring pixels is termed demosaicing. Traditional demosaicing methods can be classified into the following categories: spatial interpolation methods, frequency analysis methods, wavelet-based methods, optimization methods, heuristic methods and residual space methods. Detailed survey of each method can be found in \cite{li2008image,menon2011color,syu2018learning}. However, these methods either require hand-crafted features or depend on the empirical assumptions. In contrast, learning-based methods may implicitly utilize the intrinsic correlation among pixels. Early methods include artificial neural network~\cite{kapah2000demosaicking,wang2014multilayer}, Support Vector Machine~\cite{he2012self} and Markov Random Fields (MRF)~\cite{sun2012separable}. Recently, the CNNs have shown superior performance in demosaicing. Tan et al. \cite{tan2017color} proposed a CNN that integrated the residual learning and the prior knowledge of the Bayer pattern CFA. Syu et al. \cite{syu2018learning} utilized deep convolutional networks to address a more general demosaicing problem which can be applied to different CFA patterns. The relative advantage of deep learning methods is illustrated in Fig. \ref{fig:example:demosaicing}, which  compares several demosaicing methods. The lack of color artifacts along straight or periodic features is achieved with AI-based methods. 

\textbf{Denoising.} Denoising is, of course, an indispensable part of ISP. Methods that exploit images priors have been widely employed ~\cite{zhang2017beyond}. Commonly used methods include total variation~\cite{rudin1992nonlinear,osher2005iterative}, sparse representation~\cite{elad2006image,dong2012nonlocally}, Markov random field modeling~\cite{lan2006efficient} and non-local self-similarity~\cite{dong2012nonlocally,buades2005non,dabov2007image,buades2008nonlocal,mairal2009non}. Other methods include shrinkage fields~\cite{schmidt2014shrinkage}, trainable nonlinear reaction diffusion~\cite{chen2016trainable} and regression tree fields~\cite{schmidt2015cascades}. The disadvantages of these methods are threefold: 1) high computational cost, 2) cumbersome parameter tuning, and 3) low compatibility to different noise types or noise levels. Research in learning-based denoising algorithms has shown greater potential in performance. Early in 2008, a convolutional network was proposed and shown to achieve stronger representational power than Markov random fields~\cite{jain2009natural}. Later, multi layer perception~\cite{burger2012image} and auto-encoder approaches~\cite{xie2012image} were applied successfully to the denoising problem. \cite{ahn2017block} combined the non-local self-similarity prior with CNNs. Matched patches were first integrated before being fed into the convolutional network. \cite{zhang2018ffdnet} proposed a flexible way of deploying CNNs by using down-sampled images as well as a noise level map as input. Lefkimmiatis {\it et al.} \cite{lefkimmiatis2018universal} formulated  denoising as a constrained optimization problem and only a shallow network was required. Recently, the most popular technique is to apply residual learning to directly learn the noise from the image~\cite{zhang2017beyond,zhang2017learning,tian2019enhanced}.

\begin{figure}[htbp]
  \centering
  \subfigure[Noisy image]{
    \label{fig:example:joint_1} %
    \includegraphics[width=1.14in]{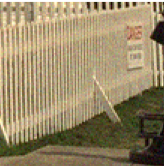}}
 \hspace{0.05in}
  \subfigure[TLSdemosaic \cite{hirakawa2006joint}]{
    \label{fig:example:joint_2} %
    \includegraphics[width=1.14in]{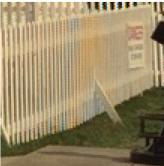}}
 \hspace{0.05in}
 \subfigure[TV \cite{condat2012joint}]{
    \label{fig:example:joint_3} %
    \includegraphics[width=1.14in]{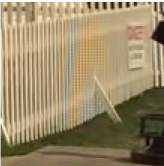}}
 \hspace{0.05in}
 \subfigure[Learning method \cite{gharbi2016deep}]{
    \label{fig:example:joint_4} %
    \includegraphics[width=1.14in]{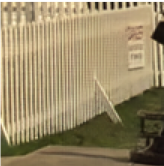}}
  \caption{Comparisons among TLSdemosaic \cite{hirakawa2006joint}, TV \cite{condat2012joint} and a learning based method \cite{gharbi2016deep} in joint demosaicing and denoising.}
  \label{fig:example:joint} %
\end{figure}

Denoising and demosaicing are typically considered separately. However, performing demosaicing prior to denoising complicates the latter by correlating the noise~\cite{park2009case}, while performing denoising first leads to degradation in the quality of the demosaiced images~\cite{danielyan2009cross}. To better cope with the coherence between the two tasks, joint demosaicing and denoising was proposed~\cite{hirakawa2006joint,park2009case,condat2012joint}. Recent research mainly focuses on the data-driven learning techniques. Khashabi {\it et al.}~\cite{khashabi2014joint} addressed the problem by learning a regression tree model. While the model was performed in the linear space, the performance was optimized in the sRGB space, which agreed with the actual pipeline in digital cameras. Klatzer {\it et al.} \cite{klatzer2016learning} defined the sequential energy minimization model, and the energy function was selected by regarding demosaicing as an inverse problem. Gharbi et. al~\cite{gharbi2016deep} proposed a deep CNN to jointly denoise and demosaic raw data, and the performance of the network was further improved by the reappearance of challenging data. Kokinos {\it et al.}~\cite{kokkinos2018deep, kokkinos2018iterative} employed a majorization-minimization framework and solved the problem in an iterative manner, where the minimization of the majorizer was addressed as a denoising problem using a neural network. Dong {\it et al.}~\cite{dong2018joint} applied a generative adversarial network (GAN) which optimized the perceptual quality of the images. Fig.~\ref{fig:example:joint} shows the comparison between two traditional methods and a deep learning based method.

As discussed in more detail below, image composition from bursts of images has shown substantial utility in improving image quality. Based on this idea, 
Ehret et al.~\cite{ehret2019joint} proposed a joint demosaicking and denoising strategy that explored the similarities between multiple images without the need for ground truth training. 

\textbf{White Balance and Contrast Enhancement.} Digital images are dependent on  illumination~\cite{barnard2002comparison}. Objects appear to be either reddish if the light source is low in color temperature, or bluish if the color temperature is high~\cite{weng2005novel}. White balance aims to `compensate' for the temperature variation and produce an image as taken under canonical illumination. The typical white balance pipeline is to estimate the illumination before image color is balanced \cite{lam2018automatic}. Popular automatic white balance approaches include the grey world method~\cite{huo2006robust}, the retinex theory method~\cite{land1977retinex,lam2005combining}, the fuzzy rule method~\cite{liu1995automatic} and the histogram stretching method~\cite{wang2011fast} and correlation-based methods~\cite{tai2012automatic}. While their performance is outstanding, practical use of correlation-based methods is limited by computational cost. A detailed survey of automatic white balance in digital photography can be found in \cite{lam2018automatic}. Some machine learning techniques have been investigated for this problem as well, including neural networks, support vector regression and ridge regression~\cite{agarwal2007machine}. Bianco {\it et al.}~\cite{bianco2015color} proposed to learn  color balance parameters using a CNN. Shi {\it  et al.}~\cite{shi2016deep} proposed to use two sub-networks to estimate the illumination. Hu {\it et al.}~\cite{hu2017fc4} proposed a fully-connected structure which avoided the information insufficiency in patch-based methods. Bianco {\it et al.} then  designed a 3-step pipeline which was able to deal with multiple illuminations~\cite{bianco2017single}. Afifi {\it et al.}~\cite{afifi2018semantic} proposed to balance the color by exploiting the semantic information of the scene.

Apart from the white balance, contrast enhancement is typically required in ISP as well to produce visually compelling images. Classical methods may be separated into local enhancement and global enhancement. Local methods apply suitable operations to different regions of the image, while the global methods are more adopted in practical systems for simpler implementation and computation~\cite{rahman2016adaptive}. Among the global enhancement methods, gamma correction is the most commonly used operation~\cite{khashabi2014joint}. Advances in gamma correction mainly focus on designing adaptive systems that determine the parameters for gamma correction based on the inputs.

\textbf{Super-resolution.} Image sharpening and upsampling are also important components of traditional ISP. Conventionally these are implemented with linear interpolation~\cite{parker1983comparison}, potentially augmented with sharpening or sparsity-based optimization. With AI-based processing, it has become common to refer to image upsampling as "super-resolution" and to consider the process as statistical estimation of high resolution features. Super-resolution is closely related to demosaicing and denoising~\cite{DBLP,zhang2017beyond}. Prior to wide-spread application of deep-learning, state-of-the-art single image super-resolution (SISR) methods exploited the similarity within an image~\cite{freedman2011image,glasner2009super,yang2013fast}. Dong {\it et al.}~\cite{dong2014learning} first proposed a CNN solution for super-resolution, and the network was shown to be another representation of sparse-coding-based super-resolution~\cite{yang2008image,yang2010image}. Since then, various techniques have been applied to super-resolution: residual learning~\cite{kim2016accurate,lim2017enhanced,zhang2018residual,wang2018esrgan,zhang2018image}, generative adversarial networks (GAN)~\cite{ledig2017photo,wang2018esrgan}, recursive networks~\cite{kim2016deeply,tai2017image}, channel attention~\cite{zhang2018image}, back propagation~\cite{haris2018deep} and dense networks~\cite{tong2017image,zhang2018residual}.

\begin{figure}[htbp]
  \centering
    \includegraphics[width=\linewidth ]{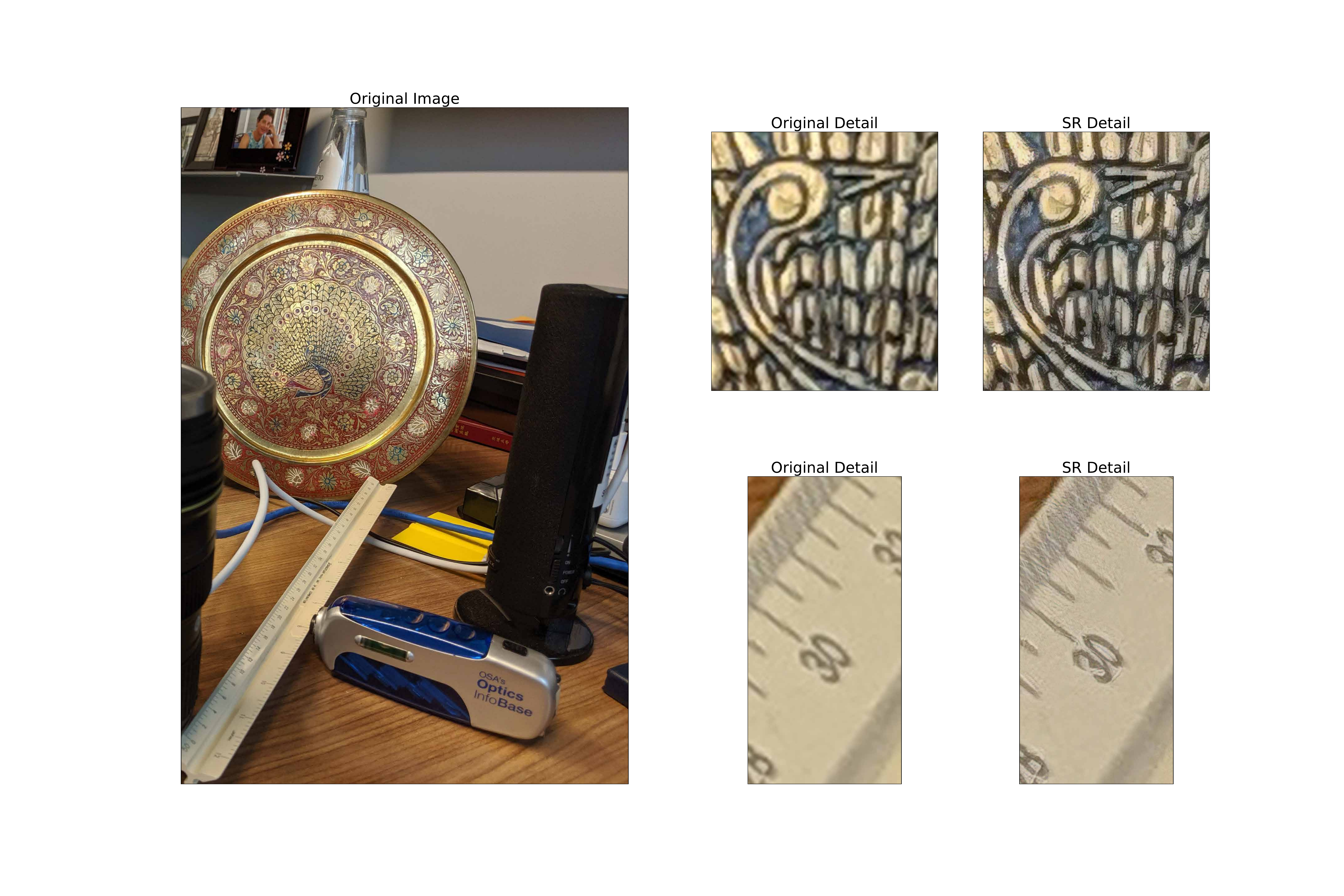}
  \caption{Singe image super-resolution using the generative adversarial network described in~ \cite{ledig2017photo}.}
  \label{fig:sisr}
\end{figure}

As discussed in \cite{Yang_2019}, single image super resolution networks may be designed to minimize image distortion (e.g., pixel-level loss) or to maximize perceptive image quality. These two metrics conflict, however \cite{Blau_2018}, cannot be simultaneously optimized. In photography, perceptual image quality has long been emphasized over pixel-level fidelity in tone mapping and other processes, so it is not surprising that this remains the case in smart camera image processing. GAN-based super resolution in particular typically adds perception-based loss functions in network training. As an example, Fig. \ref{fig:sisr} shows an image super-resolved using the GAN network described by Ledig~
\cite{ledig2017photo} {\it et al.}. The original image is shown at left, details of the image upsampled 4x by cubic spline interpolation and by the GAN network are shown at right. As described by Ledig, the GAN approach may actually be worse than simpler convolutional networks according to pixel level loss, but the visual quality is by design much higher. In the details shown here, the natural head of the peacock appears much clearer after GAN processing, but artificial objects like the lines on the ruler show clear artifacts from super-resolution processing. Of course, deep super resolution is still in its infancy. A more advanced AI algorithm could recognize the nature of the underlying object and apply super resolution algorithms appropriate to the object itself, recognizing that rulers consist of lines and peacocks consist of curves, for example. 

Where most studies of single image super-resolution are based on supervised learning and use a known blur to downgrade an image and then train a network to correct that blur, real images include field and range dependent blur. With this in mind, note that Fig.\ref{fig:sisr} is simple cell phone photo processed without regard to the original network training parameters. Despite this, deep super resolution markedly improves the visual quality of some areas of the image. In addition to training deep networks for object-dependent processing, training for range and camera-dependence may be expected to yield further improvements. 

Of course, that the fact that success of deep learning super super resolution does not suggest that cameras with poor modulation transfer are just as good as cameras with higher resolution. In fact, high quality physical resolution remains as important as ever. Deep learning super-resolution certainly expands the number of pixels that should be used to display an image, meaning that with super resolution the cell phone 12 megapixel camera used to capture Fig. \ref{fig:sisr} should be displayed using 48 megapixels. But using the sample upsampling algorithms, one may reasonably display a camera with MTF to support Nyquist sampling at 48 megapixels would reasonably be displayed at 200 megapixels. This means that deep learning super resolution has become an essential tool in effective display of photographic data, but physical resolution remains as important as ever. 
With this illustration, we clearly see how AI processing goes beyond simple improvements to traditional image processing to radically change the nature of photography. Where conventional intra-camera signal processing seeks to invert physical distortions, deep learning analyzes the sensed data to statistically estimate the original object. As we discuss in the next subsection, this potential becomes substantially more pronounced when we expand our sampling arsenal to include multiple apertures and multiple times. 

\subsection{AI for Inter-camera Image Formation}
The ability of camera arrays to efficiently capture higher space-bandwidth product than single cameras has led to numerous multiscale and hybrid designs~\cite{brady2018parallel}. These systems play vital roles in computational photography, including light field imaging, 360 VR camera and gigapixel videography.
One of the critical tasks in multiscale hybrid imaging is matching and fusing cross-resolution images from different cameras under perspective parallax. Prior to the explosion of deep learning methods, multisensor data fusion was relatively crude, but with AI methods one can reasonably assume that images captured from multiple times and multiple perspectives can be fused into a single world-view. 
 
\begin{figure}[htbp]
  \centering
  \subfigure[Template Matching \cite{briechle2001template}]{
    \label{fig:correspondence:fastncc}
    \includegraphics[width=2in]{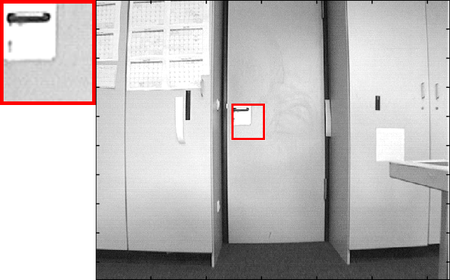}}
  \subfigure[SURF \cite{bay2006surf}]{
    \label{fig:correspondence:surf}
    \includegraphics[width=2in]{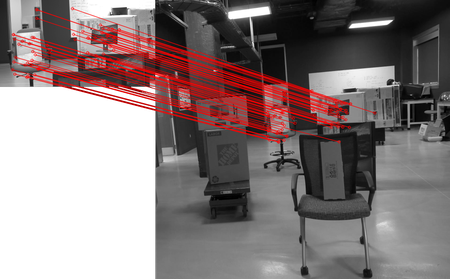}}

  \subfigure[Variational Optical Flow \cite{brox2010large}]{
    \label{fig:correspondence:browflow} 
    \includegraphics[width=2.1in]{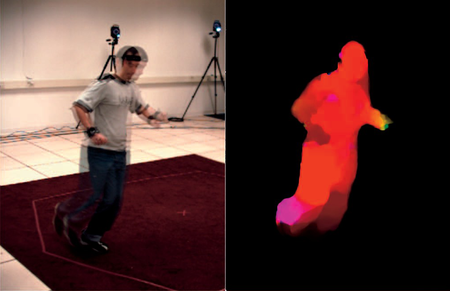}}
  \subfigure[PWC-Net Optical Flow \cite{sun2018pwcnet}]{
    \label{fig:correspondence:pwcnet}
    \includegraphics[width=1.9in]{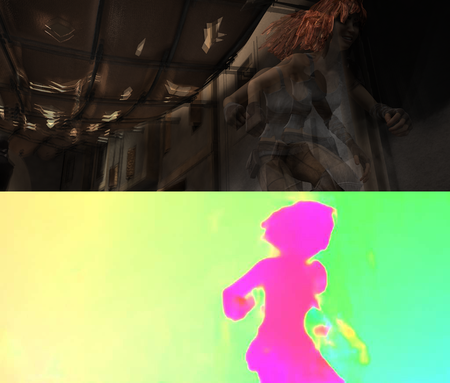}}
  \caption{Illustration of representative methods for image correspondences. For optical flow demonstration, the left/upper image is the transparent overlay of input frames, and the right/bottom image is the flow estimated.}
  \label{fig:correspondence:joint} 
\end{figure}

Conventionally, the key to information fusion from camera arrays is to build correspondences between images. Such correspondences may base on patch matching, pixel feature matching or pixel matching. Patch level correspondence can be achieved by simple template matching, but it may require a set of templates to deal with scaling and rotation, and it usually can not handle smooth regions. Feature level correspondences are obtained by matching features via feature descriptors such as Scale Invariant Features (SIFT)~\cite{lowe1999sift}, Speeded Up Robust Features (SURF)~\cite{bay2006surf}, ORB \cite{rublee2011orb} and DAISY \cite{tola2010daisy}. These approaches are robust to scaling and rotation, but the features are usually sparsely represented. Pixel level correspondences, which are usually represented as optical flow maps (or disparity maps for stereo pairs), enable dense matching. The flow maps can be calculated  via conventional methods \cite{horn1981opticalflow, lucas1981lucaskanade, black1996robustflow, brox2010large}, but neural-network-based methods \cite{dosovitskiy2015flownet,ilg2017flownet2,ranjan2017spynet,sun2018pwcnet} generally outperform their conventional ancestors. Fig. \ref{fig:correspondence:joint} demonstrates several representative methods for feature correspondence, where (a) \cite{briechle2001template} and (b) \cite{bay2006surf} are patch-level and feature-level schemes, respectively. (c) Variational Optical Flow \cite{brox2010large} and (d) PWC-Net \cite{sun2018pwcnet} are pixel-level schemes, via conventional and learning-based pipelines, respectively. It can be shown that (d) outperforms (c) in delivering more accurate and smooth flow map, even if a larger movement happens to both foreground and background objects.

Neural multi-image fusion does not necessarily require high resolution image to image correspondence. Conventional fusion algorithms align, warp  and blend images, meaning that at the pixel level the final result is still basically proportional to original data. Neural algorithms, in contrast, can generate pixel level data. This means, for example, that neural algorithms can generate color images from monochrome~\cite{zhang2016colorful}, generate 3D scenes from 2D photographs~\cite{eigen2014depth} and generate images of horses from images of zebra~\cite{isola2017image}. While such processes may in many cases generate "fake" views, they are also powerfully useful in generating best estimates of true views that traditional correspondence and blending strategies cannot achieve. Beyond simple image blending, neural systems can be used with diverse data arrays to, for example, improve infrared images using visible images, infer high resolution color images by combining low resolution color and high resolution monochrome, infer high resolution 3D from low resolution LIDAR and high resolution visible, etc. To keep the length of this review to reasonable bounds, we limit our attention in this section to three generative problems: high resolution panoramic image generation using multiple apertures, high frame rate video generation using multiple frames and novel view point generation using multiple apertures. In each case the novelty of the neural process is that it is able to estimate true scene values from generative processes. 

\subsubsection{Panoramic imaging}
\begin{figure}[htbp]
\centering
\begin{minipage}[c]{0.52\linewidth}
  \centering
  \subfigure[Input images]{
    \label{fig:brown:input}
    \includegraphics[width=0.48\linewidth]{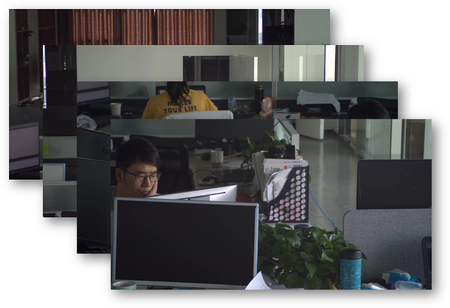}}
  \subfigure[Image with features]{
    \label{fig:brown:feature}
    \includegraphics[width=0.48\linewidth]{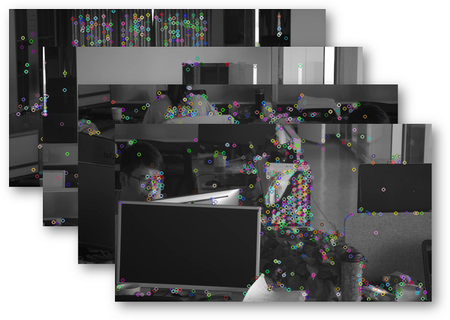}}
  \subfigure[Matched features between adjacent images]{
    \label{fig:brown:matching}
    \includegraphics[width=\linewidth]{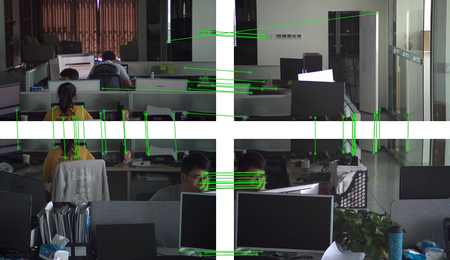}}
\end{minipage}
~ %
\begin{minipage}[c]{0.46\linewidth}
  \subfigure[Warped and aligned images]{
    \label{fig:brown:alighed}
    \includegraphics[width=0.95\linewidth]{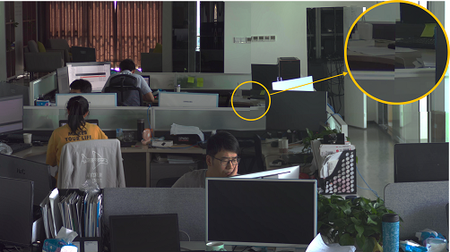}}
  \subfigure[Blended result]{
    \label{fig:brown:blended}
    \includegraphics[width=0.95\linewidth]{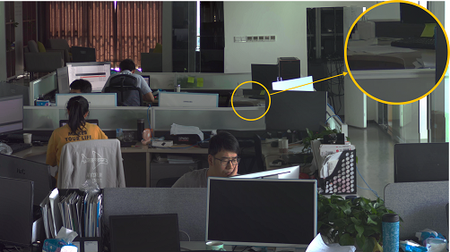}}
\end{minipage}    
  \caption{Illustration of a representative stitching pipeline based on the overlap of adjacent images \cite{brown2007automaticpano}.}
  \label{fig:brown:joint}
\end{figure}

Panoramic image formation is the most intuitive of the images captured by heterogeneous microcameras. While physical stitching by mosaicing actual prints dates to the 19$^{\text{th}}$ century, digital image stitching became popular in the 1990s~\cite{teodosio1993panoramic,mann1994virtualbellows, szeliski1994imagemosaicing, capel1998automatedmosaicing}. The most universal pipeline was first proposed by Brown \textit{et al.} in 2003~\cite{brown2003recognisingpano}, which was further improved in 2007~\cite{brown2007automaticpano}. This pipeline contains four major steps: feature extraction, feature matching, image warping and alignment, and image correction and blending. The features are first extracted and matched from the overlapping areas between adjacent images and used to infer the geometrical relationship between images. After proper alignment the images are fused into one large panorama. SIFT \cite{lowe1999sift} is often used for feature extraction and matching, inferred single homography for perspective warping, and performed multi-band blending \cite{burt1983multiband}. The pipeline is illustrated in Fig.~\ref{fig:brown:joint}. Regardless more and more tools and techniques have been introduced into the image stitching task, such as new feature descriptors \cite{bay2006surf, rublee2011orb}, non-uniform warping algorithms \cite{zaragoza2013projective, lin2015adaptive}, more parallax-robust blending methods \cite{kwatra2003graphcut} etc, the overall pipeline rarely changed. Such a pipeline computes correspondences between every image pair, and requires jointly global optimization namely bundle adjustment in warping and alignment process.

\begin{figure}[htbp]
    \centering
    \includegraphics[width=1\linewidth]{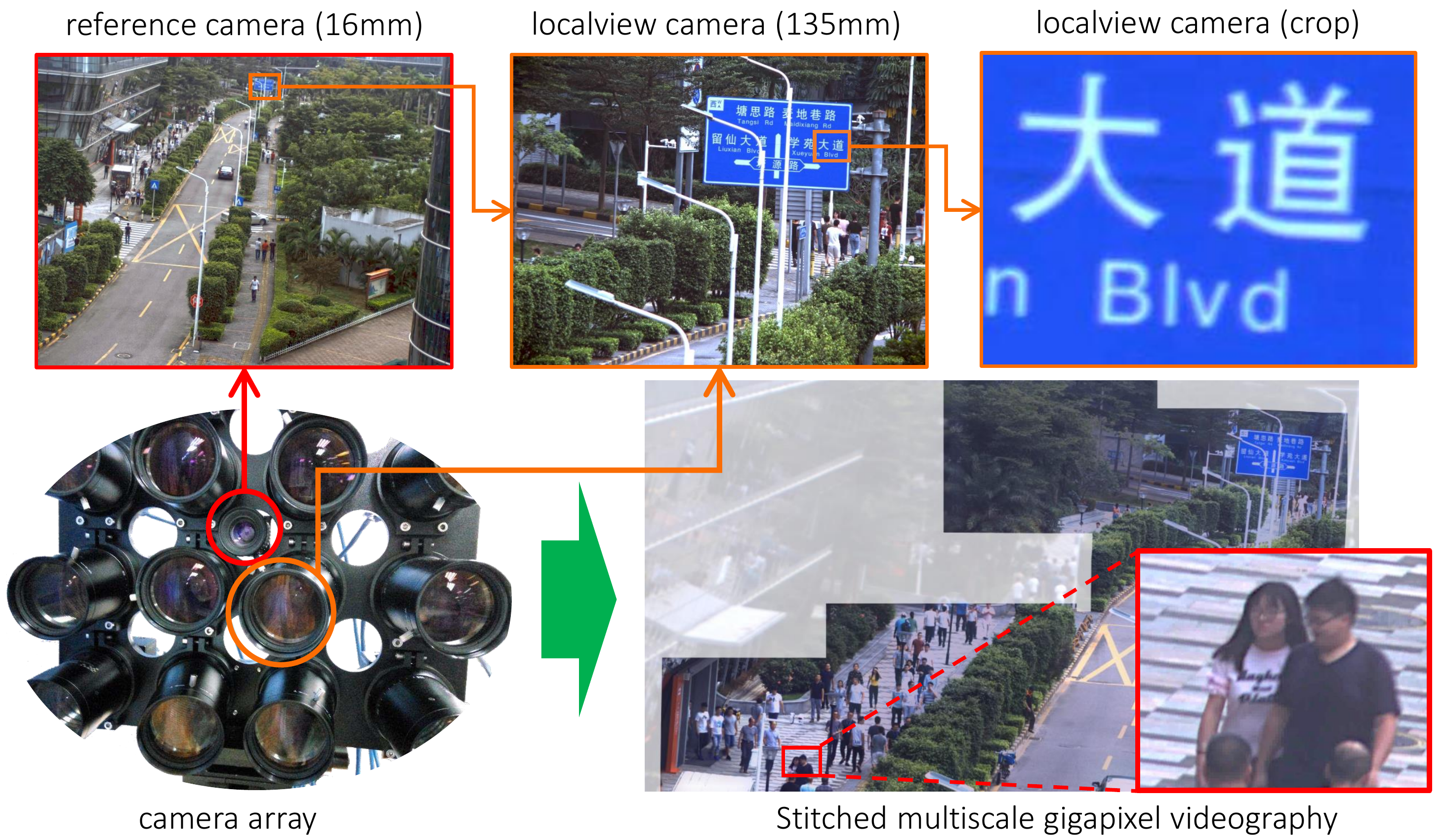}
    \caption{Illustration of cross-scale stitching scheme \cite{yuan2017multiscale}. The input images consist of a low-res large-FoV global image and a set of high-res small-FoV local images. The local images are warped and corrected according to its corresponding part in the global image. }
    \label{fig:yuan:joint}
\end{figure}

Contrary to the high computational burden required in global optimization, Yuan \textit{et al.} proposed a highly efficient local optimization based stitching pipeline \cite{yuan2017multiscale}. It inferred the geometrical relationship between small-FoV high-resolution ``local images'' from a wide-FoV low-resolution ``global image''. In this way, the overlaps among local images were no longer required, and the global consistence is maintained if each local image is warped to the geometry of the global image. An example of this scheme is illustrated in Fig.~\ref{fig:yuan:joint}. Such a cross-resolution, hierarchical pipeline differs largely from the conventional pipeline, in which the images are captured with uniform spatial sampling rate and treated equally. Moreover, the local method is naturally suitable for parallel computing, which is critical in real-time image stitching.

It is worth-noting that, in Yuan's algorithm, the cross-scale correspondences between global image and local images were calculated via patch matching \cite{yoo2009fastncc} and variational optical flow estimation, mesh-based warping, and seam-line driven blending \cite{kwatra2003graphcut}. Neural networks offer handy tools to replace each element within the pipeline. For example, the learning-based patch matching \cite{zbontar2015nnpatch,zagoruyko2015nnpatch} or descriptor matching~\cite{fischer2014descriptor} can be adopted for feature correspondence. Neural networks can also estimate a homography directly from image pair~\cite{detone2016homographynet}. Image segmentation networks \cite{long2015cnnseg,chen2017deeplab,badrinarayanan2017segnet} may be utilized to improve the result of GraphCut \cite{kwatra2003graphcut}, leading to better parallax-robust image blending methods. 

More generally, neural networks enable completely new approaches to the image stitching or fusion task. These methods are generative rather than pixel-based and have only weak reliance on pre-aligned correspondence. They allows us to organize, process, or fuse information from very different sources and finally generate new images that are more visually satisfying and accurate than conventional panoramic fusion. As outlined above, previous approaches use "warped, aligned and blended'' local images, no ``new'' pattern or image is introduced. Using neural networks, we can alternatively regard the stitched image as the ``super-resolved'' global image, in which the texture information is transferred from local images. This method utilizes RefSR networks, which would be introduced below.

Reference based superresolution, RefSR, consists of rendering a high resolution image from the view point of a low resolution image taken from a different spatial or temporal view point~\cite{SS-Net}. 
The high-resolution details in the reference image usually enable RefSR to obtain superior results compared to SISR. RefSR has been successfully applied in light-field reconstruction~\cite{PatchMatch,SR_lr_TVCG16,SS-Net} and gigapixel video synthesis~\cite{yuan2017multiscale}. 
\begin{figure}[t]
	\centering
	\includegraphics[width=0.8\linewidth]{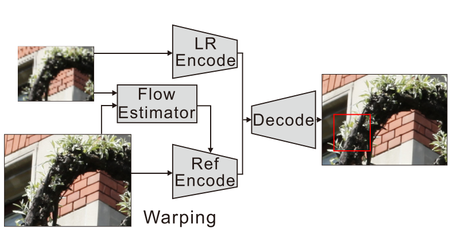}
	\caption{Illustration of representative reference-based super resolution (RefSR) scheme CrossNet \cite{zheng2018crossnet}, which performs multiscale warping for feature alignment and synthesis in an end-to-end fashion.}
	\label{fig:refsr_pipeline}
\end{figure}

RefSR boils down to two sub-problems, i.e.,  correspondence between the two images and high-resolution synthesis of the LR image. Previous approaches \cite{NLM,PatchMatch,SR_lf_VCIP15,zheng2017learning,ICCVW2017} rely on `patch-matching + patch-synthesis'. Patch-matching is based on either gradient features or CNN learned features to identify the correspondences between the down-sampled patches of HR image and LR image, which is followed by the patch averaging for image synthesis. 
Boominathan \textit{et al.}~\cite{PatchMatch} adopted a high-resolution image captured by DLSR as the reference and used a patch-based synthesis algorithm via non-local means \cite{NLM} for super-resolving the low-resolution light-field images. Wu \textit{et al.}~\cite{SR_lf_VCIP15} improved it by employing patch registration before the nearest neighbor searching, then applied dictionary learning for reconstruction. Wang \textit{et al.}~\cite{SR_lr_TVCG16} iterated the patch synthesizing step of \cite{PatchMatch} for enriching the exemplary database. Zheng \textit{et al.}~\cite{ICCVW2017} decomposed images into sub-bands by frequencies and applied patch matching for high-frequency sub-band reconstruction. In 2017, Zheng \textit{et al.}~\cite{zheng2017learning} proposed a learning based approach for the cross-resolution patch matching and synthesis, which significantly boosted the accuracy of RefSR. 
However, such schemes have fundamental limitations. Firstly, the adopted sliding averaging window blurs the output image and causes grid artifacts. Moreover, patch-based synthesis is inherently incapable of handling the non-rigid image deformation caused by viewpoint changes. Finally, the sliding window searching that is required by the patch-based matching is inherently inefficient, especially in large disparity cases.

To impose the non-rigid deformation to patch-based algorithms, the approach proposed in \cite{SR_lr_TVCG16} enriches the reference images by iteratively applying non-uniform warping before the patch synthesis. However, directly warping between the low and high-resolution images is inaccurate, and such an iterative combination of patch matching and warping introduces heavy computational burden, around $O(d^2)$ where $d$ is the disparity. For a $320 \times 520$ image, image synthesis required over 30 minutes computation. In 2018, Zheng \textit{et al.}~\cite{zheng2018crossnet} proposed an end-to-end pipeline named CrossNet to address the problem of fast and accurate high-frequency details transferring. Instead of a flow estimator to predict the optical flow map between the LR image and the reference image, Zheng {\it et al.} generated multiscale flow maps in feature domain and performed feature transferring through backward warping. The multiscale warping strategy benefits the high-frequency details transferring in a more fine-grained way. As shown in Table \ref{Table:Refsr quantitative comparison} and Fig. \ref{fig:RefSR on light field data}, RefSR algorithms outperform SISR methods by a large margin. Meanwhile, the flow-based RefSR method CrossNet~\cite{zheng2018crossnet} also leads to the superior performance compared to patch-based methods.

\begin{table}[htbp]
\begin{minipage}{\linewidth}
 \caption{
 Quantitative comparisons between SISR and RefSR on light-field dataset Flower~\cite{Flower}, under different viewpoints (1,1) and (7,7). Both patch-based and flow-based RefSR approaches are considered. }
 
 \label{Table:Refsr quantitative comparison}
 \footnotesize

\centering
\begin{tabular}{c m{2.0cm} c ccc ccc}
\toprule
 \multirow{3}{*}{Frameworks} & \multirow{3}{*}{Methods} &  \multirow{3}{*} {Scale}&\multicolumn{3}{c}{Flower (1,1)} & \multicolumn{3}{c}{Flower (7,7)} \\
\cmidrule(l{2pt}r{2pt}){4-9}
& & &PSNR & SSIM & IFC & PSNR & SSIM & IFC \\
\midrule
\multirow{3}{*}{SISR} & SRCNN \cite{SRCNN}	& $8\times$ & 28.17 & 0.77 & 0.98 & 28.25 & 0.77 & 1.00  \\
& VDSR \cite{VDSR}	& $8\times$ & 28.58 & 0.78 & 1.04 & 28.68 & 0.78 & 1.06  \\	
& MDSR \cite{MDSR}	& $8\times$ & 29.15 & 0.79 & 1.17 & 29.26 & 0.80 & 1.19  \\	\midrule				
\multirow{2}{*}{Patch-based} & PatchMatch \cite{PatchMatch} & $8\times$ & 35.26 & 0.95 & 4.00 & 30.41 & 0.85 & 2.07 \\
& SS-Net \cite{zheng2017learning}& $8\times$&37.46 & 0.97 & 4.72 & 32.42 & 0.91 & 2.95 \\
\midrule
 Flow-based & CrossNet \cite{zheng2018crossnet} & $8\times$& 40.31 & 0.98 & 5.74 & 34.37 & 0.93 & 3.45 \\

\bottomrule
\end{tabular} 
\end{minipage}	 
\end{table}

\newcommand{\figw}{0.19}
\begin{figure}[!htb]
	\setlength{\belowcaptionskip}{-10pt}
	\centering
		\begin{tabular}[c]{c}
        \footnotesize
    	\begin{minipage}[c]{1.0\linewidth}
    		
    		\begin{minipage}[c]{0.278\linewidth}
    			\vspace*{4px}
    			\subfigure{
    				\stackunder[5pt]{\includegraphics[width=0.91\linewidth]{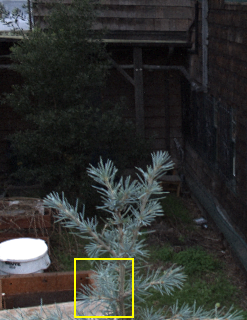}}{Ground-truth HR}
    
    			}	
    			
    		\end{minipage} 
    		\begin{minipage}[b]{.718\linewidth}
    			\begin{tabular}[c]{c c c c}
    				\subfigure{
    					\stackunder[5pt]{\includegraphics[width=\figw\linewidth]{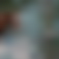}}{LR}}		& 				
    				\subfigure{
    					\stackunder[5pt]{\includegraphics[width=\figw\linewidth]{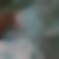}}{SRCNN\cite{SRCNN}}}	 & 
    				\subfigure{
    					\stackunder[5pt]{\includegraphics[width=\figw\linewidth]{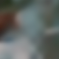}}{VDSR\cite{VDSR}}}	 &
    				\subfigure{
    					\stackunder[5pt]{\includegraphics[width=\figw\linewidth]{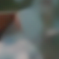}}{MDSR\cite{MDSR}}}	\\
    				\subfigure{
    					\stackunder[5pt]{\includegraphics[width=\figw\linewidth]{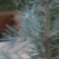}}{PatchMatch\cite{PatchMatch}}}		& 				
    				\subfigure{
    					\stackunder[5pt]{\includegraphics[width=\figw\linewidth]{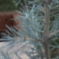}}{SSNet\cite{zheng2017learning}}}	 & 
    				\subfigure{
    					\stackunder[5pt]{\includegraphics[width=\figw\linewidth]{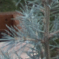}}{CrossNet\cite{zheng2018crossnet}}}	 &
    				\subfigure{
    					\stackunder[5pt]{\includegraphics[width=\figw\linewidth]{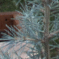}}{GT}}
    			\end{tabular}
    		\end{minipage}
	\end{minipage}
		\end{tabular}
	\caption{Qualitative comparisons between SISR and RefSR methods. SISR methods include SRCNN~\cite{SRCNN}, VDSR~\cite{VDSR} and MDSR~\cite{MDSR}. RefSR methods cover the patch-based PatchMatch~\cite{PatchMatch}, SSNet~\cite{zheng2017learning}, and the flow-based CrossNet~\cite{zheng2018crossnet}.}
	\label{fig:RefSR on light field data}
\end{figure}

\begin{figure}[htbp]
    \centering
    \subfigure[Low-res global image]{
    \includegraphics[width=0.455\linewidth]{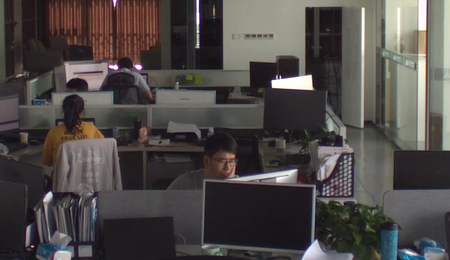}
    \label{fig:cheng:global}
    }
    \quad
    \subfigure[RefSR stitching result]{
    \includegraphics[width=0.455\linewidth]{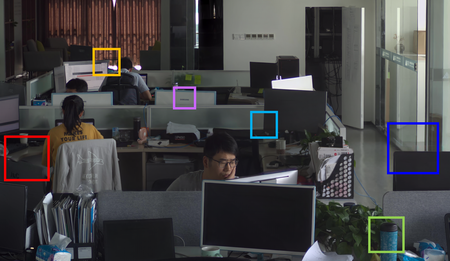}
    \label{fig:cheng:stitched}
    }
    \\
    \subfigure[De-parallax capability]{
    \begin{tabular}{c}
        \includegraphics[width=0.14\linewidth]{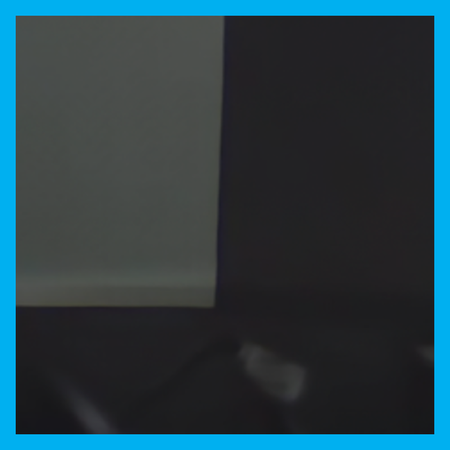} ~
        \includegraphics[width=0.14\linewidth]{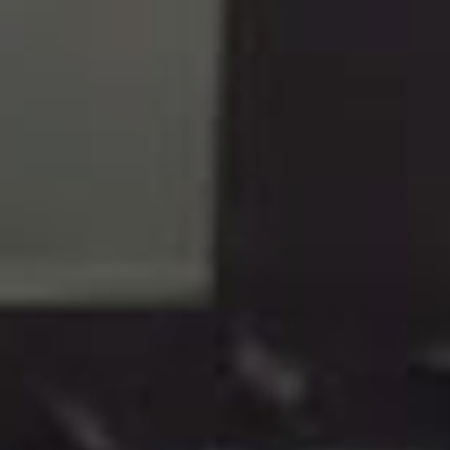} ~
        \includegraphics[width=0.14\linewidth]{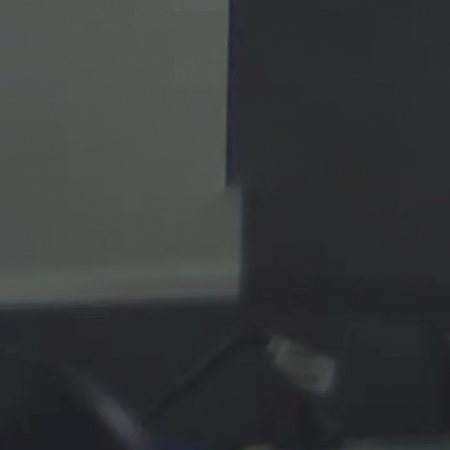}
        \quad 
        \includegraphics[width=0.14\linewidth]{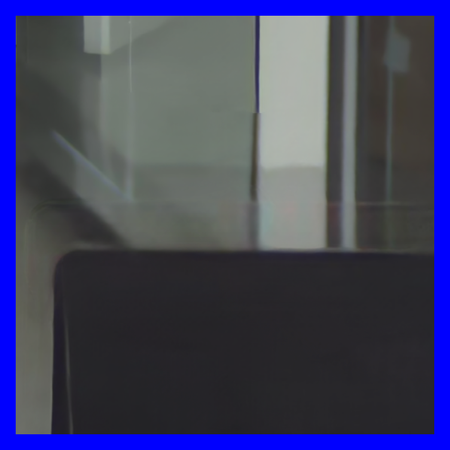} ~
        \includegraphics[width=0.14\linewidth]{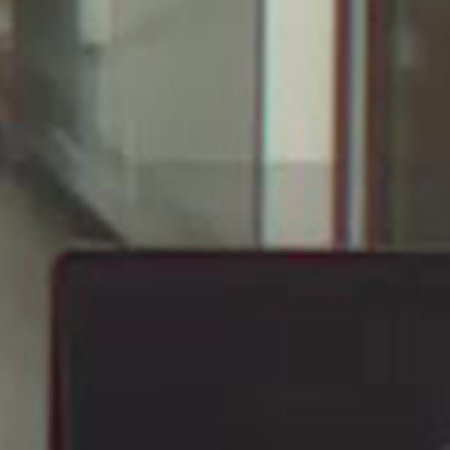} ~
        \includegraphics[width=0.14\linewidth]{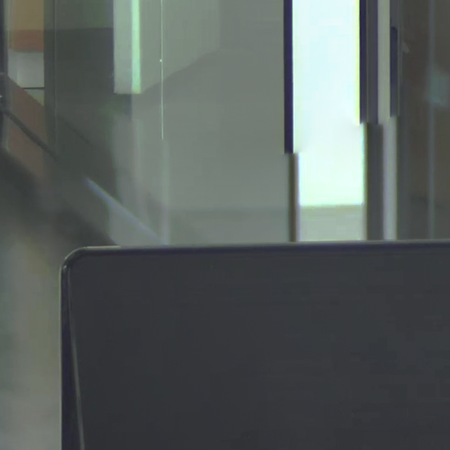}
        \\
        {\footnotesize(left to right in one set: RefSR stitching result, global image, conventional stitching result)}
    \end{tabular}
    \label{fig:cheng:parallax}
    }
    \\
    \subfigure[Recovered details]{
    \begin{tabular}{c}
        \includegraphics[width=0.14\linewidth]{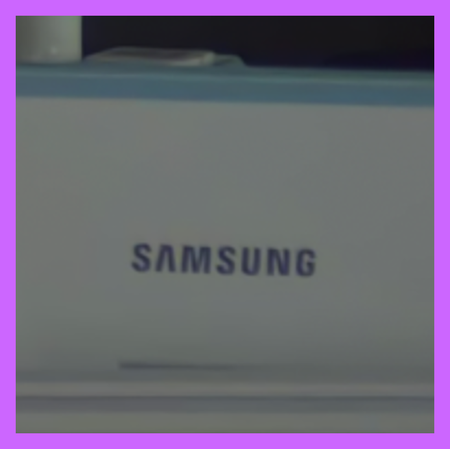} ~
        \includegraphics[width=0.14\linewidth]{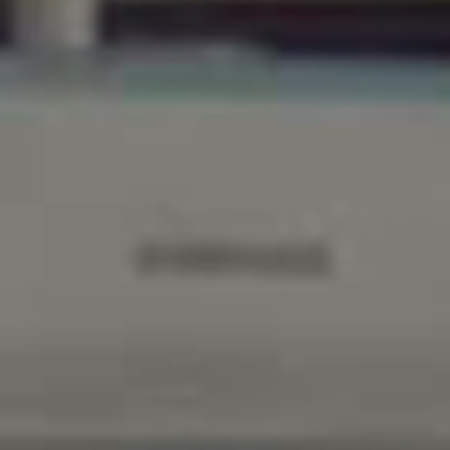} ~
        \includegraphics[width=0.14\linewidth]{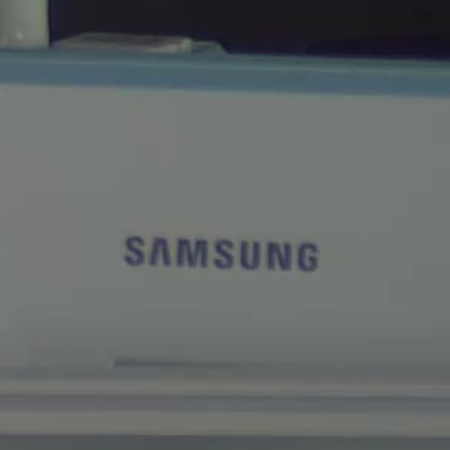}
        \quad
        \includegraphics[width=0.14\linewidth]{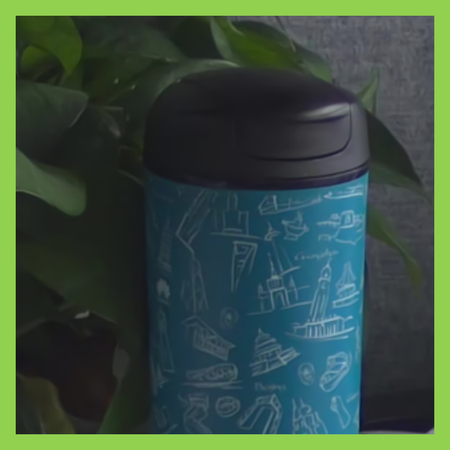} ~
        \includegraphics[width=0.14\linewidth]{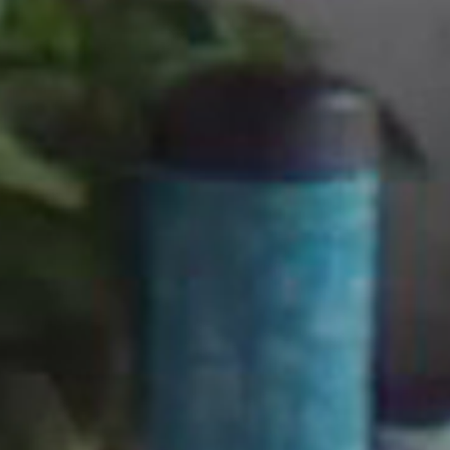} ~
        \includegraphics[width=0.14\linewidth]{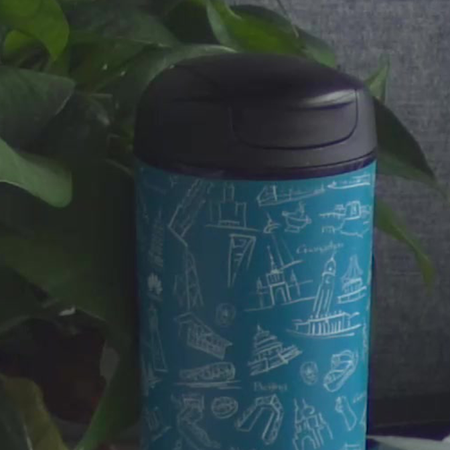}
        \\
        {\footnotesize(left to right in one set: RefSR stitching result, global image, local image)}
    \end{tabular}
    \label{fig:cheng:detail}
    }
    \\
    \subfigure[Defects]{
    \begin{tabular}{c}
        \includegraphics[width=0.14\linewidth]{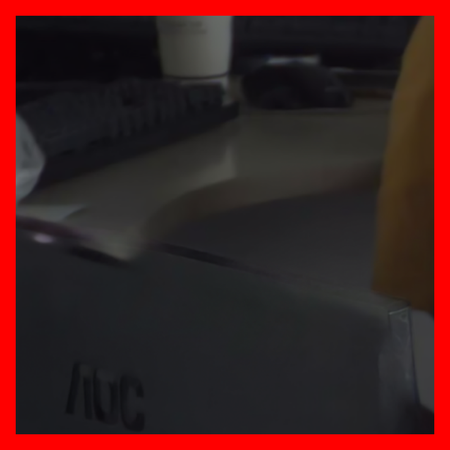} ~
        \includegraphics[width=0.14\linewidth]{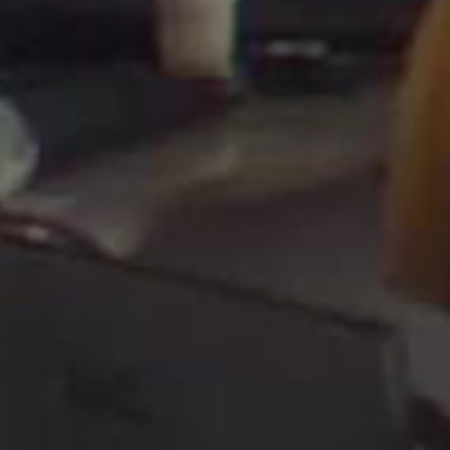} ~
        \includegraphics[width=0.14\linewidth]{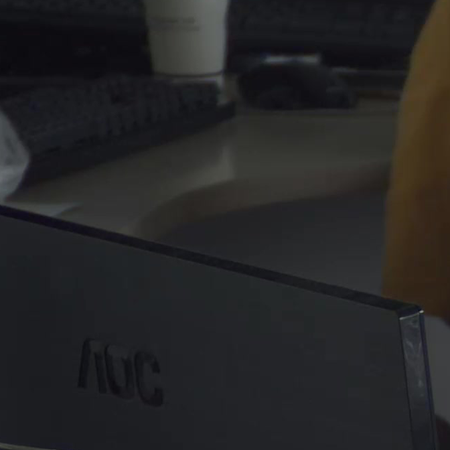}
        \quad 
        \includegraphics[width=0.14\linewidth]{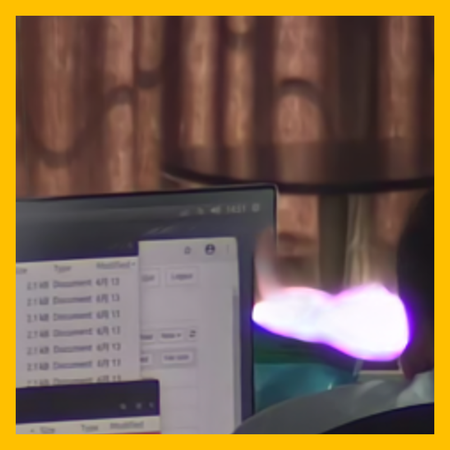} ~
        \includegraphics[width=0.14\linewidth]{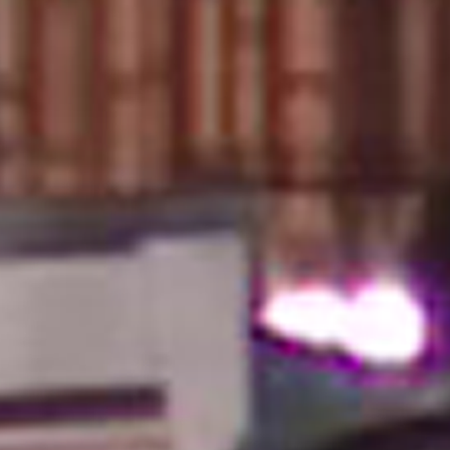} ~
        \includegraphics[width=0.14\linewidth]{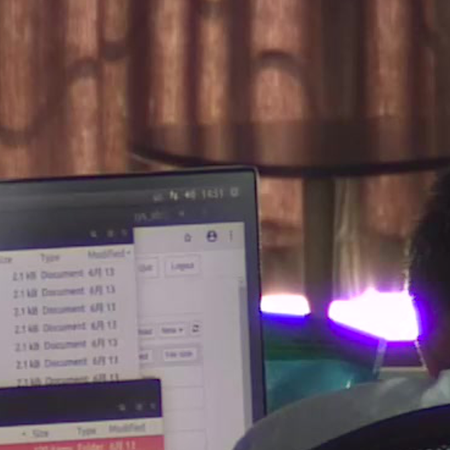}
        \\
        {\footnotesize(left to right in one set: RefSR stitching result, global image, local image)}
    \end{tabular}
    \label{fig:cheng:defect}
    }
    \caption{An demonstration of applying RefSR method in image stitching pipeline. AWnet \cite{cheng2019spatiltemporal} is used as the RefSR method.}
    \label{fig:cheng:joint}
\end{figure}

As mentioned above, the RefSR algorithms can  facilitate the stitching, which we call "RefSR stitching". An example of such a method was implemented using AWnet \cite{cheng2019spatiltemporal}, a powerful RefSR neural network with similar structure as Fig. \ref{fig:refsr_pipeline}. A result of the method is shown in Fig. \ref{fig:cheng:joint}. It took in two sets of inputs, the local images, Fig. \ref{fig:brown:input}, and a low-resolution global image, Fig. \ref{fig:cheng:global}. The resolution gap between the local images and the global image is six. The global image was firstly split into different patches, where each patch covered roughly the same area as a local image. Then all patch-local pairs were sent through the AWnet, generating super-resolved patches. These patches were fused to give the stitching result, as shown in Fig. \ref{fig:cheng:stitched}. Some zoom-in patches are shown for details with color boxes denoting their location. Fig. \ref{fig:cheng:parallax} shows the de-parallax capability of RefSR stitching. 
We can easily notice the discontinuous edges in conventional stitching results, and RefSR stitching follows the geometry of the global image, fixing that problem. In Fig. \ref{fig:cheng:detail}, RefSR stitching shows the capability to preserve textures in local images. However, RefSR stitching has its defects, which is shown in Fig. \ref{fig:cheng:defect}. Distortion and blur may occur, especially in disoccluded regions.

\newcommand{\figww}{0.15}
\begin{figure}[htbp]
  \centering
  \footnotesize
  \begin{tabular}[\figww\linewidth]{c @{\hspace{3pt}} c @{\hspace{3pt}} c @{\hspace{3pt}}c @{\hspace{3pt}}c @{\hspace{3pt}}c}
    \subfigure{\stackunder[3pt]{\includegraphics[width=\figww\linewidth]{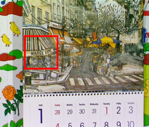}}{Calendar}}
    & \subfigure{\stackunder[3pt]{\includegraphics[width=\figww\linewidth]{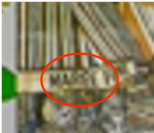}}{Bicubic}} 
    & \subfigure{\stackunder[3pt]{\includegraphics[width=\figww\linewidth]{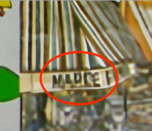}}{DBPN\cite{haris2018deep}}} 
    & \subfigure{\stackunder[3pt]{\includegraphics[width=\figww\linewidth]{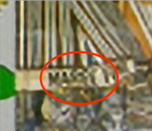}}{VSR\cite{kappeler2016video}}} 
    & \subfigure{\stackunder[3pt]{\includegraphics[width=\figww\linewidth]{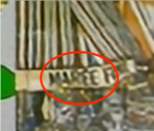}}{VESPC\cite{caballero2017real}}}
    & \subfigure{\stackunder[3pt]{
    \includegraphics[width=\figww\linewidth]{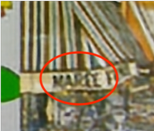}}{$B_{123}$+T\cite{liu2017robust}}}
      \\
       &\subfigure{\stackunder[3pt]{\includegraphics[width=\figww\linewidth]{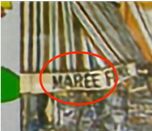}} {DRDV\cite{tao2017detail}}}
       &\subfigure{\stackunder[3pt]{\includegraphics[width=\figww\linewidth]{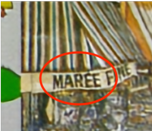}}{FRVSR\cite{sajjadi2018frame}}}
       &\subfigure{\stackunder[3pt]{\includegraphics[width=\figww\linewidth]{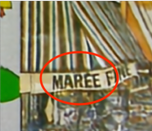}} {{VSRDU\cite{jo2018deep}}}}
       &\subfigure{\stackunder[3pt]{\includegraphics[width=\figww\linewidth]{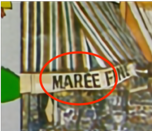}} {RBPN/6\cite{haris2019recurrent}}}
       &\subfigure{\stackunder[3pt]{\includegraphics[width=\figww\linewidth]{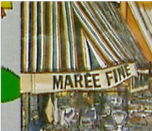}}{GT}}
  \end{tabular}{}
  \caption{Visual comparisons between different multi-frames super-resolution algorithms.}
  \label{fig:videosr_visual_comparison} 
\end{figure}

Video super-resolution (VideoSR) expands the idea of image-based RefSR in several ways. First, the target of VideoSR is to reconstruct a high-quality frame depending on a sequence of LR frames. Second, LR frames suffer from both camera motions and object motion, but there does not exist resolution gaps between them. Most existing VideoSR methods follow the pipeline of aligning multiple frames to build correspondences and then fusing image details for a high-quality output. The key to construct correspondences lies in the utilization of motion compensation, easing the difficulty of locating corresponding regions.

Similar to RefSR, most previous works~\cite{farsiu2004fast,liu2011bayesian,liao2015video,ma2015handling,kappeler2016video,caballero2017real,makansi2017end,xue2017video,tao2017detail,liu2017robust,sajjadi2018frame, hyun2018spatio, zhan2019video, haris2019recurrent } compensated inter-frame motion by estimating optic flows or applying patch-matching~\cite{takeda2009super,tian2018tdan, jo2018deep}. In 2004, Farsiu \textit{et al.}~\cite{farsiu2004fast} proposed an alternate approach using $L_1$ norm minimization and robust regularization based on a bilateral prior to deal with different data and noise models. Later, Takeda \textit{et al.}~\cite{takeda2009super} proposed an adaptive enhancement and spatio-temporal upscaling framework without explicit accurate motion estimation for handling more complex motions.  Ma \textit{et al.}~\cite{ma2015handling} proposed an EM framework to guide residual blur estimation to address ubiquitous motion blur. 
In 2015, with the success of CNNs in many vision tasks, Liao \textit{et al.}~\cite{liao2015video} and Kappeler \textit{et al.}~\cite{kappeler2016video} are the pioneers to apply CNNs into VideoSR field. Liao \textit{et al.}~\cite{liao2015video} designed a convolutional network for fast VideoSR via SR draft ensemble generation and its optimal reconstruction. Kappeler \textit{et al.}~\cite{kappeler2016video} further studied the architecture design for VideoSR task. Later, inspired by the novel flow estimation network FlowNet~\cite{dosovitskiy2015flownet} and the differentiable warping module~\cite{jaderberg2015spatial}, some studies~\cite{caballero2017real, makansi2017end, xue2017video, tao2017detail} equipped such modules to combine motion compensation and HR reconstruction via a trainable end-to-end framework to improve performance. In 2017, Liu \textit{et al.}~\cite{liu2017robust} considered the temporal receptive filed to perform multi-branches temporal aggregation for higher performances. To further enhance the visual result and efficiency, in 2018, Sajjadi \textit{et al.}~\cite{sajjadi2018frame} proposed a recurrent network that used the former inferred SR image to super-resolve the latter frame, which helped to maintain temporal consistent and reduce computation. Kim \textit{et al.}~\cite{hyun2018spatio} proposed a spatial-temporal transformer network(STTN) to capture long-range temporal dependencies and establish correspondences across several frames. In 2019, Haris \textit{et al.}~\cite{haris2019recurrent} proposed a recurrent back-projection network by adding the residual features extracted from neighbor frames to recover missing details on LR frame. In addition, some novel patch-based methods were proposed in 2018. Tian \textit{et al.}~\cite{tian2018tdan} proposed a temporal deformable alignment network (TDAN) to perform alignment adaptively in feature domain. Jo \textit{et al.}~\cite{jo2018deep} introduced a network generating dynamic upsampling filters and a residual image to reconstruct HR image without explicit motion compensation.

\subsubsection{Temporal Superresolution}
While high frame rate videos are demanded in a variety of scenarios \cite{jiang2018super,janai2017slow}, the temporal sampling rate of the sensors in commercial cameras are usually limited. Thus, video interpolation, i.e., generating new intermediate frames given a sequence of frames taken within a certain period, has attracted the attention of many researchers.

Early studies~\cite{krishnamurthy1999frame,herbst2009occlusion} mainly focused on optical flow based solutions~\cite{barron1994performance}, which first performed flow estimation and occlusion reasoning, followed by frame interpolation by sampling from the estimated flow. The performance of such methods highly depends on the flow estimation, which can be improved by techniques such as motion compensation~\cite{brox2010large, xu2011motion} and correspondences interpolation~\cite{revaud2015epicflow}. Apart from optical flow based methods, Meyer {\it et al.}~\cite{Meyer2015PhasebasedFI,meyer2018phasenet} modeled the motion of signals using phase-shift, which was efficient and robust to appearance change. However, motion artifacts, appearance change, and high-frequency detail restoration were still challenging. 

\begin{figure}[htbp]
	\centering
	\includegraphics[width=1.0\linewidth]{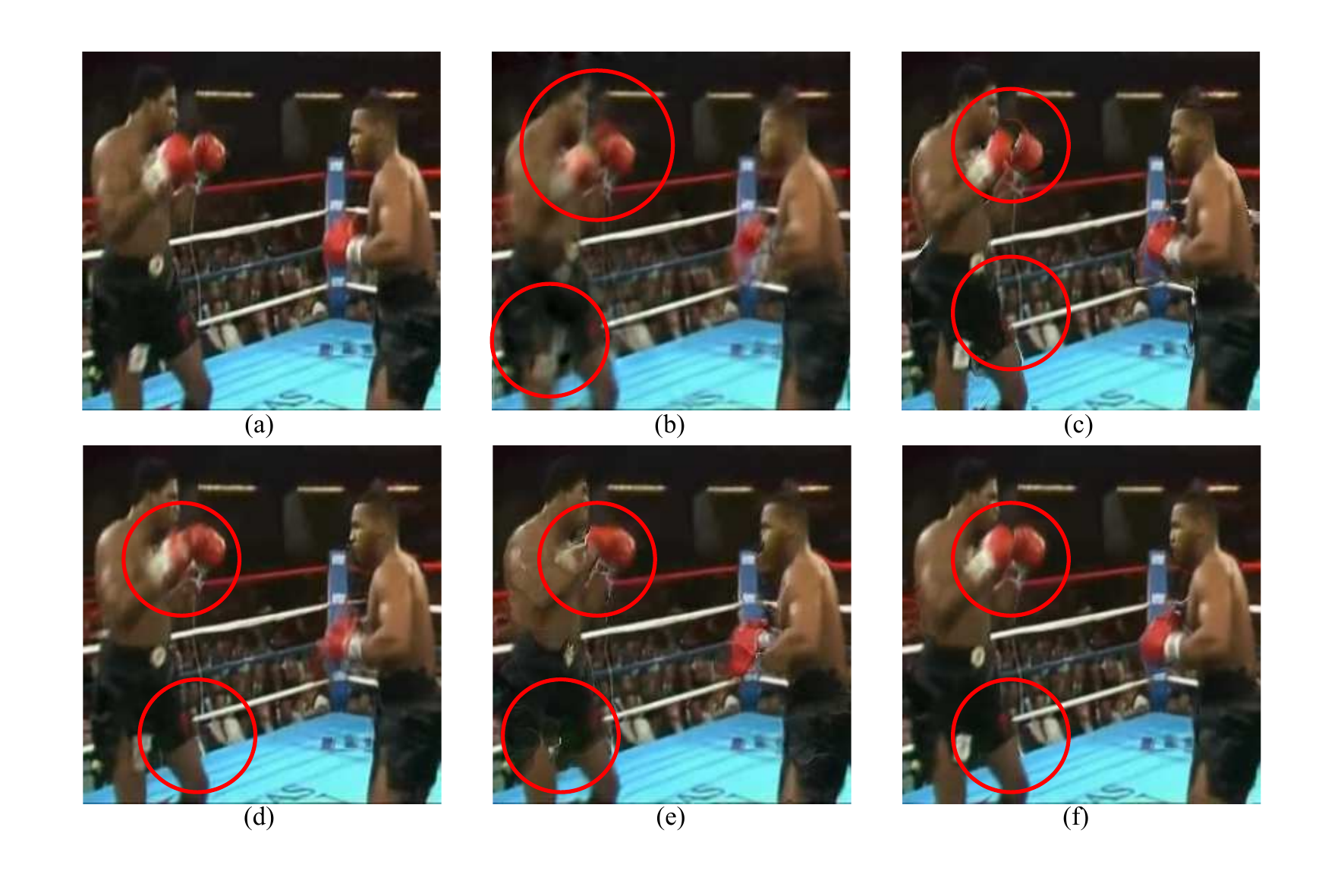}
	\caption{Visual comparison of video interpolation results generated through representative non-learning and learning based methods. (a) Ground Truth, (b) Phase-Based \cite{Meyer2015PhasebasedFI}, (c) FlowNet2 \cite{ilg2017flownet2}, (d) SepConv \cite{niklaus2017video2}, (e) DVF \cite{liu2017video}, and (f) SloMo \cite{jiang2018super}.}
	\label{fig:VideoInterpolationCMP}
\end{figure}

With the emergence of advanced deep learning technologies, many deep network architectures \cite{dosovitskiy2015flownet,ilg2017flownet2,ranjan2017optical,sun2018pwc} have been proposed to directly regress more accurate, dense and fine-grain flow, boosting the performance of optical flow based interpolation methods. \cite{liu2017video,bao2018memc} integrated motion sub-network into the architecture, enabling end-to-end training for flow estimation and occlusion reasoning. Xue {\it et al.}~\cite{xue2017video} proposed a motion representation called task-oriented flow and used mask-based synthesis to resolve occlusion. Different from the two-step method, Niklaus {\it et al.}~\cite{niklaus2017video1} skipped the intermediate flow and directly regressed spatially-adaptive convolution blur kernel for each pixel, and the efficiency is further improved by predicting separable kernels~\cite{niklaus2017video2}. These blur kernel-based methods achieved high-quality results when the motion was limited to kernel size. Regarding multiple intermediate frames interpolation, Jiang {\it et al.}~\cite{jiang2018super} used bi-directional optical flow to construct linear approximation of arbitrary diverse intermediate frames. For intuitive comparison, we show some visual results form representative methods \cite{Meyer2015PhasebasedFI,ilg2017flownet2,niklaus2017video2,liu2017video,jiang2018super} in Fig. \ref{fig:VideoInterpolationCMP}.

With neural network's ability to fuse information from multiple sources, we can go beyond single-video temporal super-resolution. In 2019, Ming {\it et al.} designed and implemented AWnet \cite{cheng2019spatiltemporal}, which could fuse a high-spatial-resolution-low-frame-rate (HSR-lFR) video and a low-spatial-resolution-high-frame-rate (LSR-HFR) video and generate a high-spatial-resolution-high-frame-rate (HSR-HFR) video. The HSR-lFR video and LSR-HFR video are first synchronized, and then the frame pairs are processed one by one. Within the AWnet, the flow from the LSR frame to the HSR frame is estimated by a FlowNet, and the HSR is warped to fit the geometry of the LSR frame. After that, a FusionNet would generate a mask and a set of dynamic filters to fuse the warped HSR frame and LSR frame. AWnet can successfully fuse a 4K@30FPS video and a 720p@240FPS video from a dual camera system composed by two iPhone 7, and the resulting 4K@240FPS video is realistic and visually plausible, surpassing both state-of-the-art video interpolation method and VideoSR method.

\subsubsection{Viewpoint Generation}
\begin{figure}[htbp]

\centering
    \subfigure[Depth \cite{chaurasia2013depth}]{
    \label{fig:synth:depth}
    \begin{tabular}{c}
        \includegraphics[width=0.2785\linewidth]{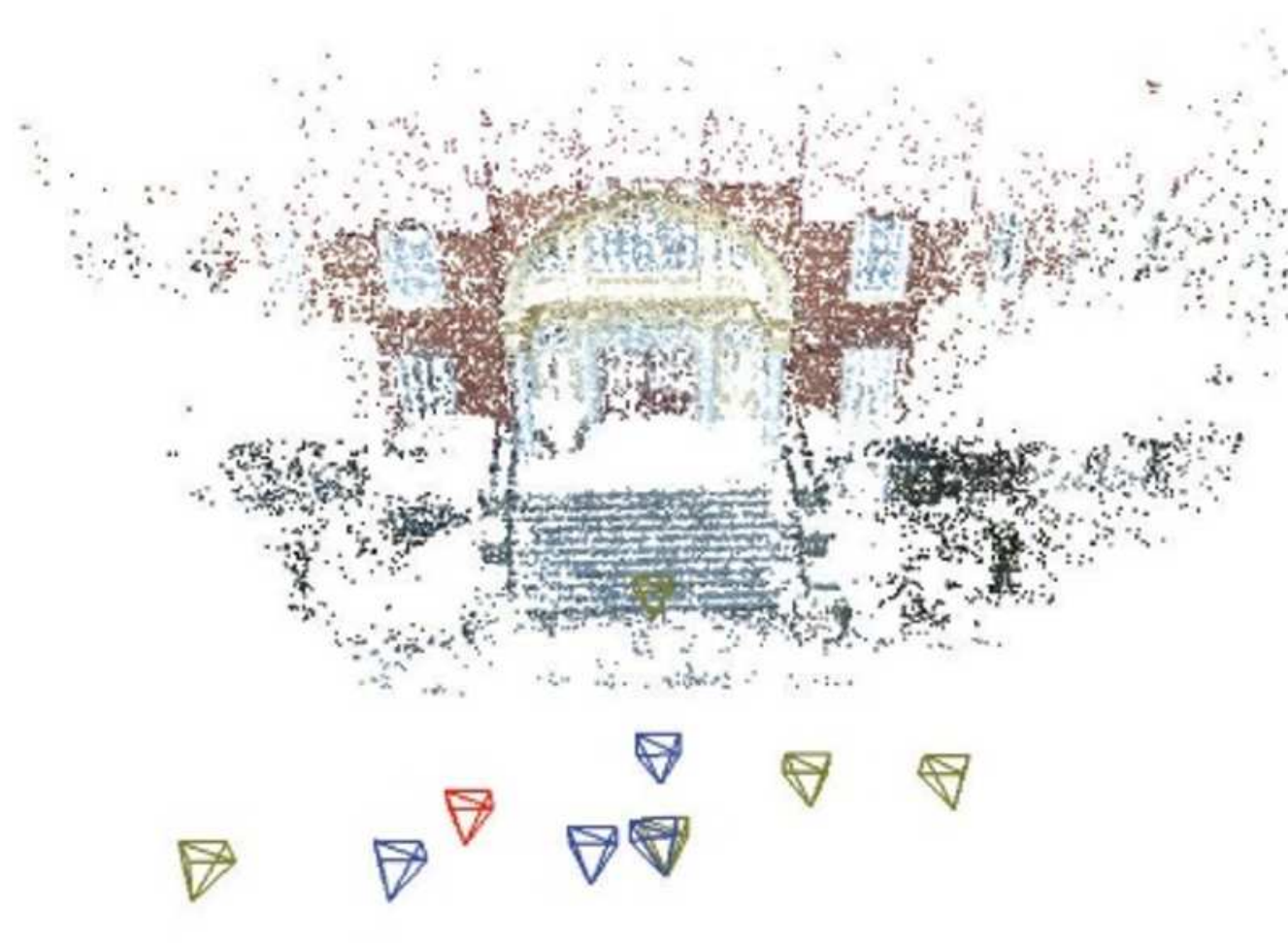} \\
        \includegraphics[width=0.2785\linewidth]{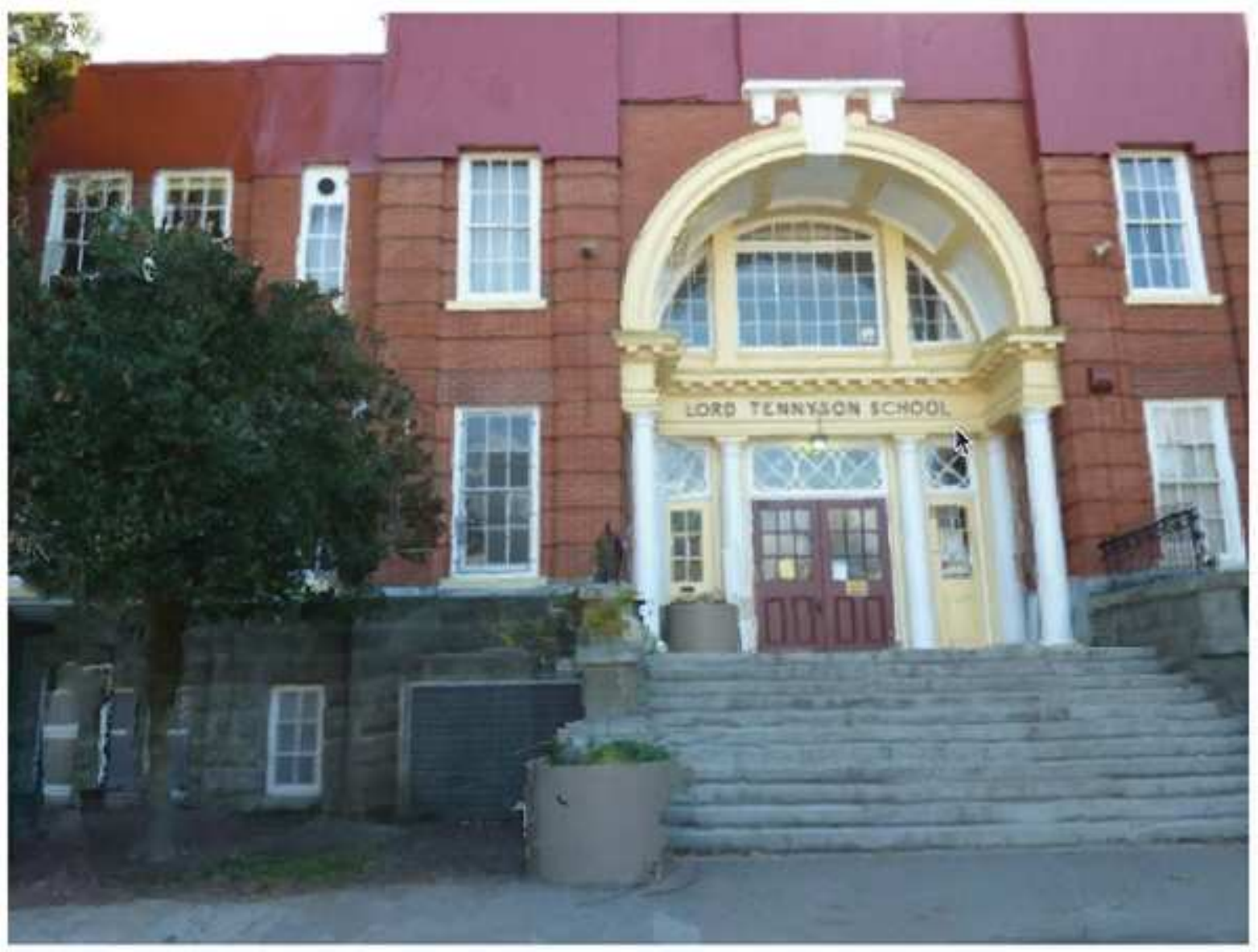}
    \end{tabular}
    }
    \subfigure[Point cloud \cite{goesele2010pointcloud}]{
    \label{fig:synth:pointcloud}
    \begin{tabular}{c}
        \includegraphics[width=0.288\linewidth]{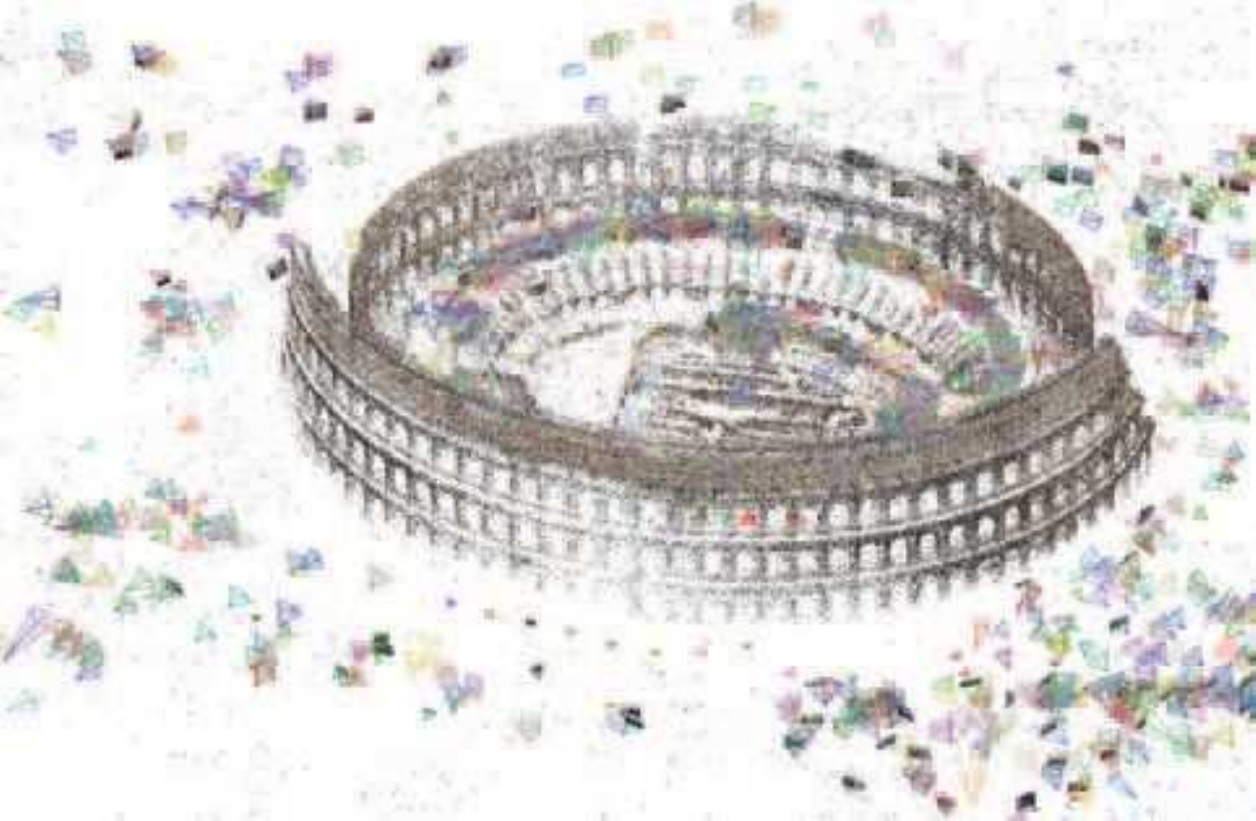} \\
        \includegraphics[width=0.288\linewidth]{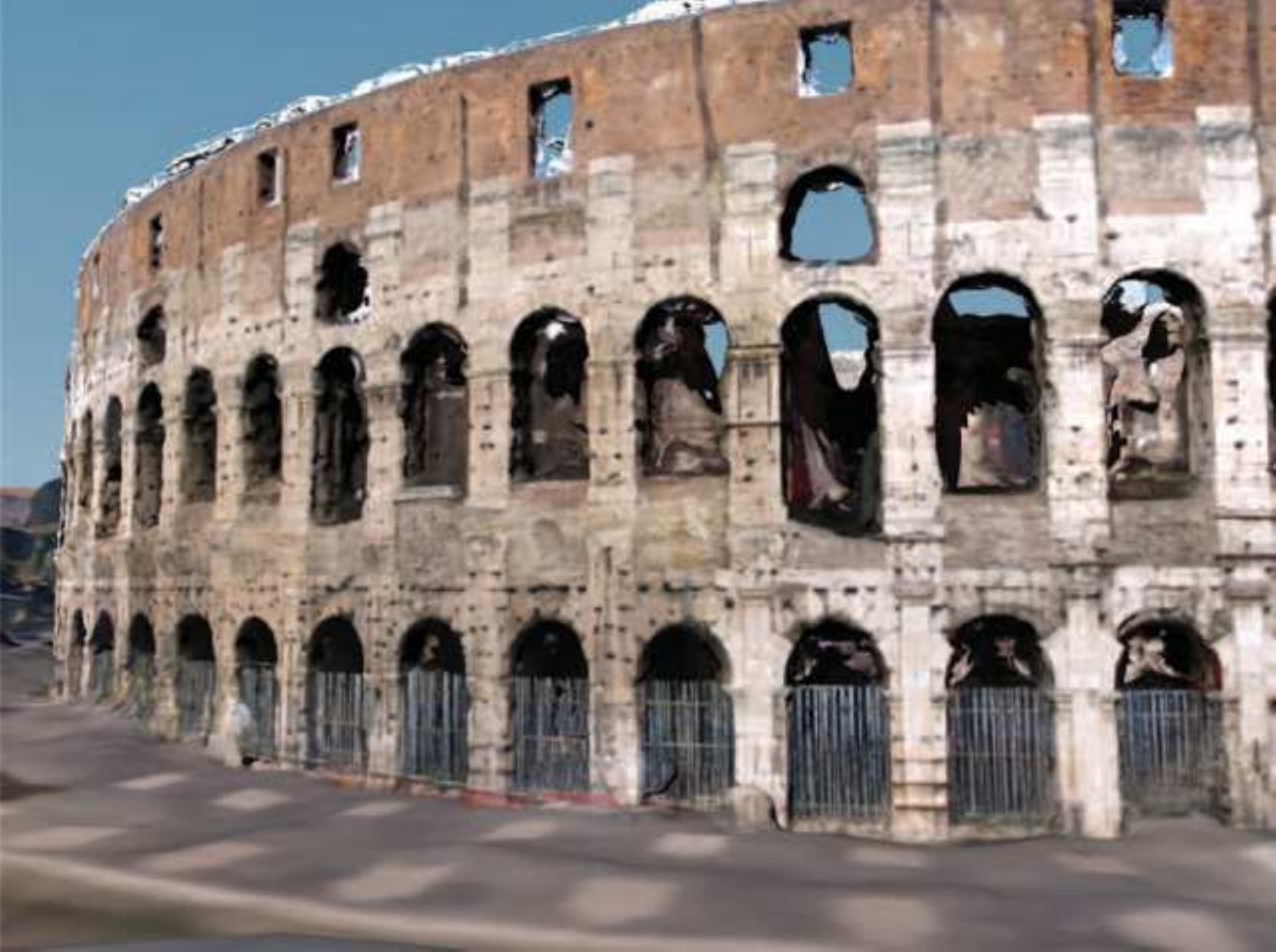}
    \end{tabular}
    }
    \subfigure[Depth from single image \cite{niklaus20193d}]{
    \label{fig:synth:sidepth}
    \begin{tabular}{c}
        \includegraphics[width=0.2835\linewidth]{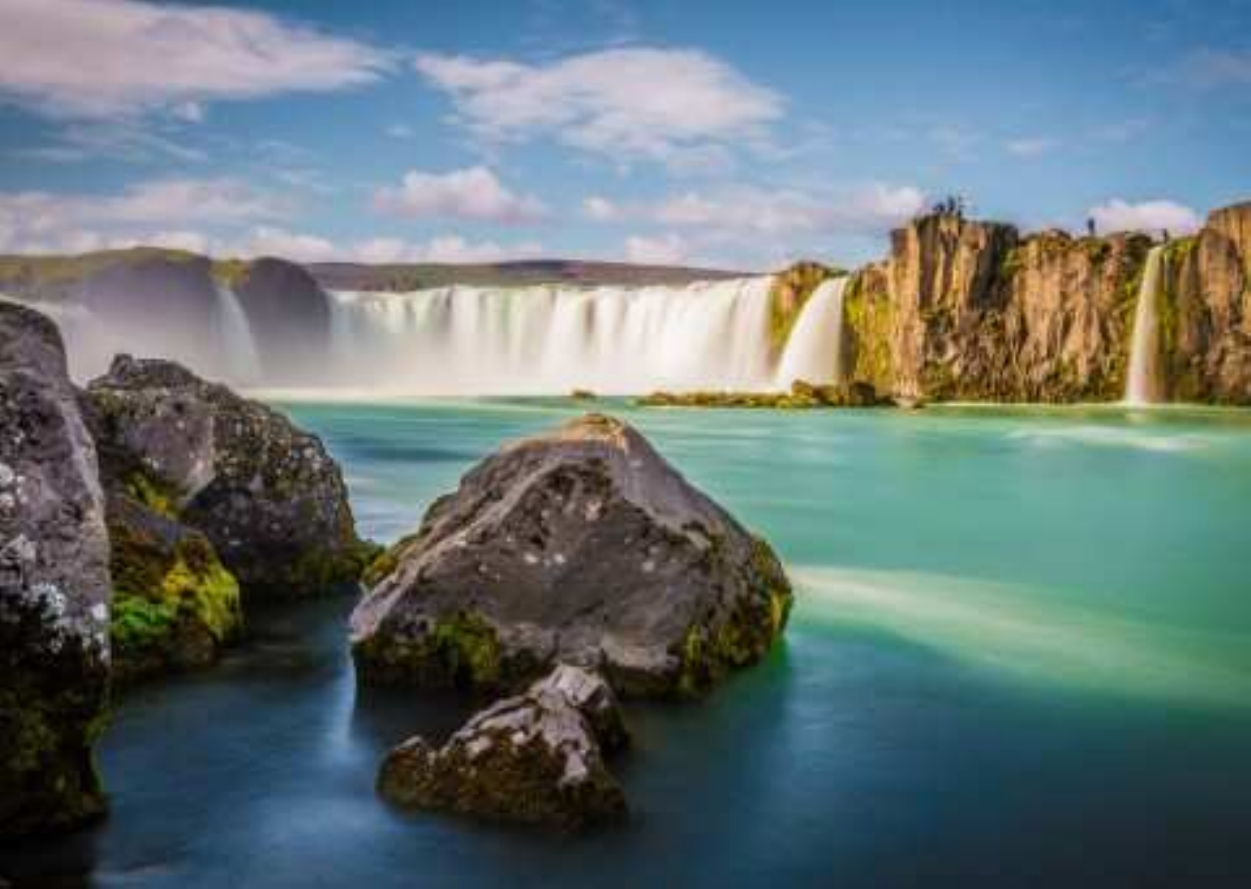} \\
        \includegraphics[width=0.2835\linewidth]{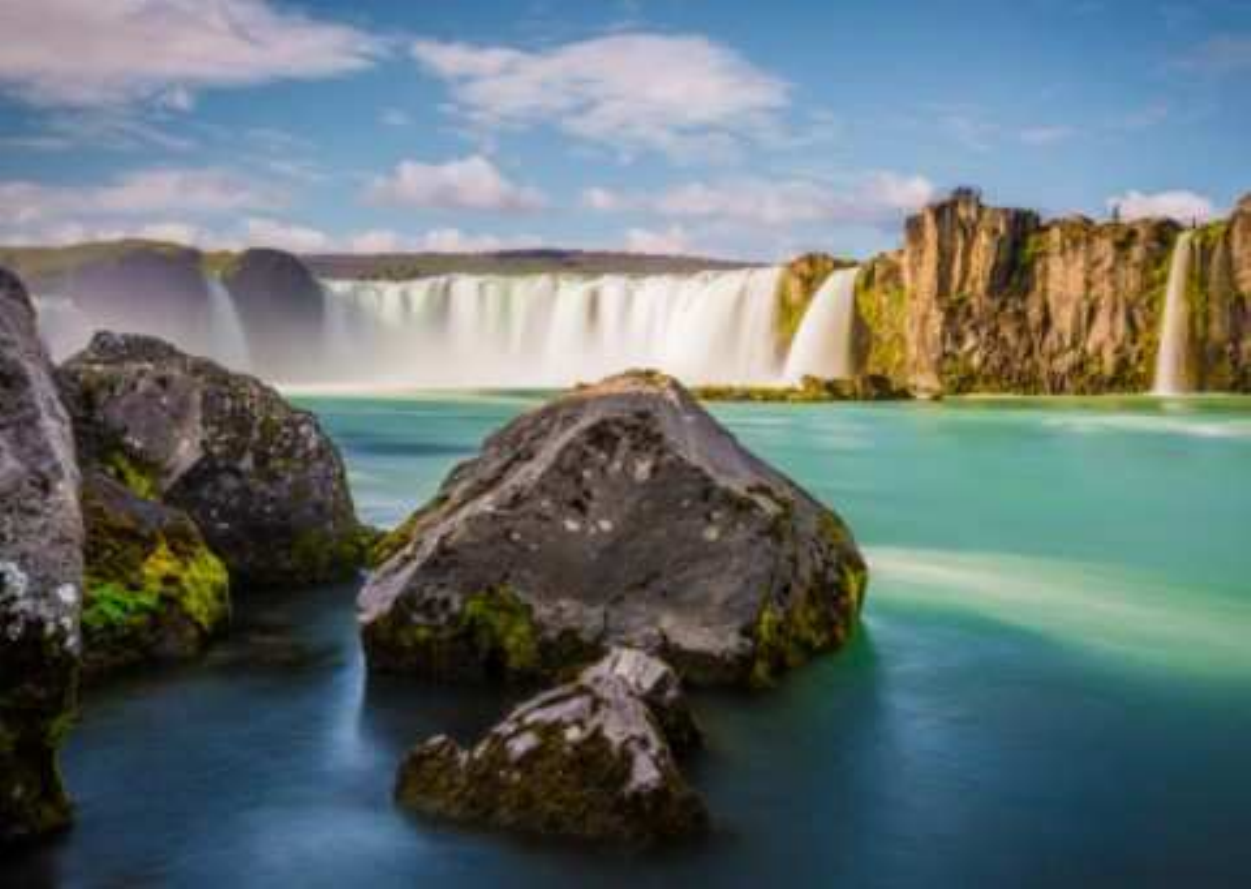}
    \end{tabular}
    }\\
    \subfigure[Light field \cite{ng2005light}]{
    \label{fig:synth:lightfield}
    \begin{tabular}{c}
        \includegraphics[width=0.2555\linewidth]{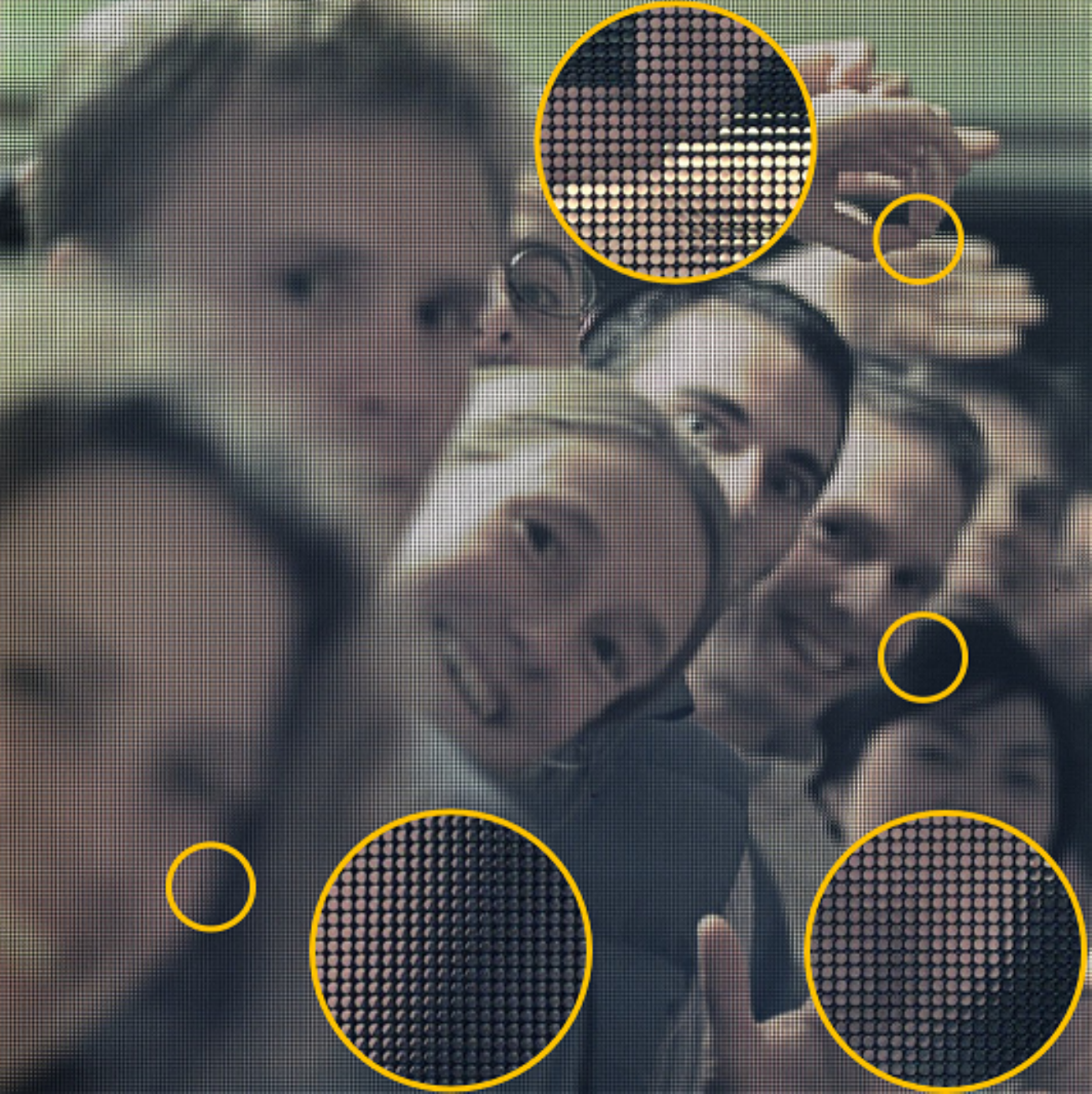} \\
        \includegraphics[width=0.2555\linewidth]{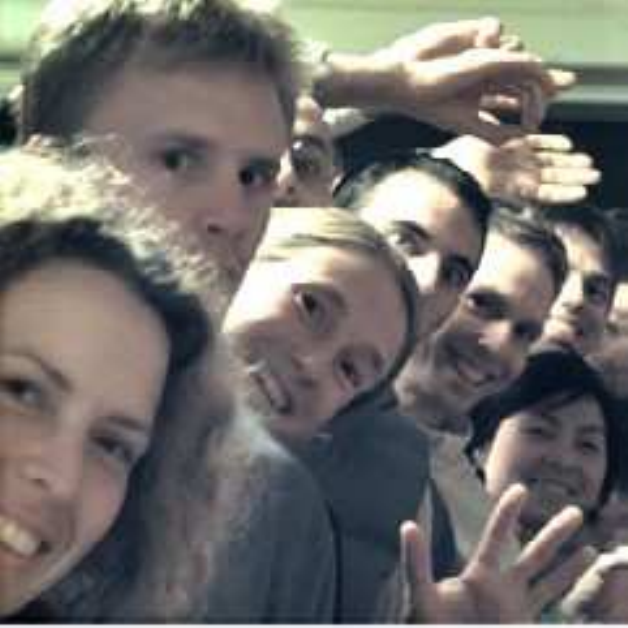}
    \end{tabular}
    }
    \subfigure[End-to-end \cite{flynn2016deepstereo}]{
    \label{fig:synth:endtoend}
    \begin{tabular}{c}
        \includegraphics[width=0.228\linewidth]{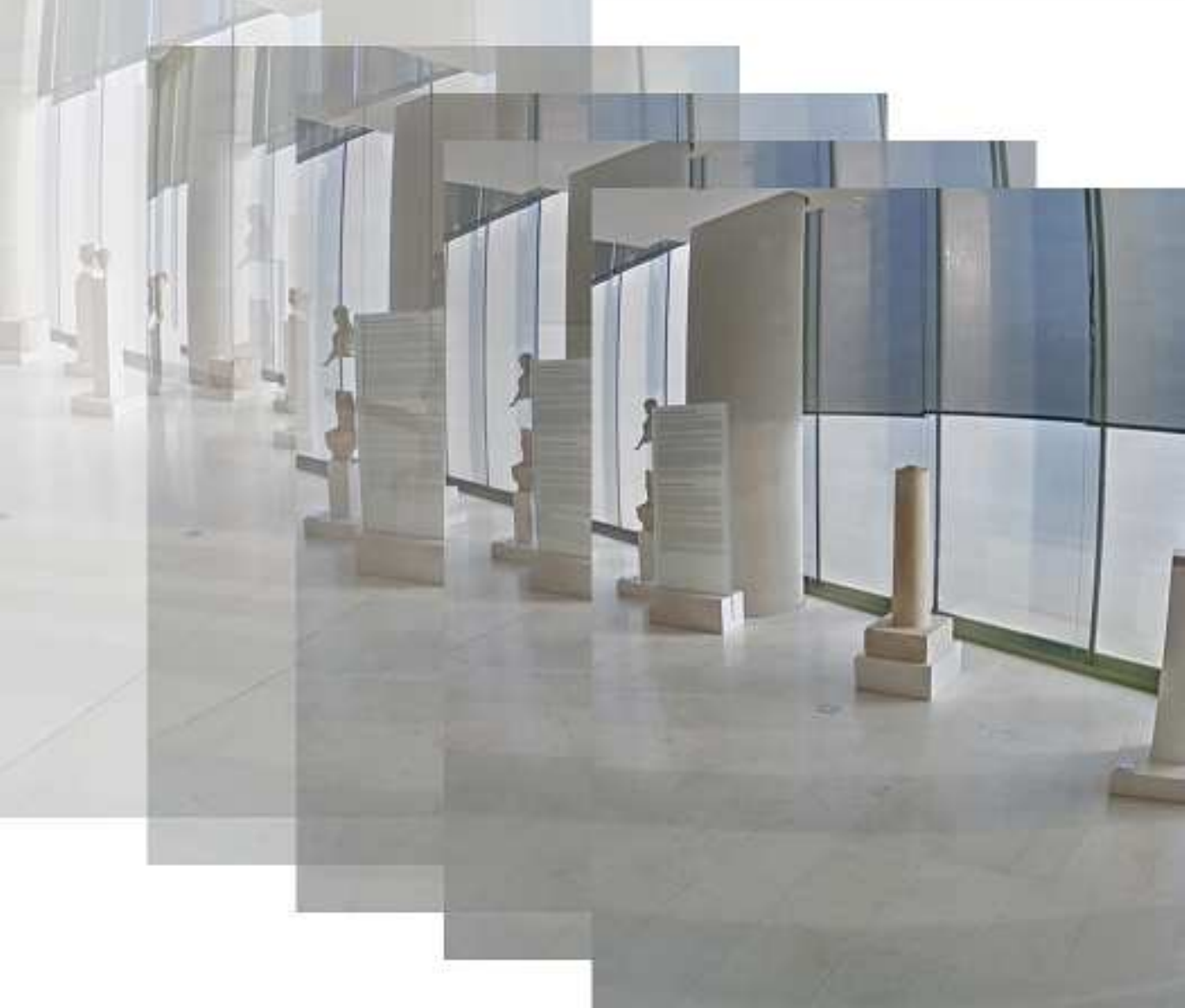} \\
        \includegraphics[width=0.228\linewidth]{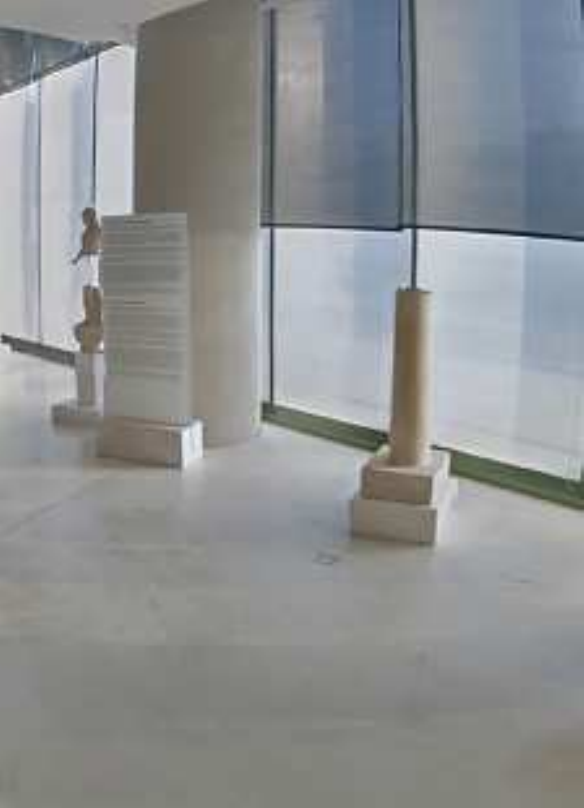}
    \end{tabular}
    }
    \subfigure[MPI \cite{flynn2019deepview}]{
    \label{fig:synth:mpi}
    \begin{tabular}{c}
        \includegraphics[width=0.3665\linewidth]{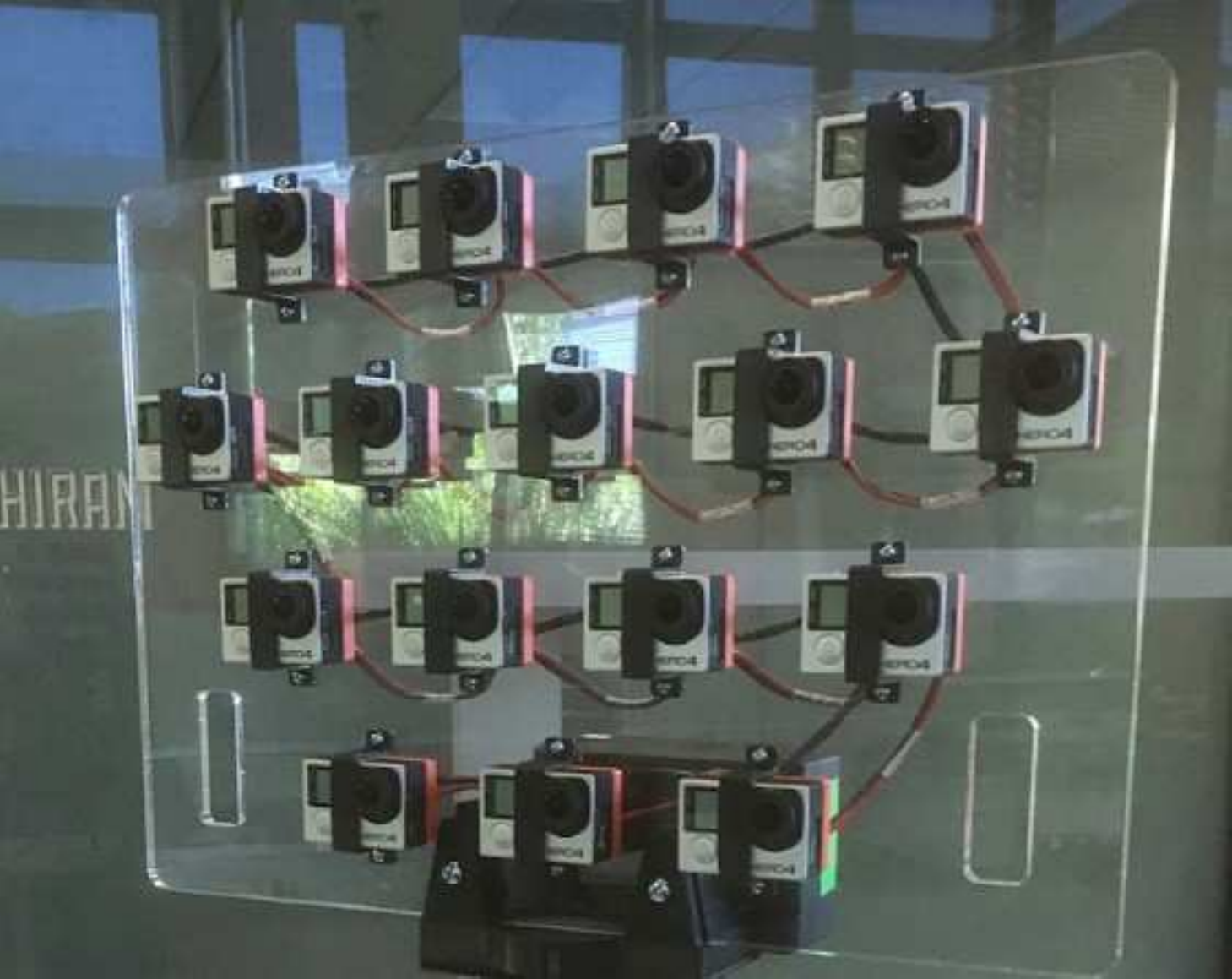} \\
        \includegraphics[width=0.3665\linewidth]{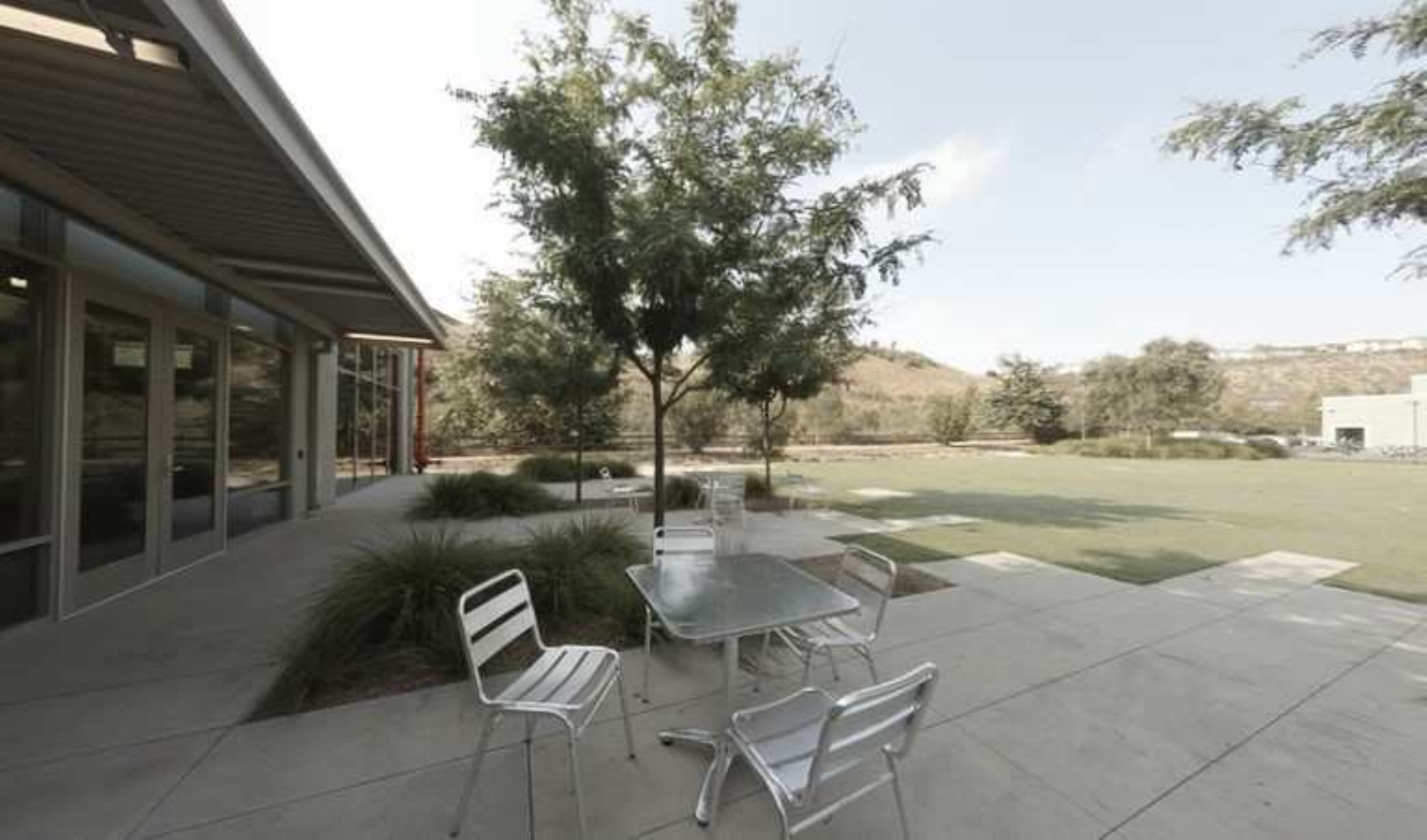}
    \end{tabular}
    }
    
    \caption{Inputs and results of some representative view synthesis algorithms. In each sub-figure, the title implies the intermediate result of that algorithm, the upper images shows the input camera array or the input images, and the lower images shows the output.}
    \label{fig:synth:joint}
\end{figure}

View synthesis is defined as rendering new views from a set of observations (views) of a scene \cite{seitz1995viewsynth}. Such an image-based rendering (IBR) approach plays vital role in applications such as virtual reality, movie taking, sport broadcasting, etc. Many studies have considered this problem, Fig. \ref{fig:synth:joint} shows the inputs and output of some representative works. Looking into these view synthesis methods, the overall trend is that learning-based methods using end to end processing are gaining popularity. 

Conventional view synthesis approaches generally rely on depth-based rendering methods. Oh \textit{et al.} rendered new view by warping image based on depth \cite{oh2009holefilling}. Chaurasia \textit{et al.} followed similar pipeline but achieved shape-preserving warping by pre-segment the image into superpixels \cite{chaurasia2013depth}. Disparity, which is closely related to depth, is also used to guide warping \cite{chen1993view, scharstein1996stereo}. In addition to simple view synthesis, some approaches use  intermediate representations, such as 3D models \cite{goesele2010pointcloud, shan2013visual} and light fields~\cite{levoy1996lightfield,ng2005light}. There are also works that render new views directly from input views without inferring intermediate results \cite{seitz1995viewsynth, seitz1996viewmorphing, fitzgibbon2005image}. However, most conventional methods take several views as input, infer an intermediate result (usually depth map) that contains 3D geometric information of the scene, and then render the new views using input views together with the intermediate result. The rendering process usually includes warping input views, fusing warped views, and filling occluded pixels. Fig. \ref{fig:synth:depth} shows the input and output of an algorithm using depth as an intermediate result \cite{chaurasia2013depth}. In the upper figure, the 4 blue camera symbols denote the viewpoint of input images, and the red symbol denotes the viewpoint of the output image. The lower figure is the synthesized view. Fig. \ref{fig:synth:pointcloud} shows the input and output of the algorithm developed by Goesele \textit{et al.} \cite{goesele2010pointcloud}. The camera symbols in the upper figure denote the viewpoint of input images. Fig. \ref{fig:synth:lightfield} shows the input and the output of the Lytro camera \cite{ng2005light}, whose algorithm utilizes the light field as intermediate result. As the upper image shows, the microlens array in front of the sensor plane forms a unique pattern, and a light field can be inferred from that input, then rendered to the target view.

As neural networks show superior performance in both high-level and low-level vision tasks \cite{ronneberger2015unet, krizhevsky2012alexnet, liu2015deep, long2015cnnseg, dosovitskiy2015flownet}, researchers increasingly rely on neural networks for view synthesis. 
Firstly, neural systems provide better tools to replace individual elements in conventional pipeline. Single-image depth estimation neural networks are introduced into the view synthesis task \cite{hoiem2005automaticpopup, saxena2008make3d}, leading to better results and wider application scenarios because stereo pairs are no longer needed. To handle the vague edges and inaccurate value in single-image-generated depth, some conventional techniques like superpixel are still used in the pipeline to maintain object integrity. Meanwhile, neural networks' capacity for image inpainting \cite{xie2012inpainting,yang2017inpainting} is also used in view synthesis pipeline. Lim {\it et al.}~\cite{lim2016holefilling} use a neural network to do complementary inpainting, handling the occluded regions in warped image.

Moreover, Niklaus \textit{et al.} introduce multiple neural networks into the image synthesis pipeline \cite{niklaus20193d}. The input image is first fed to a complex depth estimating network, which contains multiple refinement processes and explicitly uses semantic information. With refinement and the aid of semantic information, the network makes sharper edges and preserves object shape. Then the estimated depth and original input image turns into a point cloud, which goes through a color- and depth- inpainting network, filling the missing areas. The inpainted point cloud is then rendered to new views, which are more visually satisfying than previous results. A sample input and output pair is shown in Fig. \ref{fig:synth:sidepth}, where the input and output are the upper and lower image, respectively.

Neural networks can also enhance methods with intermediate results other than depth map. StereoMagnification \cite{zhou2018stereomagnification} and DeepView \cite{flynn2019deepview} use multiplane images (MPI) as the intermediate results. The MPI is generated by a neural network, with a stereo pair (StereoMagnification) or a set of rectified images (DeepView) as input, and the MPI can form the synthesized view with simple and back-propagatable math operation. This property enables the researchers to train these networks in an end-to-end manner, which is different from \cite{niklaus20193d}, in which the depth estimation network and the inpainting network are trained separately, using a computer generated dataset providing ground-truth depth maps. The end-to-end training makes it easier to build field shot training sets, which may works better when it's hard to generate simulated scenario with computer graphic techniques. The input camera array and one rendered view are shown in Fig. \ref{fig:synth:mpi}. 

Lastly, end-to-end neural view synthesis methods with no intermediate results is also  possible. DeepStereo \cite{flynn2016deepstereo} takes in several input images and generates a synthesized view directly, without introducing handcrafted features or intermediate representations. DeepStereo shows promising results, but it is not flexible as these learning-based method mentioned above, for the view point of the output view is fixed for one network. Further researches can be done to eliminate this drawback. Fig. \ref{fig:synth:endtoend} shows the input images and a rendered view.

\section{Conclusion}
\label{sec:conc}

The history of imaging technology is punctuated by the tension between analog and digital processing. Since it is impossible to obtain the sharp high frequency sensitivity of coherent focus without a lens, analog processing will forever be a central part of optical cameras. As computational power and algorithms improve, however, the continuing trend is for electronic computation to play an increasing large role in optical imaging. In modern cameras electronic components already tend to be larger in size, weight, power and cost than optics, the use of AI to reduce electronic SWaP and increase image quality is the major opportunity to improve cameras. 

We refer to "electronic computation" rather than "digital processing" here (1) in recognition that the the structure of electronic processors in the form of massively parallel graphical processing units (GPU) and tensor processing units (TPU) components has played a central role in the emergence of deep learning-based image processing and (2) in anticipation that quasi-analog physical neural networks may yet play a future role in image processing. Whatever form the processing system takes, however, simplified compressive sampling strategies and delayed image processing as discussed in \ref{sec:ds} seems likely to play a key role in reducing electronics SWaP and enabling ever growing capacity for image data capture.

As advanced computational coding, control and estimation algorithms emerge, the optimal optical structure of cameras must evolve. Physical sampling strategies, such as focus and exposure control, color filtering and temporal sampling are based partly on maximizing captured data quality and partly on minimizing post capture computational loads. While we have seen here that AI technologies are increasing adept at fusing multiple aperture and multiframe data, one may reasonably wonder why not simply stay with traditional single aperture interlaced sensor designs. The faults in the traditional design include physical optics and mechanical limitations. From an optical perspective, it is extremely challenging to design large-aperture, wide-field, chormatically-corrected diffraction-limited lenses. Breaking the optical system in to parallel microcameras dramatically simplifies lens design and enables piece-wise diffraction limited performance over arbitraryly large aperture size, field of view and spectrum. Micro camera arrays also enable the use of small aperture focal mechanisms that are substantially less expensive and faster than traditional systems. Robust multiaperture fusion also calls into question traditional interlaced sampling as in the Bayer color filter array~\cite{bayer1976color} or the 
single aperture light field camera~\cite{ng2005light}. Optimization of optical sampling specific to a given field, time and spectrum enables camera design may move away from interlaced sampling strategies. Already, numerous camera systems utilize dual sensors combining a clear monochrome system for low light measurement and an RGB filtered system for color. As multiaperture fusion continues to improve, such systems might be replaced with monochrome, blue and red systems in which the optics and sensor geometry of the color channels can be independently optimized. Similarly, multiapeture light field systems in which the focus and pointing of each subsensor is independently controlled are more attractive than interlaced light field cameras when neural processing is used for image data fusion. 

Despite the continuing improvements in computational capacity, careful consideration of processing strategy remains central to camera design. Where, for example, one might naively assume that a system might use camera data to estimate the physical scene and then re-project the physical scene to estimate new view points, {\it ad hoc} intermediate states such as multiplane images have become essential components of smart camera data processing. Since physical scene estimation is not a necessary component of most camera data visualization and exploitation applications, we anticipate that novel data structures and visualization strategies will remain a central component of high performance cameras. 

In just the last five years, for the first time ever, practical camera systems have begun to move away from the physical image reporting and toward computational scene estimation. Of course, computational image formation opens the door to "fake" images based on generative networks~\cite{isola2017image} and may require AI to ensure image fidelity~\cite{guera2018deepfake}, but assuming designers stay mostly honest computational image formation offers potential for revolutionary improvements in image quality and information rates as described above. While processing platforms for neural image processing will continue to rapidly improve and new sampling strategies and network architectures will emerge, we certainly anticipate that the transition from physical to computational image formation is permanent and will expand on the basic strategies outlined in this review.

\bibliography{refs}

\end{document}